\documentclass[lettersize,journal]{IEEEtran}
\usepackage{booktabs}
\usepackage{diagbox}
\usepackage{makecell}
\usepackage{array}
\usepackage{graphicx,amssymb,amsmath}
\usepackage{multicol}
\usepackage[noadjust]{cite} 
\usepackage[justification=centering]{caption}
\usepackage[font=small]{caption}
\usepackage{setspace}
\usepackage{subfigure}
\usepackage{graphicx}
\usepackage{float}
\usepackage {url}
\usepackage{stfloats}
\usepackage[normal]{threeparttable}
\usepackage{amsthm,pifont}
\usepackage{flushend}
\usepackage{cases,subeqnarray}
\usepackage{bm,multirow,bigstrut}
\usepackage{amsmath, amsthm, amssymb}
\usepackage{textcomp}
\usepackage{latexsym,bm}
\usepackage{booktabs}
\usepackage{mathtools}
\usepackage{dsfont}
\usepackage{extarrows}
\usepackage{epsfig}
\usepackage{epsfig}
\usepackage{epstopdf}
\usepackage[table]{xcolor}
\usepackage{colortbl} 
\usepackage{multirow, hhline}
\usepackage{xcolor}
\usepackage{array} 
\usepackage{algorithmicx,algorithm}
\usepackage{hyperref}
\usepackage{tikz}
\usepackage{makecell}
\usepackage{enumitem}
\usepackage{comment}
\usepackage{pifont}
\usepackage{graphicx}
\usepackage{tabularray}
\usetikzlibrary{shapes.geometric, arrows, positioning}
    
\theoremstyle{plain}

\theoremstyle{plain}

\usepackage{amsmath}

\usepackage[edges]{forest}
\usetikzlibrary{shadows}

\usetikzlibrary{shapes.geometric, arrows}

\definecolor{lightblue}{rgb}{0.68, 0.85, 0.90}
\definecolor{lightgreen}{rgb}{0.56, 0.93, 0.56}
\definecolor{lightpurple}{rgb}{0.75, 0.58, 0.92}

\tikzstyle{box} = [rectangle, rounded corners, minimum width=3cm, minimum height=1cm,text centered, draw=black, fill=lightblue]
\tikzstyle{boxgreen} = [rectangle, rounded corners, minimum width=3cm, minimum height=1cm,text centered, draw=black, fill=lightgreen]
\tikzstyle{boxpurple} = [rectangle, rounded corners, minimum width=3cm, minimum height=1cm,text centered, draw=black, fill=lightpurple]
\tikzstyle{line} = [draw, -latex]

\newcommand{\redcircle}{\tikz[baseline=-0.5ex] \draw[gray,fill=red,thin] (0,0) circle (0.7ex);} 
\newcommand{\greencircle}{\tikz[baseline=-0.5ex] \draw[gray,fill=green,thin] (0,0) circle (0.7ex);}

\IEEEoverridecommandlockouts
\begin{document}
\title{ Mixture of Experts for Decentralized Generative AI and Reinforcement Learning in Wireless Networks: A Comprehensive Survey}

\author{
Yunting Xu, Jiacheng Wang, Ruichen Zhang, Changyuan Zhao, Dusit Niyato,~\IEEEmembership{Fellow,~IEEE}, Jiawen Kang, \\ Zehui Xiong, Bo Qian, Haibo Zhou~\IEEEmembership{Fellow,~IEEE}, Shiwen Mao,~\IEEEmembership{Fellow,~IEEE}, Abbas Jamalipour,~\IEEEmembership{Fellow,~IEEE}, \\ 
Xuemin Shen,~\IEEEmembership{Fellow,~IEEE}, Dong In Kim,~\IEEEmembership{Life Fellow,~IEEE}
 
\thanks{Y. Xu, J. Wang, R. Zhang, C. Zhao, and D. Niyato are  with the College of Computing and Data Science, Nanyang Technological University, Singapore, 639798 (e-mail: yunting.xu@ntu.edu.sg, jiacheng.wang@ntu.edu.sg, ruichen.zhang@ntu.edu.sg, zhao0441@e.ntu.edu.sg, dniyato@ntu.edu.sg).}
\thanks{J. Kang is with the School of Automation, Guangdong University of Technology, Guangzhou, China, 510006 (e-mail: kavinkang@gdut.edu.cn).}
\thanks{Z. Xiong is with the School of Electronics, Electrical Engineering and Computer Science (EEECS), Queen's University Belfast, Belfast, BT7 1NN, U.K. (z.xiong@qub.ac.uk).}
\thanks{B. Qian is with the Information Systems Architecture Science Research Division, National Institute of Informatics, Tokyo 101-8430, Japan (e-mail: boqian@ieee.org).}
\thanks{H. Zhou (Corresponding Author) is with the School of Electronic Science and Engineering, Nanjing University, Nanjing, China, 210023 (e-mail: haibozhou@nju.edu.cn).}
\thanks{S. Mao is with the Department of Electrical and Computer Engineering, Auburn University, Auburn, USA (e-mail: smao@ieee.org).}
\thanks{A. Jamalipour is with the School of Electrical and Computer Engineering, University of Sydney, Australia (e-mail: a.jamalipour@ieee.org).}
\thanks{X. Shen is with the Department of Electrical and Computer Engineering, University of Waterloo, Canada (email: sshen@uwaterloo.ca).}
\thanks{D. I. Kim is with the Department of Electrical and Computer Engineering, Sungkyunkwan University, Suwon 16419, South Korea (email: dongin@skku.edu).}

}

\maketitle
\begin{abstract}
Mixture of Experts (MoE) has emerged as a promising paradigm for scaling model capacity while preserving computational efficiency, particularly in large-scale machine learning architectures such as large language models (LLMs).
Recent advances in MoE have facilitated its adoption in wireless networks to address the increasing complexity and heterogeneity of modern communication systems.
This paper presents a comprehensive survey of the MoE framework in wireless networks, highlighting its potential in optimizing resource efficiency, improving scalability, and enhancing adaptability across diverse network tasks.
We first introduce the fundamental concepts of MoE, including various gating mechanisms and the integration with generative AI (GenAI) and reinforcement learning (RL). Subsequently, we discuss the extensive applications of MoE across critical wireless communication scenarios, such as vehicular networks, unmanned aerial vehicles (UAVs), satellite communications, heterogeneous networks, integrated sensing and communication (ISAC), and mobile edge networks. Furthermore, key applications in channel prediction, physical layer signal processing, radio resource management, network optimization, and security are thoroughly examined. Additionally, we present a detailed overview of open-source datasets that are widely used in MoE-based models to support diverse machine learning tasks. Finally, this survey identifies crucial future research directions for MoE, emphasizing the importance of advanced training techniques, resource-aware gating strategies, and deeper integration with emerging 6G technologies.

\end{abstract}

\IEEEpeerreviewmaketitle

\section{Introduction}

\subsection{Background}
In recent years, the integration of artificial intelligence (AI) and machine learning technology in wireless communication systems has undergone substantial development, notably characterized by the deployment of large generative AI (GenAI) models like Generative Pre-trained Transformer (GPT) \cite{chen2024big, achiam2023gpt, 10884770}.  As the complexity of evolutionary mobile networks escalates and the demand for enhanced wireless performance intensifies, the utilization of large GenAI models has shown significant potential in improving system efficiency and optimizing overall wireless performance \cite{friha2024llm, wang2022transformer, zhang2024handover, xu2024unleashing}.  
These GenAI models with billions of parameters possess the ability to process tremendous amounts of wireless data and extract intricate patterns of network behavior, which become instrumental in advancing technologies such as channel state information (CSI) prediction \cite{liu2024llm4cp}, orthogonal frequency division multiplexing (OFDM) symbol decoding \cite{rajagopalan2023transformers}, and transmit power optimization \cite{lee2024llm}. The employment of GenAI models has effectively addressed the increasing heterogeneity and dynamism of modern networks, facilitating more efficient resource management, capacity expansion, and enhancement of user experiences \cite{bariah2024large, xu2024large, qiu2024large}.

However, despite the substantial potential of GenAI models, their large-scale parameter size presents significant challenges for practical deployment in wireless environments. Prominent examples include LLaMA from Meta, which comprises approximately 65 billion parameters \cite{touvron2023llama}, PaLM from Google with 540 billion parameters \cite{chowdhery2023palm}, and GPT-4, which exceeds trillions of parameters \cite{baktash2023gpt}.
Such immense sizes necessitate considerable computational and storage resources, making it difficult for resource-constrained wireless infrastructures such as access points (APs), edge servers, and mobile devices to directly support models of this magnitude \cite{qu2024mobile, hadi2024large, stojkovic2024towards, zhou2024large}.
To alleviate the challenge of resource limitations while guaranteeing computational accuracy, distributed deployment and parallel computing strategies have been widely investigated for large-scale GenAI models \cite{ zhouhao2024large, sandler2018mobilenetv2, jiao2019tinybert}. Nonetheless, the distributed design often lacks cooperation among devices, leading to limited scalability and inefficiency \cite{shen2024large, lin2024efficient, schuster2022confident}. Addressing these constraints necessitates a unified framework that combines the strengths of diverse paradigms, emphasizing localized specialization and cooperative mechanisms, especially in heterogeneous and dynamic wireless environments.

\begin{table*}[ht]
\centering
\renewcommand\arraystretch{1.2}
\caption{Summary of Related Survey}
\label{table1}
{\fontsize{8.5pt}{10pt} \selectfont
\begin{tabular}{|m{1cm}| m{11.5cm}|l|l|m{1.5cm}|}
\hline
\textbf{Ref.} & \textbf{Description} & \textbf{MoE} & \textbf{GenAI} & \textbf{Wireless Networks}\\ \hline
\cite{yuksel2012twenty} & Reviews on the two-decade evolution of MoE, covering foundational concepts, key advancements, and applications in model selection and expert structure. & \ding{51} & \ding{55} & \ding{55} \\ \hline
\cite{vats2024evolution} & Traces the development of MoE, highlighting key advances and its role in deep learning, with a focus on expert specialization and optimization techniques. & \ding{51} & \ding{55} & \ding{55} \\ \hline
\cite{masoudnia2014mixture} & Categorizes MoE models into mixture of implicitly localised experts (MILE) and mixture of explicitly localised experts (MELE), discussing their advantages, challenges, and applications in regression and classification tasks. & \ding{51} & \ding{55} & \ding{55}\\ \hline
\cite{liu2024survey} & Explores optimization techniques for MoE inference, focusing on architectural, system, and hardware-level approaches to improve efficiency and scalability. & \ding{51} & \ding{55} & \ding{55} \\ \hline

\cite{mcintosh2023google} & Investigates the transformative impact of  MoE and multimodal learning on GenAI models, focusing on the advanced methodologies of MoE in enhancing GenAI. & \ding{51} & \ding{51} & \ding{55} \\ \hline
\cite{cai2024survey} & Introduces the MoE framework, categorizes MoE models, and explores the applications of different gating mechanisms in LLMs.  & \ding{51} & \ding{51} & \ding{55}\\ \hline
\cite{wan2023efficient} &  Reviews techniques for enhancing the efficiency of LLMs, focusing on model-centric, data-centric, and framework-centric methods to reduce resource demands.  & \ding{51} & \ding{51} & \ding{55} \\ \hline

\cite{van2024generative} &  Investigates the applications of GenAI in physical layer communications, such as channel estimation and signal classification, and compares GenAI’s benefits over traditional AI models.
 & \ding{55} & \ding{51} & \ding{51}\\ \hline
\cite{cao2023comprehensive} & Provides a comprehensive history and survey of AI-generated content (AIGC), focusing on GenAI models such as GANs and VAEs, and discusses their potential applications in wireless communication networks. & \ding{55} & \ding{51} & \ding{51}\\ \hline
\cite{zhouhao2024large} & Surveys the use of LLMs in telecommunications, including network optimization, traffic management, and security, as well as the integration of GenAI in sixth generation mobile networks. & \ding{55} & \ding{51}& \ding{51} \\ \hline
\cite{xu2024integration} & Explores the integration of MoE and GenAI in the Internet of Vehicles (IoVs), focusing
on traffic management, autonomous driving, and collaborative decision-making. & \ding{55} & \ding{51}& IoV \\ \hline

This survey & Focuses on the efficient training and deployment of MoE frameworks with advanced GenAI approaches for various wireless network scenarios and wireless technologies. & {\color{blue}\ding{51}} & {\color{blue}\ding{51}} & {\color{blue}\ding{51}}\\ \hline
\end{tabular}
}
\end{table*}

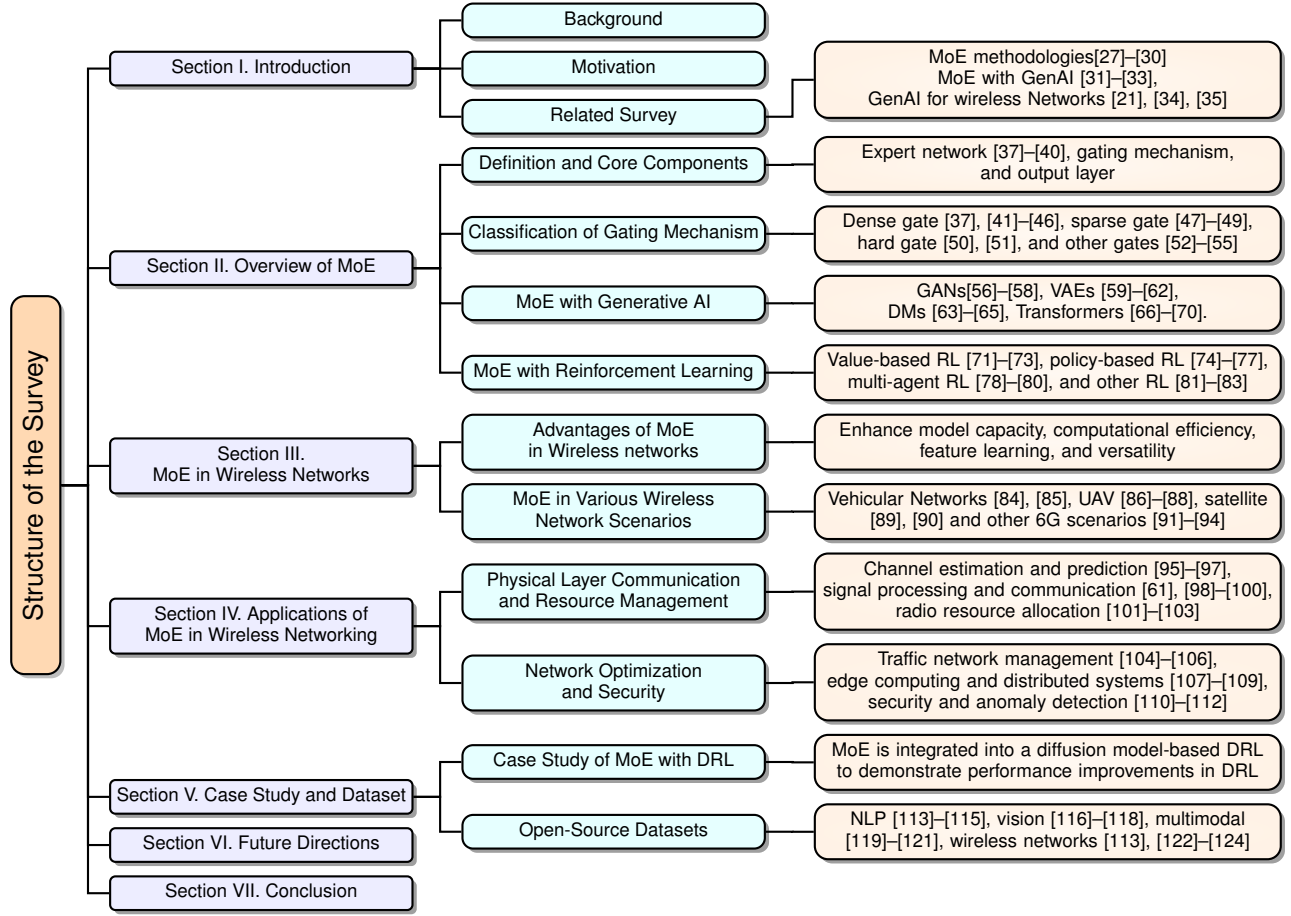
\begin{figure*}[htb]
\centering
\scriptsize
\tikzset{
    my node/.style={
        draw=black,
        thick,
        font=\sffamily,
        drop shadow,
        minimum height=0.65cm,
        minimum width=5cm,
        rounded corners=5pt, 
        fill=orange!30,
    },
    my node level 1/.style={
        my node,
        inner color=blue!7,
        outer color=blue!7,
        minimum width=5cm,
        minimum height=0.37cm,
        rounded corners=2,
    },
    my node level 2/.style={
        my node,
        inner color=cyan!10,
        outer color=cyan!10,
        minimum width=2cm,
        rounded corners=4,
        minimum height=0.37cm,
        align=center, 
    },
    my node level 3/.style={
        my node,
        inner color=orange!5,
        outer color=orange!10,
        minimum height=0.5cm,
        minimum height=0.4cm,
        minimum width=3cm,
        align=center,
    },
}
    \begin{forest}
        for tree={%
            my node,
            l sep+=10pt,
            grow'=east,
            edge={black, thick},
            parent anchor=east,
            child anchor=west,
            tier/.option=level,
            edge path={
            \noexpand\path [draw, \forestoption{edge}] (!u.parent anchor) -- +(10pt,0) |- (.child anchor)\forestoption{edge label};
            },
            if level=1{ 
                edge path={
            \noexpand\path [draw, \forestoption{edge}] (!u.south) -- +(10pt,0) |- (.child anchor)\forestoption{edge label};
            },
            }{},
            if level=1{my node level 1}{},
            if level=2{my node level 2}{},
            if level=3{my node level 3}{},
            align=center, 
            baseline, 
        }
        [\normalsize Structure of the Survey, rotate=90
        [Section I. Introduction, minimum width=4.0cm,
        [Background, minimum width=4.0cm]
        [Motivation, minimum width=4.0cm]
        [Related Survey, minimum width=4.0cm  [{ MoE methodologies\cite{yuksel2012twenty, vats2024evolution, masoudnia2014mixture, liu2024survey} \\
        MoE with GenAI \cite{mcintosh2023google, cai2024survey, wan2023efficient}, \\  GenAI for wireless Networks \cite{van2024generative, cao2023comprehensive, zhouhao2024large} 
        }, minimum width=7.0cm,  minimum width=3.0cm, xshift=-2.0cm, text width=6.0cm, yshift = 0.77cm
        ] ]
        ]
        [Section II. Overview of MoE, minimum width=4.0cm, base=midpoint, yshift = -0.07cm
        [Definition and Core Components, minimum width=4.0cm, yshift = -0.147cm
        [{Expert network \cite{collobert2001parallel, zhang2022mixture, shen2024jetmoe, li2022branch}, gating mechanism, \\ and output layer}, text width=6.0cm, xshift=-2.0cm, ]
        ]
        [Classification of Gating Mechanism, minimum width=4.0cm, [{Dense gate \cite{jacobs1991adaptive, jordan1994hierarchical, collobert2001parallel, dou2023loramoe, wu2024mixture, pan2024dense, eigen2013learning},  sparse gate  \cite{shazeer2017outrageously, du2022glam, lepikhin2020gshard}, \\ hard gate \cite{fedus2022switch, riquelme2021scaling }, and other gates \cite{nguyen2024expert, li2023ac, li2024hierarchical, ma2018modeling} }, text width=6.0cm, xshift=-2.0cm, yshift = 0.14cm]
        ]
                [MoE with Generative AI,  minimum width=4.0cm,[{GANs\cite{chai2023improved, park2018megan, zhu2023exploring}, VAEs  \cite{shi2019variational, yu2021mixture, saidutta2021joint, yi2024variational}, \\ DMs \cite{balaji2022ediff, ganjdanesh2024mixture, park2024switch}, Transformers \cite{zhong2022meta, li2024uni, qu2024llama, dai2024deepseekmoe, artetxe2021efficient}.
        }, text width=6.0cm, xshift=-2cm, yshift = -0.005cm,  base=midpoint,
        ] ]
                [MoE with Reinforcement Learning, minimum width=4.0cm, [{Value-based RL \cite{obando2024mixtures, tao2022double, zheng2019self},  policy-based RL  \cite{khamassi2006combining, danket2024mixture, deycomparing, ren2021probabilistic}, \\ multi-agent RL \cite{meng2024new, zhang2024optimizing,nguyen2025csaot}, and other RL \cite{triantafyllidis2023hybrid, wang2022learning, cheng2023multi} }, text width=6.0cm, xshift=-2.0cm, yshift = 0.14cm]
        ] 
        ]
        %
        [Section III. \\ MoE in Wireless Networks, minimum width=4.0cm, base=midpoint,
                [Advantages of MoE \\ in Wireless networks, minimum width=4.0cm, [{ Enhance model capacity, computational efficiency, \\ feature learning, and versatility },  text width=6.0cm, xshift=-2cm, yshift = -0.145cm, base=midpoint]]
                [MoE in Various Wireless \\ Network Scenarios, minimum width=4.0cm, [{Vehicular Networks \cite{vyas2023federated, yao2025mixture}, UAV  \cite{kawamura2023hierarchical, chen2024giant, wong2024addressing}, satellite \\ \cite{zhang2024generative, loyola2002combining} and other 6G scenarios \cite{liu2024meta, wang2024optimizing, liu2024multimodal, li2024theory}},  text width=6.0cm, xshift=-2cm, yshift = -0.145cm, base=midpoint]
        ]    
        ]
        [Section IV. Applications of \\ MoE in Wireless Networking, minimum width=4.0cm, base=midpoint,
        [Physical Layer Communication \\and Resource Management, minimum width=4.0cm, yshift = 0.142cm, [{Channel estimation and prediction \cite{senevirathna2006channel, lopez2020channel, jaiswal2023leveraging}, \\ signal processing and communication \cite{van2024mean, fischer2022mixture, fischer2023sparsely, saidutta2021joint}, \\ radio resource allocation \cite{zecchin2020team, ma2022demand, du2024mixture}
        },  text width=6.0cm, xshift=-2cm, yshift = 0.00cm, base=midpoint, ]]
        [Network Optimization \\ and Security, minimum width=4.0cm, yshift = -0.15cm, [{Traffic network management \cite{chattopadhyay2022mixture, jiang2024interpretable, jiang2024hybrid}, \\  edge computing and distributed systems \cite{sarkar2023edge, xue2024wdmoe, yuan2024efficient}, \\
        security and anomaly detection \cite{wang2024fighting, ilias2024convolutional, zhao2024enhancing}
        },   text width=6.0cm, xshift=-2cm, ]
        ]
        ]
        [Section V. Case Study and Dataset, minimum width=4.0cm,
        [Case Study of MoE with DRL, minimum width=4.0cm, [{ MoE is integrated into a diffusion model-based DRL \\ to demonstrate performance improvements in DRL
        }, text width=6.0cm, xshift=-2cm, base=midpoint, yshift = -0.01cm]
        ]
        [Open-Source Datasets, minimum width=4.0cm, [{NLP \cite{wang2018glue, raffel2020exploring, maatouk2023teleqna}, vision \cite{deng2009imagenet, caesar2020nuscenes, xia2018dota},  
        multimodal \\ \cite{lin2014microsoft, schuhmann2022laion, miech2019howto100m}, wireless networks \cite{o2018over, alkhateeb2019deepmimo, wang2018glue, zhang2023citysim} }, text width=6.0cm, xshift=-2cm, yshift = 0.14cm  ]]
        ]
        [Section VI. Future Directions, minimum width=4.0cm, ]
        [Section VII. Conclusion, minimum width=4.0cm, ]
        ]
    \end{forest}
\caption{ The structure of this survey.
}
\label{fig:my_tikz}
\end{figure*}

\subsection{Motivation}

The framework of mixture of experts (MoE) has emerged as an effective paradigm to address diverse computational demands and resource constraints for wireless communication systems \cite{yuksel2012twenty}.
The fundamentals of MoE are to decompose a large model into multiple expert modules, each functioning as an independent sub-network specialized for specific inputs \cite{jacobs1991adaptive}. The expert modules are dynamically managed by a gating mechanism that selectively activates only a subset of experts based on the characteristics of the input data, rather than activating all modules simultaneously. Compared with traditional AI, GenAI models require more computational resources for training due to their complex mission objectives and inference procedures. 
State-of-the-art GenAI models, such as DeepSeek \cite{dai2024deepseekmoe}, GLaM \cite{du2022glam}, and Switch Transformer \cite{fedus2022switch}, have effectively utilized MoE frameworks to expand model capacity while minimizing computational costs.
The dynamic expert selection and sparse activation mechanism of MoE can effectively mitigate computational complexity and improve energy efficiency for GenAI models, enabling resource-constrained wireless infrastructures to enhance scalability and reduce resource consumption in complex wireless task \cite{jordan1994hierarchical}.

Another significant advantage of MoE architecture lies in its ability to perform distributed computing and enable collaborative strategies over local experts \cite{li2024theory}.
The decentralized implementation of MoE experts enables independent processing, inference, and fine-tuning of local data, enhancing specialization by effectively capturing the unique characteristics of personal input \cite{dai2024deepseekmoe}.
Through using gating mechanisms, the collaborative strategies inherent in the MoE architecture empower edge nodes and wireless devices to collaborate in dynamic and adaptive manner \cite{shazeer2017outrageously}. Moreover, these collaborative weighting and activation mechanisms present extensibility for multimodal and multi-task processing \cite{xu2024survey, mustafa2022multimodal, zhangGAI3}. By integrating multimodal data in wireless environments, such as signal strength, antenna gain, and channel state, MoE can significantly enhance network perception and optimization capabilities based on a shared expert pool to accommodate the complex wireless modalities \cite{ma2018modeling}. For multi-task processing in heterogeneous wireless networks, different tasks often exhibit interdependencies, where tasks such as power control \cite{xu2023federated} and beamforming \cite{zhang2025ris} aim to enhance signal-to-noise ratio (SNR), and tasks such as spectrum \cite{qian2020leveraging} and channel allocation \cite{zhang2020new} may be constrained by limited resource. With the dynamic expert selection and task-specific delegation, MoE provides a robust solution to the challenges of task conflicts and relationship modeling, achieving significant improvements in terms of efficiency and adaptability for wireless tasks that require simultaneous optimization of multiple metrics \cite{hazimeh2021dselect}.



\subsection{Comparisons with Related Surveys and Contributions}

The framework of MoE has demonstrated substantial potential for optimizing wireless resources, improving energy efficiency, and supporting heterogeneous network management. This paper aims to provide a comprehensive survey on the fundamentals of MoE, along with its applications in GenAI and a broad range of wireless scenarios and technologies. Table \ref{table1} presents an in-depth comparison of related surveys, emphasizing MoE, GenAI, and their implementation in wireless networks.
These surveys have primarily focused on summarizing the evolution and optimization of MoE. For instance, the authors in \cite{yuksel2012twenty} reviewed the historical development of MoE, covering fundamental concepts, statistical properties, and practical applications across various domains. The authors in \cite{vats2024evolution} investigated MoE’s core principles, including expert specialization and routing mechanisms with a focus on the field of deep learning. The authors in \cite{masoudnia2014mixture} classified MoE frameworks into mixture of implicitly localized experts (MILE) and mixture of explicitly localized experts (MELE), discussing their applications in regression and classification tasks. Furthermore, the authors in \cite{liu2024survey} explored the optimization of MoE across system and hardware levels, investigating techniques such as expert selection, load balancing, and distributed computing strategies.
For GenAI, existing studies have contributed to enhancing the efficiency of large models with the framework of MoE. For example, the authors in \cite{mcintosh2023google} explored the transformative impact of MoE and multimodal learning on GenAI, discussing the role of MoE in improving training and load balancing for GenAI tasks. The authors in \cite{cai2024survey} and \cite{wan2023efficient} reviewed training techniques for large language models (LLMs), such as fine-tuning, load balancing strategies, gradient routing optimization, and efficient expert pruning.


Additionally, several studies have investigated the integration of GenAI and wireless communication networks. The authors in \cite{van2024generative} surveyed the application of GenAI in physical layer communications, underscoring its role in improving channel estimation, equalization, signal classification, and emerging technologies such as reconfigurable intelligent surface (RIS) and beamforming. Similarly, the authors in \cite{cao2023comprehensive} offered a review of AI-generated content (AIGC), exploring GenAI models and their potential applications in wireless communication networks, while the authors in \cite{zhouhao2024large} discussed the implementation of LLMs into telecommunications, focusing on network optimization, traffic management, and security in sixth generation (6G) mobile networks.  Furthermore, the authors in \cite{xu2024integration} incorporated MoE and GenAI models in the Internet of Vehicles (IoV), addressing challenges in traffic management, autonomous driving, and collaborative decision-making for vehicular networks.

However, existing studies lack sufficient investigation into the integrated potential of MoE and advanced GenAI technologies within the broader context of wireless communication systems.
This paper fills this gap by comprehensively exploring the diverse application scenarios of MoE in wireless networks, such as vehicular networks \cite{vyas2023federated, yao2025mixture}, unmanned aerial vehicle (UAV) networks \cite{kawamura2023hierarchical, chen2024giant, wong2024addressing}, satellite networks \cite{zhang2024generative, loyola2002combining}, and other fundamental 6G scenarios \cite{liu2024meta, wang2024optimizing, liu2024multimodal, li2024theory}. Through in-depth analysis, we demonstrate the capabilities of MoE in solving complex wireless tasks, such as channel estimation \cite{senevirathna2006channel, lopez2020channel, jaiswal2023leveraging}, physical layer signal processing \cite{van2024mean, fischer2022mixture, fischer2023sparsely, saidutta2021joint},
radio resource allocation \cite{zecchin2020team, ma2022demand, du2024mixture}, traffic management \cite{chattopadhyay2022mixture, jiang2024interpretable, jiang2024hybrid}, distributed edge computing \cite{sarkar2023edge, xue2024wdmoe, yuan2024efficient}, and network security \cite{wang2024fighting, ilias2024convolutional, zhao2024enhancing}.
The dynamic expert selection and sparse activation mechanisms of MoE have been developed to improve resource utilization, enhance computational efficiency, and ensure adaptability in heterogeneous wireless environments.
The main contributions are summarized as follows.

\begin{itemize}
    \item We provide a foundational overview of MoE methodologies, covering classical MoE models as well as their integration with GenAI and reinforcement learning (RL), highlighting MoE's significant capabilities in addressing complex computational demands and optimizing resource utilization.
    \item We illustrate the advantages of applying MoE frameworks in heterogeneous and dynamic wireless systems by extensively examining diverse scenarios such as vehicular networks, UAV networks, satellite communications, and emerging 6G scenarios.
    
    \item We thoroughly investigate innovative applications of MoE in wireless technologies, including channel estimation, physical layer signal processing, radio resource allocation, traffic optimization, distributed computing, and network security, demonstrating MoE’s versatility and effectiveness across a broad range of complex tasks.

    \item We present a detailed case study that demonstrates the performance advantages of MoE, along with a comprehensive review of publicly available datasets that are widely utilized in MoE-based models to facilitate further research and experimentation in wireless networking.

\end{itemize}

\begin{figure*}
\centering
\includegraphics [width=0.98\textwidth]{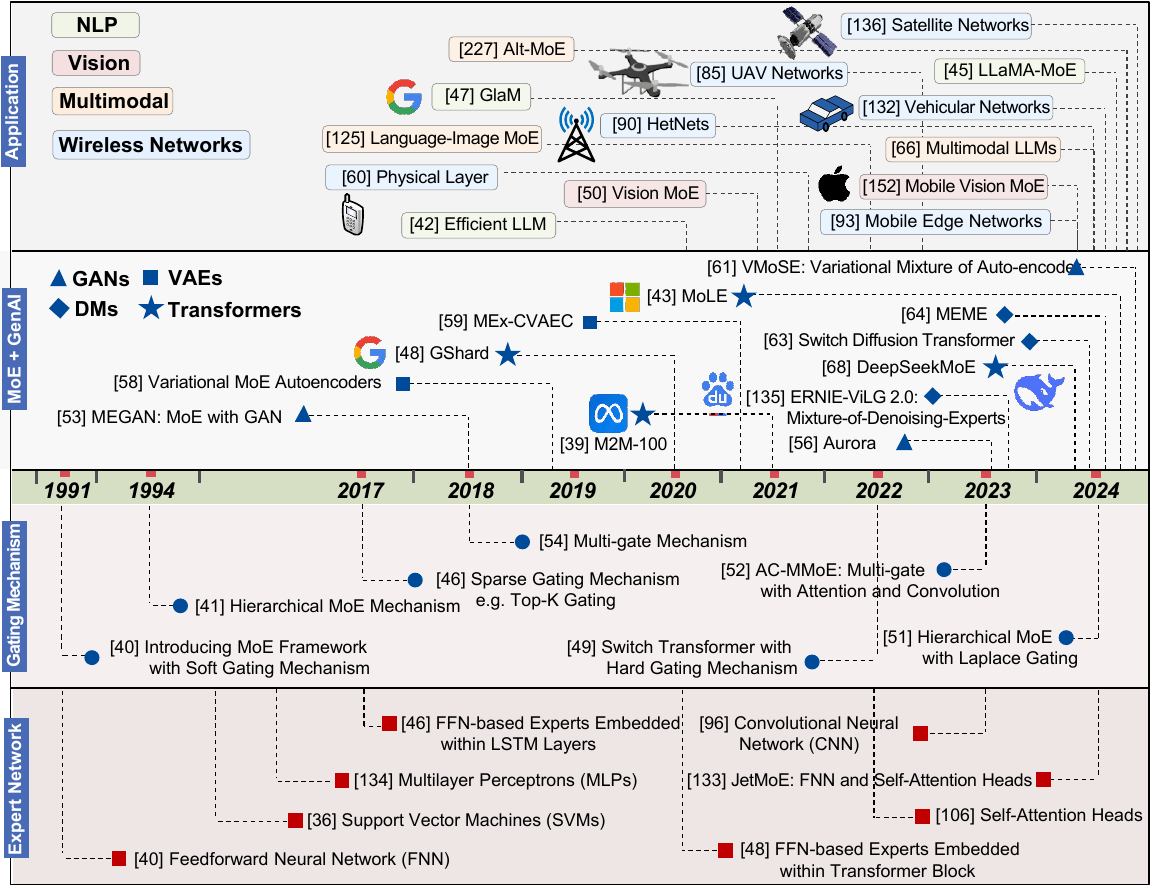} 
\captionsetup{justification=justified,format=plain}
\caption{ A comprehensive timeline illustrating the evolution of MoE models, structured across four dimensions: expert networks, gating mechanisms, MoE integrated with GenAI, and MoE applications. The timeline presents diverse expert networks such as FNNs, SVMs, MLPs, CNNs, and self-attention heads, along with the development of dense, sparse, hard, and other variant gating mechanisms. The framework of MoE is widely integrated with GenAI models, such as GANs, VAEs, DMs, and Transformers. Recent applications span NLP, computer vision, multimodal tasks, and wireless networks, demonstrating the versatility of MoE in modern AI and communication systems.
}
\label{Section2_Development}
\end{figure*}

The rest of this paper is organized as follows. Section II provides an overview of MoE, including its components, gating mechanisms, and the integration with GenAI and RL.
Section III presents a comprehensive investigation of MoE for diverse wireless networks. 
Section IV focuses on the practical applications of MoE in wireless technologies.
In Section V, we provide a case study on the integration of MoE with an RL task and summarize open-source datasets that support MoE research. Section VI discusses the future research directions. Finally, the paper is concluded in Section VI. The overall structure of this survey is illustrated in Figure \ref{fig:my_tikz}.

\section{Overview of Mixture of Experts}

Initially introduced in the early 1990s to overcome the limitations of single-architecture models \cite{jacobs1991adaptive}, MoE has undergone substantial advancements in modularity and composition, allowing for integration with diverse expert networks, sophisticated gating mechanisms, and recent advanced GenAI models and RL methods. These developments have established MoE as a foundational technique across various domains, including natural language processing (NLP), computer vision, multimodal learning, and wireless networks. Figure \ref{Section2_Development} presents a comprehensive timeline that traces the evolution of MoE expert networks, gating strategies, and the combination with GenAI, alongside the growing breadth of applications. Building on this foundation, this section presents the definition and core components of MoE, provides an overview of gating mechanisms, and examines its integration with GenAI models and RL methods.

\subsection{Definition and Core Components of MoE}

MoE is a machine learning framework that incorporates multiple specialized models, referred to as experts, each handling a subset of input data within the training process \cite{masoudnia2014mixture}. As illustrated in Figure \ref{Section2_GenAI} (a) and (b), compared with classical deep neural networks (DNNs) that directly process an entire dataset, the MoE framework leverages a gating network to allocate dynamically training samples to the most relevant experts \cite{bishop2012bayesian}. 
The core components of MoE consist of the expert network, gating scheme, and output layer. 

\begin{itemize}
    \item \textbf{Expert Nework}:
    MoE decomposes a model into smaller, more manageable subnetworks, enabling each expert to specialize in distinct data distributions or feature representations.
    The expert of MoE is typically a trainable network with variations in architecture, depth, or hyperparameters, ranging from conventional machine learning models to advanced DNNs. Examples of expert networks include support vector machines (SVMs) \cite{collobert2001parallel}, multilayer perceptrons (MLPs) \cite{ebrahimpour2007face}, self-attention heads \cite{zhang2022mixture}, and hybrid structures that integrate feedforward networks (FNNs) with self-attention heads \cite{shen2024jetmoe}. In more complex scenarios, entire language models can also serve as the expert network within the MoE framework \cite{li2022branch}.
    \item \textbf{Gating Mechanism}:
    A gating mechanism is responsible for selecting appropriate experts to deal with input samples. Typically implemented as a lightweight model, a gating network computes a probability distribution over the available experts, determining which experts should be activated for model training and inference, thereby reducing computational overhead. Through joint training with expert networks, the gating network continuously improves expert selection, enhancing task specialization and overall model efficiency. Various gating mechanisms have been proposed to optimize expert selection, including dense, sparse, and hard gating, as well as more advanced schemes such as hierarchical gating and multi-gate mechanisms \cite{cai2024survey}. 
    \item \textbf{Output Layer}:
    The output layer aggregates the predictions from activated experts and computes the final output as their combination. The most common approach is a weighted sum, where the weights are determined by the gating network to appropriately scale expert contributions based on their relevance to the input \cite{radford2021learning}. While this method facilitates smooth integration, more advanced implementations incorporate additional approaches to enhance the output quality. For instance, normalization techniques \cite{wei2024skywork} and residual connections \cite{wu2022residual} are employed to regulate the scale of expert outputs, preventing any single expert from dominating the aggregation process.
    
\end{itemize}

Compared to single architectures designed for uniform data, the expert-structured MoE framework improves specialization and training efficiency for diverse data or tasks, particularly in domains requiring high model capacity and adaptability \cite{zhou2022mixture}.


\subsection{ Classification of Gating Mechanism} 

The gating mechanism plays a pivotal role in MoE frameworks, serving as the decision maker that dynamically activates the most appropriate experts to process a specific input.
We classify the gating mechanism into four types, including dense gating, which assigns weights to all experts; sparse gating, which selects only a subset of experts for enhancing efficiency; hard gating, deterministically allocating each input to one expert; and variant strategies, designed to enhance flexibility and specialization in expert selection.

\subsubsection{Dense Gating Mechanism}

The dense gating mechanism activates all expert networks by assigning probability weights at each iteration. These weights are typically computed using a softmax function, producing a probability distribution over all available experts. The final output is derived as a weighted sum of all expert outputs, ensuring that each expert contributes to the final prediction. Formally, for an MoE framework consisting of $N$ experts $\{E_{1}, \ldots, E_{N} \}$,  the output of the dense MoE layer is given by
\begin{equation}
\mathcal{M}_{\text {dense }}\left(\mathbf{x};\Theta,\left\{\boldsymbol{\theta}_n\right\}_{n=1}^N\right)=\sum_{n=1}^N G(\mathbf{x};\Theta)_n E_n\left(\mathbf{x};\boldsymbol{\theta}_n\right),
\end{equation}
where $\mathbf{x}$ denotes the input data and $E_n\left(\mathbf{x};\boldsymbol{\theta}_n\right)$ is the output of the $n$-th expert parameterized by $\boldsymbol{\theta}_n$. $G(\mathbf{x} ; \Theta)_n$ represents the gating weight for the $n$-th expert, with $\Theta$ denoted as the parameters of the gating network. If the softmax function is used to compute the gating weight, $G(\mathbf{x}; \Theta)_n$ is expressed as
\begin{equation}
G(\mathbf{x};\Theta)_n \triangleq \operatorname{Softmax} \left(G(\mathbf{x};\Theta)\right)_n = \frac{\exp \left(G(\mathbf{x};\Theta)_n\right)}{\sum_{i=1}^N \exp \left(G(\mathbf{x};\Theta)_i\right)}.
\end{equation}

Dense gating has been adopted in multiple early investigations \cite{jacobs1991adaptive, jordan1994hierarchical, collobert2001parallel}, and has more recently been applied to large AI models such as LoRAMoE \cite{dou2023loramoe}, MoLE \cite{wu2024mixture}, and DS-MoE \cite{pan2024dense}. Since the dense gating mechanism activates all experts for each input instance, they have proved effective in tasks demanding comprehensive expert engagement, especially under high uncertainty or diverse feature distributions \cite{eigen2013learning}. However, every expert receives gradients during back-propagation, causing the total computational cost and memory usage to scale linearly with the number of experts $N$. While this scheme simplifies expert selection and enhances model expressiveness, it  usually incurs substantial inter-expert communication overhead and computational costs, making it impractical for large-scale applications.\cite{wang2024scaling}.


\subsubsection{Sparse Gating Mechanism} 
Sparse gating mechanism uses the Top-K expert selection approach to limit the number of activated experts per input to $K \ll N$, which tremendously reduces the computational overhead \cite{shazeer2017outrageously}. The gating weight of the $n$-th expert $G(\mathbf{x};\Theta)_n$ using the Top-K approach is computed as
\begin{equation}
\begin{gathered}
G(\mathbf{x};\Theta)_n =\operatorname{Softmax} \left(\operatorname{Top-K}\left(G(\mathbf{x};\Theta) \right)\right)_n, \\
\operatorname{Top-K}\left(G(\mathbf{x};\Theta)\right)_i= \begin{cases}G(\mathbf{x};\Theta)_i, & G(\mathbf{x} ; \Theta)_i \text { is among the } \\ & \text{top-K values of } G(\mathbf{x};\Theta),~~~ \\
-\infty, & \text {otherwise. }\end{cases}
\end{gathered}
\end{equation}

Based on the sparse activation mechanism, only the selected experts participate in forward and backward passes, thereby reducing the training floating point operations (FLOPs) and parameter updates by an approximate factor of $N/K$.  Examples of sparse gating implementations include GShard \cite{lepikhin2020gshard} and GLaM \cite{du2022glam}, which significantly improve efficiency by activating a small subset of experts.
Empirical results of sparse gating from GShard demonstrate more than 2$\times$ improvements in training throughput and substantial reductions in memory consumption compared to dense gating mechanism. However, these gains come at the cost of a slight reduction in model capacity and the introduction of a load-balancing loss, which is required to ensure uniform expert utilization in sparse setups.
 
\subsubsection{Hard Gating Mechanism} 

Unlike soft and sparse gating, the hard gating mechanism deterministically selects a single expert per input based on the highest gating probability. The output weights of the hard gating mechanism are expressed as
\begin{equation}
G(\mathbf{x};\Theta)_n= \begin{cases}1, & n=\arg \max _i G(\mathbf{x} ; \Theta)_i, \\ 0, & \text {otherwise. }\end{cases}
\end{equation}

Consequently, the final output of MoE layer is provided by the output of the highest-weighted expert as
\begin{equation}
\begin{split}
\mathcal{M}_{\text {hard }}\left(\mathbf{x};\Theta,\left\{\boldsymbol{\theta}_n\right\}_{n=1}^N\right)=E_{i^*}\left( \mathbf{x};\boldsymbol{\theta}_{i^*} \right), \\
i^*=\arg \max _i G(\mathbf{x} ; \Theta)_i. \quad \quad \quad
\end{split}
\end{equation}

The hard gating mechanism substantially reduces computational overhead, making it particularly suitable for resource-constrained environments. Despite its simplicity, the hard gating mechanism, as utilized in Switch Transformer \cite{fedus2022switch} and DeepSpeed-MoE \cite{rajbhandari2022deepspeed}, has demonstrated competitive performance even when activating only one expert per sample. Nonetheless, the hard gating mechanism may confronted with the challenge of imbalanced expert utilization, where certain experts may be over-utilized while others remain underused, potentially limiting the model's capacity utilization and expressive generalization.

\subsubsection{Other Variant Gating Mechanisms} 

Beyond conventional gating mechanisms, several advanced and specialized gating strategies have been developed to further optimize expert selection. Multi-gate MoE (MMoE) frameworks employ multiple gating networks to handle different tasks, enabling task-specific routing that enhances model specialization \cite{ma2018modeling, li2023ac}. The expert networks of MMoE are shared across all tasks, while each task is assigned an independent gating network that separately determines the optimal weighted combination of experts. For an MMoE framework with $T$ gates, the $t$-th output of the MoE layer is computed as 
\begin{equation}
\mathcal{M}_{\text {multi-gate }}^{t}\left( \mathbf{x};\Theta_{t},\left\{\boldsymbol{\theta}_n\right\}_{n=1}^N 
 \right)=\sum_{n=1}^N G_{t}(\mathbf{x}; \Theta_{t})_n E_n\left(\mathbf{x};\boldsymbol{\theta}_n\right).
\end{equation}

This framework enables efficient parameter sharing among tasks while allowing task-specific specialization through dedicated gating mechanisms. Meanwhile, hierarchical gating, as a variant of the gating mechanism, introduces multi-level decision-making, where experts are partitioned into subgroups, and the gating functions are performed in multiple stages to refine expert selection \cite{nguyen2024expert, li2024hierarchical}. The output of a two-level hierarchical gating mechanism is expressed as
\begin{equation}
\begin{split}
    &\mathcal{W}_{\text {hier}}\left(\mathbf{x};\Theta^{high},  \left\{\Theta_m^{low}\right\}_{m=1}^M, \left\{\boldsymbol{\theta}_{m,n}\right\}_{n=1}^{N_m}\right)= \\
    &~\sum_{m=1}^M G^{high}(\mathbf{x} ; \Theta^{high})_m \sum_{n = 1}^{N_m} G_m^{low}(\mathbf{x} ; \Theta_m^{low})_n E_{m, n}\left(\mathbf{x};\boldsymbol{\theta}_{m,n}\right),
\end{split}
\end{equation}
where the high-level gating $G^{high}(\mathbf{x} ; \Theta^{high})$ first selects expert group $m$ from $M$ available groups for the input $\mathbf{x}$. Within the selected group $m$, the low-level gating $G_m^{low}(\mathbf{x} ; \Theta_m^{low})$ further weights the expert output among $N_m$ experts, with the $n$-th expert output denoted as $E_{m, n}\left(\mathbf{x};\boldsymbol{\theta}_{m,n}\right)$. The hierarchical gating mechanism is well-suited for tasks requiring multi-level decision-making and is particularly effective in wireless network scenarios with strong task correlations. 

\textbf{Mitigation of Expert Collapse and Overfitting:} Although gating mechanisms effectively regulate expert activation and manage computational resource allocation, issues such as expert collapse and model overfitting persist and require further mitigation, particularly under limited training data. Expert collapse occurs when a small number of experts dominate training, leading to under-utilization of other experts, thereby diminishing model diversity and capacity. Several strategies have been proposed to address the challenge of expert collapse. Load-balancing losses, such as auxiliary entropy-based losses, encourage a balanced distribution of expert activations \cite{riquelme2021scaling}. For example, a router Z-loss is introduced to enable a more uniform output proportion of the gating network \cite{zoph2022st}. Given $B$ input data samples, the Z-loss is formulated as
\begin{equation}
L_Z=\frac{1}{B} \sum_{b=1}^B\left(\log \sum_{n=1}^N \exp \left(G\left(\mathbf{x}_b ; \Theta\right)_n\right)\right)^2,
\end{equation}
where $\mathbf{x}_b$ is the $b$-th input data and the loss function helps the $G\left(\mathbf{x}_b ; \Theta\right)$ values converge towards a more balanced distribution. The load-balancing strategies effectively mitigating expert collapse by penalizing excessive reliance on individual experts. thereby enhancing the stability of model training.
Additionally, overfitting problem arises when experts overly adapt to limited training data, reducing their ability to generalize effectively. Regularization methods such as dropout, early stopping, and weight decay are commonly integrated into MoE frameworks to prevent overfitting \cite{shen2023mixture}. More recently, knowledge distillation \cite{shu2024llava} and expert pruning techniques \cite{hou2025lightweight} have emerged as effective approaches to maintain expert diversity, especially under data-scarce conditions. These mechanisms collectively enhance the robustness of MoE, ensuring better generalization across diverse tasks and environments.

\textbf{Lessons Learned:} 
Gating mechanisms serve as a critical component of the MoE framework, responsible for regulating expert activation and managing the allocation of computational resources. These mechanisms differ significantly in their trade-offs among computational efficiency, representational capacity, and system complexity. Specifically, dense gating activates all experts for every input, ensuring full expert participation and rich expressiveness, but at the cost of substantial computational and memory overhead. Sparse gating, such as Top-k selection, reduces training FLOPs and memory consumption by activating only a subset of experts per input, which is widely adopted in large-scale models such as GShard and GLaM. However, it usually incurs a slight reduction in model capacity and requires effective load-balancing strategies to mitigate expert underutilization. 
Hard gating deterministically selects a single expert for each input, achieving maximal computational efficiency and making it suitable for resource-constrained scenarios. Nonetheless, hard gating introduces high gradient variance and potential expert imbalance that impairs the stability of the training process. Advanced gating strategies, such as multi-gate and hierarchical gating, support task-specific or multi-level expert selection, enabling finer specialization and better adaptation to heterogeneous inputs, but also increase architectural and training complexity.
Overall, each gating strategy offers distinct advantages depending on the target deployment scenarios and resource constraints. As such, the design and deployment of gating mechanisms are pivotal for scaling MoE frameworks to meet the diverse requirements of modern machine learning applications, particularly in dynamic and resource-sensitive environments such as wireless networks.

\begin{figure*}
\centering
\includegraphics [width=\textwidth] 
{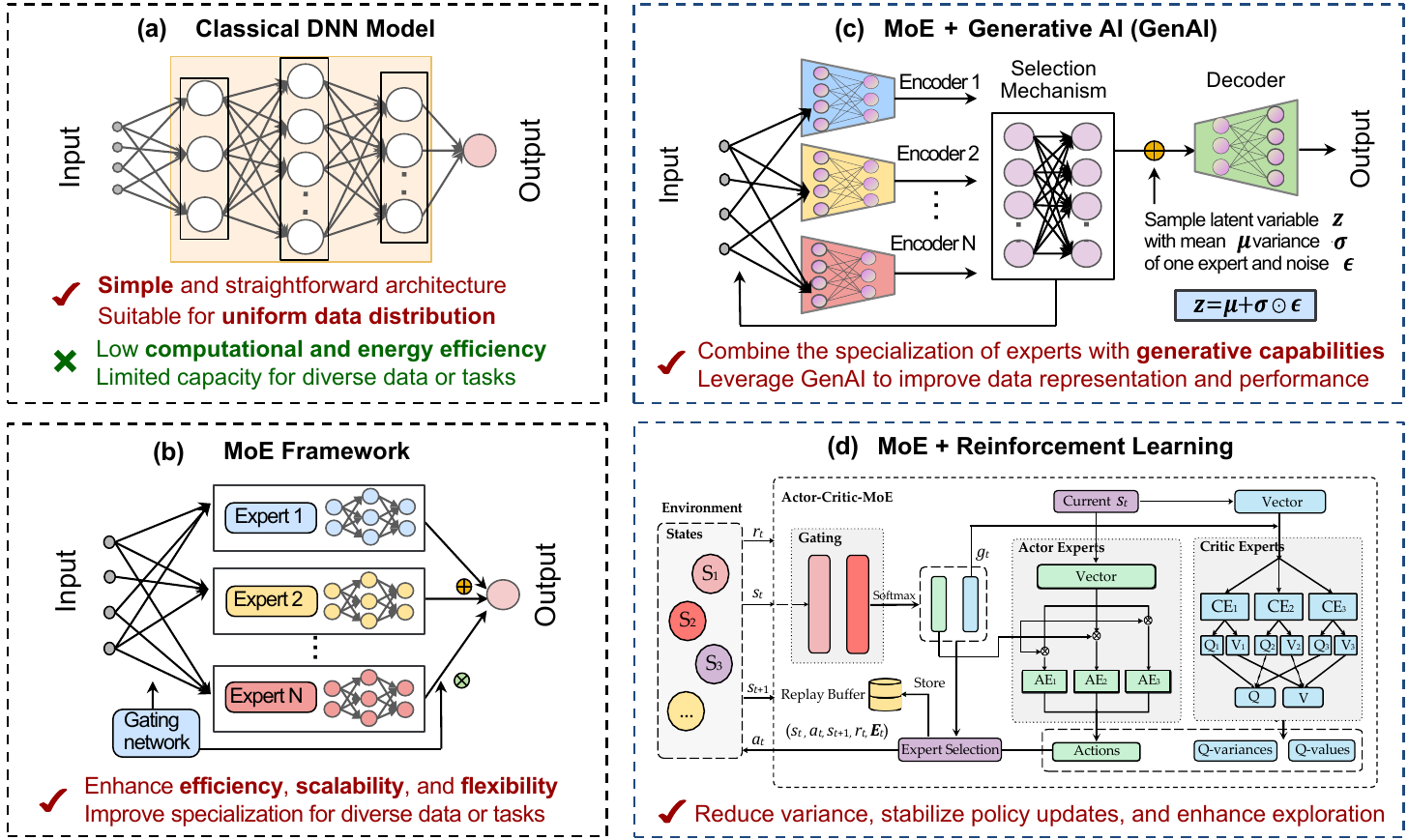} 
\captionsetup{justification=justified,format=plain}
\caption{ Illustration of three learning architectures. (a) \textbf{Classical DNN model}: A straightforward architecture for uniform data distribution but limited by low computational and energy efficiency. (b) \textbf{MoE framework}: Incorporate multiple expert networks coordinated by a gating mechanism for diverse data or tasks. 
(c) \textbf{MoE with GenAI}: Combine the specialization of MoE with generative capability, illustrated by an example of MoE integrated with multiple encoder experts of variational autoencoder (VAE) \cite{saidutta2021joint}. (d) \textbf{MoE with RL}: Employ multiple actor experts (AEs) and critic experts (CEs) to enhance the training performance of the actor-critic (AC) algorithm \cite{zheng2019self}. 
}
\label{Section2_GenAI}
\end{figure*}

\subsection{MoE with Generative AI}

The employment of MoE framework in GenAI models has garnered increasing attention, leveraging MoE experts with generative capabilities to enhance data synthesis, representation learning, and adaptive processing. Recent advancements have demonstrated the effectiveness of MoE in improving the performance of GenAI models, including generative adversarial networks (GANs), variational autoencoders (VAEs), diffusion models (DMs), and Transformer-based architectures \cite{zhang2024optimizing}. This section explores the integration of MoE and GenAI, highlighting key advancements in optimizing generative performance across diverse applications.
Figure~\ref{Section2_GenAI}(c) illustrates an MoE-enhanced GenAI framework based on VAE, and a summary of related approaches is provided in Table~\ref{table:summary_GenAI}.


\subsubsection{Generative Adversarial Networks}

GANs are widely used for high-quality image synthesis, consisting of a generator that synthesizes realistic data samples and a discriminator that differentiates between real and synthetic inputs \cite{creswell2018generative}. Despite their effectiveness in generation capability, conventional GANs often suffer from mode collapse, limited sample diversity, and difficulties in modeling complex multimodal distributions \cite{gurumurthy2017deligan}. The integration of MoE into GANs provides a scalable solution by incorporating multiple specialized generators or expert modules within the generator \cite{hoang2017multi}. The gating network in MoE dynamically selects the most relevant expert, improving mode diversity, mitigating mode collapse, and enhancing generalization across heterogeneous data distributions. Recent advancements have explored the integration of MoE framework into GANs to enhance generative modeling across different domains. In the context of language generation, MoE-GAN \cite{chai2023improved} incorporates multiple expert generators for language feature statistics alignment, which offers fine-grained learning fidelity and improves the diversity and coherence of generated text. Additionally, MEGAN \cite{park2018megan} applies MoE to image generation by employing a set of generators and a gating network adaptively selecting suitable generators for each input, thereby capturing distinct data modalities. Extending this concept to text-to-image synthesis, a sparse MoE-based GAN framework is introduced to leverage both textual embeddings and latent variables for specialized feature generation, thus providing more specialized and semantically aligned images \cite{zhu2023exploring}.

\subsubsection{Variational Autoencoders}

VAEs are probabilistic generative models that encode input data into a latent distribution and reconstruct data from a sampled latent variables \cite{shi2019variational}. While the encoding-decoding structure of VAEs is effective in data compression and generation, conventional VAEs often struggle with capturing complex representation in the latent space. Exploiting MoE in VAEs enables specialized latent space modeling by dynamically allocating data samples to different encoder or decoder experts, improving reconstruction capability and representation diversity. For instance, in anomaly detection, employing a set of convolutional VAEs as experts, together with a convolutional gating network to select experts for modeling manifold-specific patterns, has been shown to improve detection accuracy \cite{yu2021mixture}. In communication scenarios, an introduction of multiple VAE encoders to mitigate representation discontinuities under bandwidth constraints has led to improved robustness across varying channel conditions \cite{saidutta2021joint}. Furthermore, in multi-modal recommendation, a mixture of variational stochastic auto-encoders \cite{yi2024variational} is leveraged to select and fuse modality-specific latent experts, which effectively enhances generalization for different modalities under incomplete input conditions.

\subsubsection{Diffusion Models}
DMs comprise a forward process, where random noise is gradually added to the original data, and a reverse process that reconstructs structured data through a multi-step denoising process \cite{wu2023cddm}. Conventional DMs use a single set of parameters across all denoising steps, limiting their adaptability to different noise levels \cite{feng2023ernie}. MoE-based diffusion models overcome this constraint by employing specialized experts for different denoising stages based on distinct noise characteristics \cite{lee2024multi}. Additionally, MoE enhances computational efficiency through sparse expert activation, reducing the need for full-network inference at every denoising step \cite{luo2024staleness}. 
Recent developments have explored the integration of MoE into diffusion models to improve both generation quality and computational efficiency. For instance, eDiff-I, a text-to-image diffusion model proposed by NVIDIA, has introduced an ensemble of expert denoisers for different generation stages, thereby improving text alignment and image fidelity in text-to-image synthesis \cite{balaji2022ediff}. Meanwhile, DiffPruning \cite{ganjdanesh2024mixture} reduces the inference cost of image generation by assigning clustered timestep intervals to pruned expert networks, while Switch-DiT \cite{park2024switch} integrates timestep-specific characteristics within sparse MoE layer to accelerate convergence.

\subsubsection{Transformers}
Transformers have emerged as the prevalent architecture in GenAI models, demonstrating remarkable capabilities in sequential modeling through self-attention mechanisms and encoder-decoder architectures \cite{daxberger2023mobile}.  However, as the model sizes continue to grow, the computational demands of Transformers increase substantially, posing challenges for efficient training and deployment, particularly in LLM applications \cite{xi2024maasformer}. A widely adopted solution is to replace the standard feedforward networks (FFNs) in Transformer blocks with MoE layers, where a gating mechanism dynamically selects a sparse subset of smaller FNN for each input token, thereby reducing computation while maintaining model capacity \cite{zhong2022meta}. Moreover, MoE also facilitates expert specialization by enabling different experts to handle distinct input patterns or tasks, enhancing performance across language, vision, and multimodal domains \cite{li2024uni}.
The effectiveness of MoE-based Transformers has been validated in recent models such as LLaMA-MoE \cite{qu2024llama} for scalable language modeling, DeepSeek-MoE \cite{dai2024deepseekmoe} for cross-layer expert allocation, and MoME \cite{artetxe2021efficient} for flexible multimodal integration.

\textbf{Lessons Learned:} 
The integration of MoE into GenAI models has demonstrated significant improvements in computational efficiency, representation learning, and adaptive specialization throughout the generative process. A key insight is that generative models often involve highly structured procedures, such as multi-step denoising \cite{balaji2022ediff, ganjdanesh2024mixture, park2024switch}, sequential decision-making \cite{song2024neuromorphic, zecchin2024cell}, or modality-conditioned synthesis \cite{zhu2023exploring, li2024uni}. In these settings, MoE allows GenAI models to activate only the most relevant experts at each stage, enabling fine-grained refinement of the generation process and adaptive utilization of model capacity. The resulting specialization improves fidelity and stability without increasing computational complexity, particularly in models designed to handle heterogeneous data types 
Furthermore, with the emergence of new generative paradigms such as Generative Flow Networks (GFlowNets) \cite{khoramnejad2025generative}, MoE provides a scalable and modular framework that supports expert specialization, enhances generation quality, and addresses the increasing demands of advanced GenAI models.

\begin{table*}
\centering
\caption{ Summary of Generative AI model with Mixture of Experts Integration. \\  \textbf{Note:} Red circles indicate the limitations of conventional GenAI and green circles highlight the advantages of MoE} 
\label{table:summary_GenAI}
\renewcommand{\arraystretch}{1.2}
{\fontsize{8.5pt}{10pt} \selectfont
\begin{tabular}{|m{1.7cm}<{\centering}|m{0.75cm}<{\centering}|m{2.25cm}<{\centering}|m{5.3cm}|m{5.95cm}<{\centering}|}
\hline
\textbf{Model} & \textbf{Ref.} & \textbf{Task} & $\quad\quad\quad\quad\quad\quad$\textbf{Insight} & \textbf{GenAI Limits \& MoE Advantages} \\ \hline
\multirow{4}{3cm}{$~~~$ GANs}
&\cite{park2018megan} & Language Generation & Improve representation and coherence via expert generators and feature alignment & \multirow{1}{5.95cm}{\vspace{-0.2in}
\begin{itemize}[leftmargin=*]
\item [\redcircle]  Mode collapse, limited sample diversity and difficulties in multimodal distributions
\vspace{1.2pt} 
\item [\greencircle] Multi GAN generators improve generation diversity and captures distinct data modes
\vspace{1.2pt} 
\item [\greencircle] Generator selection enhances semantic alignment and generation efficiency 
\end{itemize}}

\\ \cline{2-4}  

& \cite{park2018megan}  & $~$ Image $~$ Generation  & Use gating network to dynamically select specialized generators &  

\\ \cline{2-4}  

& \cite{zhu2023exploring}  &  Text-to-image Generation  & Apply sparse gating and FNN experts for semantically aligned feature synthesis &  

\\ \hline
    
\multirow{6}{3cm}{$~~~$ VAEs}
&\cite{yu2021mixture} & Anomaly Detection &  Use convolutional VAEs for manifold-specific modeling & \multirow{7}{5.95cm}{
\vspace{-0.35in}
\begin{itemize}[leftmargin=*]
\item [\redcircle] Struggle with complex representation in the latent space
\vspace{2.2pt}
\item [\greencircle] Enable expert specialization through multiple encoding and decoding modules
\vspace{2.2pt}
\item [\greencircle] Enhance model expressiveness by mitigating latent discontinuities
\end{itemize}
}

\\ \cline{2-4}

& \cite{saidutta2021joint}  & Channel Coding  & Multi encoders to handle representation discontinuities under bandwidth limits &  
\\ \cline{2-4}

& \cite{yi2024variational} & Multi-modal Recommendation  & Fuse modality-specific experts based on uncertainty-aware stochastic sampling & 
\\ \hline

\multirow{5}{3cm}{$~$ Diffusion \\$~$ Models} 
&\cite{balaji2022ediff}  & Text-to-image Generation & Train specialized denoisers for different denoising intervals to align text and image fidelity & \multirow{5.3}{5.95cm}{ \vspace{-0.26in}
\begin{itemize}[leftmargin=*]
\item [\redcircle] Fixed parameters across denoising steps hinder adaptability \\ 
\vspace{1.9pt} 
\item [\greencircle] Employ specialized experts for different denoising stages, adapting to distinct noise characteristics at each time step \\ \vspace{1.9pt}
\item [\greencircle] Sparse experts reduce inference cost while maintaining generation quality
\end{itemize}
}

\\ \cline{2-4}

&  \cite{ganjdanesh2024mixture} & $~$Image$~$ Generation & Assign timestep intervals to pruned experts for reducing inference overhead &  
\\ \cline{2-4}

& \cite{park2024switch} & $~$Image$~$ Generation & Employ timestep-aware MoE for adaptive routing and improved convergence  & 
\\ \hline

\multirow{5}{3cm}{Transformers} 
&\cite{qu2024llama}  & Language Generation & Construct attention and MLP-based MoE for instructed LLMs with post-training & \multirow{5.3}{5.95cm}{ \vspace{-0.2in}
\begin{itemize}[leftmargin=*]
\item [\redcircle]  Substantial computational requirements and limited adaptability to diverse modalities\\ 
\vspace{1.9pt} 
\item [\greencircle]  Replace FFNs with MoE layers to reduce computation and maintain model
capacity\\ \vspace{1.86pt} 
\item [\greencircle] Enable experts to handle inputs across language, vision, and multimodal domain
\end{itemize}
}

\\ \cline{2-4}

&  \cite{dai2024deepseekmoe} & Language Generation & Employ expert segmentation and shared experts to maximize specialization &  
\\ \cline{2-4}

& \cite{artetxe2021efficient} & Multimodal Modeling & Combine modality-specific MoE in vision and language tasks  & 
\\ \hline

\end{tabular}
}




\end{table*}

\subsection{MoE with Reinforcement Learning}

\begin{table*}
\centering
\caption{Summary of Reinforcement Learning Methods with Mixture of Experts Integration. \\  \textbf{Note:} Red circles indicate the limitations of conventional RL and green circles highlight the advantages of MoE}
\label{table:summary_RL}
\renewcommand{\arraystretch}{1.2}
{\fontsize{8.5pt}{10pt} \selectfont
\begin{tabular}{|m{1.86cm}<{\centering}|m{0.75cm}<{\centering}|m{1.9cm}<{\centering}|m{5.5cm}|m{5.95cm}<{\centering}|}
\hline
\textbf{Methodology} & \textbf{Ref.} & \textbf{Algorithm} & $\quad\quad\quad\quad\quad\quad$\textbf{Insight} & \textbf{RL Limits \& MoE Advantages} \\ \hline
\multirow{3}{3cm}{Value-based \\ RL} 
&\cite{obando2024mixtures} & DQN & Improve Q-value estimation via expert selection in multiple Q-networks & \multirow{1}{5.95cm}{\vspace{-0.186in}
\begin{itemize}[leftmargin=*]
\item [\redcircle] Overestimation bias, limited generalization, non-stationarity issues 
\item [\greencircle] Employ multi-value networks for more accurate estimation and stable training 
\end{itemize}}

\\ \cline{2-4}  

& \cite{tao2022double}  & Double DQN (DDQN)  & Enhance dynamic decision-making with weighted DDQN experts &  

\\ \hline

\multirow{6}{3cm}{Policy-based \\ RL} 
&\cite{khamassi2006combining} & AC & Adaptive actor and critic network selection in dynamic environments & \multirow{7.22}{5.95cm}{
\vspace{-0.186in}
\begin{itemize}[leftmargin=*]
\item [\redcircle] Inefficient policy exploration, high policy gradient variance, and limited adaptability in complex policy scope
\vspace{2.2pt}
\item [\greencircle] MoE is used for multiple actors, while critic can be either single or multiple
\vspace{2.2pt}
\item [\greencircle] Reduce variance in policy gradients, stabilize policy updates, and improve policy exploration efficiency
\end{itemize}
}

\\ \cline{2-4}

& \cite{danket2024mixture}  & PPO  & Enhance PPO with MoE-based actors for demand-aware capacity planning &  
\\ \cline{2-4}

& \cite{deycomparing} & DDPG & Mixture Gaussian actors while retaining a standard single critic  & 
\\ \cline{2-4}

& \cite{ren2021probabilistic} & SAC & Employ probabilistic MoE in Actor while retaining a standard SAC critic & 
\\ \hline

\multirow{5}{3cm}{Multi-Agent\\ RL} 
&\cite{meng2024new}& Multi-agent AC (MAAC) & Use an MoE framework to dynamically optimize each agent’s policy & \multirow{5.3}{5.95cm}{ \vspace{-0.2in}
\begin{itemize}[leftmargin=*]
\item [\redcircle] Non-stationary agent behaviors, limited scalability, and inefficient coordination \\ 
\vspace{2.2pt} 
\item [\greencircle] MoE acts as a unified multi-agent structure or be used within each agent \\ \vspace{2.2pt}
\item [\greencircle] Mitigate non-stationarity, achieve high adaptability, and improve coordination 
\end{itemize}
}

\\ \cline{2-4}

&  \cite{zhang2024optimizing}   & MAPPO & Dynamic policy network selection for multi-agent coordination &  
\\ \cline{2-4}

& \cite{nguyen2025csaot} & MAPPO &  Integrate mixture of policy networks within each agent & 
\\ \hline

\multirow{5}{3cm}{Other RL \\ Approach} 
&\cite{triantafyllidis2023hybrid} & HRL & Employ a high-level gating network to activate low-level experts for sub-tasks such as pushing, grasping, and inserting & \multirow{1}{5.95cm}{ 
\vspace{-0.26in}
\begin{itemize}[leftmargin=*] 
\item [\redcircle] Inefficient hierarchical policy exploration
\item [\greencircle] Improve multi-tier coordination and decision-making
\end{itemize}
}

\\ \cline{2-5}

& \cite{wang2022learning}  & Meta-RL  & Mixture of representative experts to enhance generalization for unseen tasks &  
\multirow{1}{6.1cm}{ 
\vspace{-0.2in}
\begin{itemize}[leftmargin=*] 
\item [\redcircle] Limited adaptability to novel distributions
\item [\greencircle] Improve generalization across environments
\end{itemize}
}

\\ \cline{2-5}

& \cite{cheng2023multi} & Multi-task RL & Apply attention-based experts for adaptive policy learning across diverse tasks  & 
\multirow{1}{6.1cm}{ 
\vspace{-0.2in}
\begin{itemize}[leftmargin=*] 
\item [\redcircle] Task interference and negative transfer
\item [\greencircle] Enhance specialization for distinct tasks 
\end{itemize}
}

\\ \hline

\end{tabular}
}
\end{table*}

An integration of MoE with RL has demonstrated significant potential in enhancing value function estimation, policy learning, and decision-making efficiency. 
Figure~\ref{Section2_GenAI}(d) illustrates the integration of multiple actor and critic experts within a deep RL (DRL) framework.
The following explores the MoE framework across various RL paradigms, including value-based, policy-based, and multi-agent RL, as well as its applications in other RL approaches. A summary of RL methods with MoE integration is presented in Table~\ref{table:summary_RL}.

\subsubsection{MoE for Value-based RL}

In state or state-action value-based RL methods, such as Temporal Difference learning \cite{tesauro1995temporal}, Q-learning \cite{watkins1992q}, and Deep Q-Networks (DQN) \cite{mnih2015human}, as well as the critic component in actor-critic (AC) frameworks including Deep Deterministic Policy Gradient (DDPG) \cite{lillicrap2015continuous} and Twin Delayed DDPG (TD3) \cite{fujimoto2018addressing}, accurate value estimation is essential for effective policy optimization. However, these conventional methods face several challenges, such as
value overestimation bias, limited generalization across dynamic environments, and difficulties in adapting to non-stationary conditions. MoE has been introduced as an advanced value function network, leveraging multiple specialized expert networks to approximate the value function under diverse conditions \cite{obando2024mixtures}. By incorporating a gating network, MoE adaptively selects the most suitable expert, thereby enhancing estimation accuracy, mitigating overestimation bias, and improving learning stability \cite{tao2022double}.
For instance, TD3 reduces the overestimation bias of DDPG by maintaining two Q-value estimators and selecting the minimum value during policy updates. MoE extends this approach by introducing multiple value function experts, allowing the model to dynamically adjust to different environmental conditions and state distributions \cite{zheng2019self}. This multi-critic networks enabled by MoE not only improves robustness but also facilitates better decision-making in complex reinforcement learning settings.

\subsubsection{MoE for Policy-based RL}
For policy-based RL methods and the actor component in AC frameworks, such as Reinforce \cite{williams1992simple}, Proximal Policy Optimization (PPO) \cite{schulman2017proximal}, and Soft Actor-Critic (SAC) \cite{haarnoja2018soft}, policy networks effectively learn decision-making strategies that adapt to diverse environmental states. However, these methods face several challenges, including inefficient policy exploration, high variance in policy gradients, and limited adaptability in optimizing policies under complex environmental or reward landscapes \cite{liu2024mdpo}. To address these limitations, MoE provides an adaptive policy representation structure, where multiple expert networks learn distinct strategies for different environmental conditions \cite{li2020video}. Through gating mechanisms, MoE selects the most appropriate expert for distinct state distributions, effectively reducing variance in policy gradients and improving stability during policy updates. Moreover, by distributing policy learning across multiple experts, MoE encourages diverse policy exploration, enabling the discovery of more effective strategies and preventing premature convergence to suboptimal policies. MoE has been integrated into the actor-network of PPO \cite{danket2024mixture}, DDPG \cite{deycomparing}, and SAC \cite{ren2021probabilistic}, where a probabilistic gating mechanism dynamically assigns different environmental conditions or state distributions to specific actor experts. This enables some actor networks to focus on high-risk exploratory strategies while others refine exploitation-based policies, thereby reducing gradient variance, stabilizing policy updates, and improving exploration efficiency.

\subsubsection{MoE for Multi-Agent RL}

MoE significantly enhances multi-agent RL by improving coordination strategies, scalability, and adaptability in systems involving multiple interacting agents \cite{meng2024new}. One of the primary challenges in multi-agent RL is the non-stationary nature of the joint action decisions, where dynamically changing agent behaviors influence the learning process \cite{gomes2025multi}. MoE mitigates this issue by allowing all agents to share the same set of policy networks while selecting specialized experts for different interaction patterns \cite{cui2023chatlaw}. For example, in the proposed MAPPO-MoE algorithm \cite{zhang2024optimizing}, rather than relying on a single agent policy, each agent utilizes a set of different agent policy networks, where a probabilistic gating mechanism dynamically selects the most suitable experts based on the current state and interaction context. This approach employs centralized training and decentralized execution, facilitating adaptive decision-making for diverse agent behaviors while enhancing exploration efficiency by promoting diverse strategy selection. 
Furthermore, MoE enhances multi-agent RL scalability by integrating multiple expert networks within each agent, enabling robust coordination and adaptive policies for the decision-making of multi-agent systems \cite{nguyen2025csaot}.


\subsubsection{MoE for Other RL Approaches}
MoE can be further integrated into various advanced RL paradigms, such as hierarchical RL (HRL) \cite{triantafyllidis2023hybrid}, Meta-RL \cite{liu2024meta}, and Multi-task RL \cite{willi2024mixture}. 
In HRL, MoE facilitates hierarchical decision-making by selecting sub-policies operating at different decision levels, where higher-level experts focus on strategic planning and long-term decisions, while lower-level experts handle fine-grained control and real-time adaptation \cite{esteban2019hierarchical, chow2022mixture}.
In Meta-RL, MoE enables efficient knowledge transfer to new conditions through a set of pre-trained experts, expediting learning in non-stationary environments. By dynamically selecting experts with relevant prior knowledge, MoE enhances sample efficiency and accelerates policy adaptation, making it suitable for applications where rapid learning is essential for the tasks \cite{wang2022learning}. 
Additionally, in multi-task RL, the MoE framework allows for efficient specialization across different tasks by allocating task-specific expert networks to different distinct domains, ensuring that each expert optimizes a specific objective \cite{cheng2023multi}. While leveraging shared representations for all tasks, MoE-based multi-task RL not only mitigates task interference but also facilitates stable learning across potentially conflicting task distributions. 

\textbf{Lessons Learned:} Integrating MoE into RL methods has led to significant improvements, including reducing overestimation bias in value-based RL \cite{obando2024mixtures, tao2022double, zheng2019self}, high variance in policy gradients \cite{khamassi2006combining, danket2024mixture, deycomparing, ren2021probabilistic}, and non-stationarity in multi-agent systems \cite{meng2024new, zhang2024optimizing,nguyen2025csaot}. Additionally, MoE facilitates scalability in hierarchical \cite{triantafyllidis2023hybrid} and multi-task \cite{cheng2023multi} RL, enabling structured decision-making and shared generalization across related tasks \cite{wang2022learning}.
The capability of MoE combined with different RL paradigms can be further extended to address the challenges in heterogeneous wireless environments.  For example, diverse wireless technologies such as cellular \cite{liu2024meta} and UAV access \cite{zhao2025generative} can benefit from robust policy adaptation and scalable decision-making with MoE-based RL approaches.
However, whiles MoE offers substantial benefits, it introduces challenges such as increased computational overhead, expert collapse where a few experts dominate learning while others remain underutilized, and the difficulty of designing adaptive gating mechanisms.
Addressing these issues necessitates further investigations into efficient training schemes, adaptive expert allocation strategies, and scalable architectures that preserve diversity and stability across varied RL learning tasks.

\section{ Scenarios of Mixture of Experts in Wireless Networks }

With the ability to enhance model capacity and computational efficiency, MoE has been widely adopted in wireless networks. This section highlights the advantages of MoE in wireless networks and explores its applications across various communication scenarios.

\subsection{Advantages of MoE in Wireless Networks}

In wireless communications, the ongoing evolution of wireless network technologies has brought about significant improvements in network capacity and data transmission rates \cite{zhangcell2}. However, with the rapid proliferation of wireless devices and the expanding deployment of wireless infrastructures, wireless networks are becoming increasingly heterogeneous and complicated, posing substantial challenges for efficient resource allocation and network management \cite{puspitasari2023emerging}. Additionally, network components such as edge servers and APs are typically resource-constrained, facing critical limitations to support a wide range of wireless applications \cite{zhangcell3}.
The adoption of MoE frameworks in wireless communications provides substantial advantages. 

\begin{itemize}
    \item \textbf{Increases model capacity}: Traditional machine learning models struggle to efficiently capture the diverse characteristics of wireless communication environments.
    MoE extends model capacity by introducing multiple expert modules that specialize in different aspects of wireless communication, such as adaptive modulation, interference cancellation, and predictive channel estimation. This modular approach allows for a more effective representation of complex wireless features \cite{artetxe2021efficient}.
    \item \textbf{Enhances computational efficiency}:
    Wireless networks usually operate under stringent resource constraints, such as limited bandwidth, power, and storage capacity. The expert selection mechanism in MoE ensures that only a subset of expert networks is engaged for each task, beneficial for resource-constrained environments such as mobile edge networks and IoT deployments, where lightweight yet high-performance models are essential for real-time decision-making and low-latency processing \cite{wu2024mixture}.
    \item \textbf{Enables dynamic feature learning}:
    Wireless networks exhibit high dynamism, characterized by continuously changing channel conditions, user mobility, and interference patterns. 
    MoE employs multiple experts that can adapt to varying network conditions while maintaining high representation capability \cite{fan2021beyond}. 
    The adaptability of MoE enables dynamic feature learning in applications such as spectrum allocation, beamforming optimization, and interference management, which require real-time responsiveness to fluctuations in network conditions.
    \item \textbf{Versatility in multi-task and multimodal specialization}: MoE experts are specialized to address specific wireless communication tasks or adapt to varying operating conditions \cite{lei2024alt}. For example, specific expert networks are tailored for high- or low-SNR regimes in adaptive modulation classification (AMC), thereby optimizing their performance for each sub-task and significantly improving accuracy over general-purpose models \cite{gao2023moe}. Additionally, experts are designed to handle different input modalities, such as radar, LiDAR, or visual data \cite{zong2024mova}. Each expert focuses exclusively on its designated modality, achieving superior performance compared to models that process mixed-modal data without differentiation. Furthermore, MoE experts can be specialized based on general wireless network characteristics, such as network density, topology, propagation conditions, and traffic patterns \cite{wong2024addressing}, which enables the MoE framework to adapt to heterogeneous network environments and facilitate more efficient context-aware decision-making \cite{souza2021regularized}. 
\end{itemize}

\begin{figure*}
\centering
\includegraphics [width=0.95\textwidth]{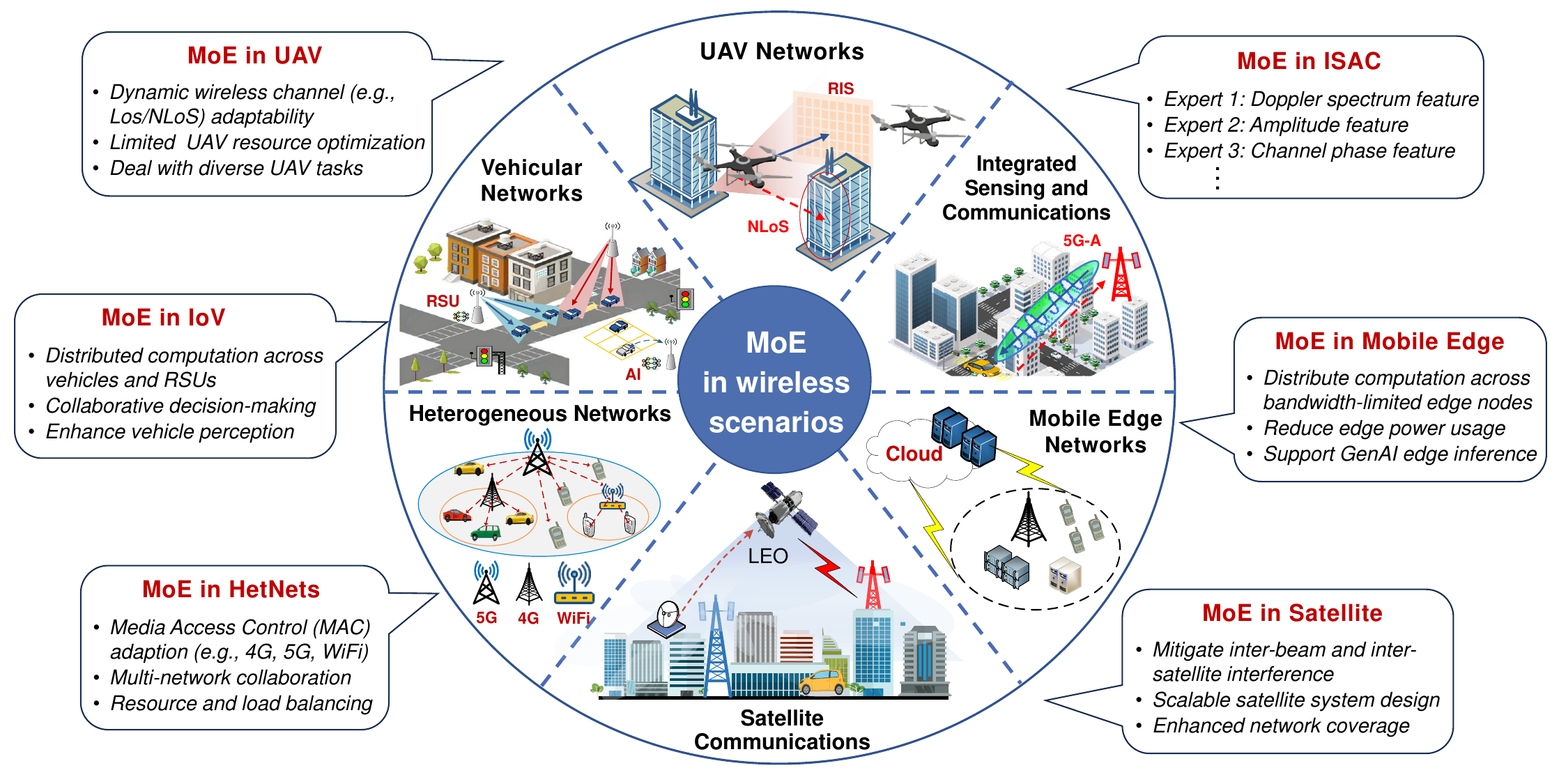} 
\captionsetup{justification=justified,format=plain}
\caption{Illustration of MoE integration across diverse wireless network scenarios, including vehicular networks, UAV networks, heterogeneous networks (HetNets), satellite networks, integrated sensing and communications (ISAC), and mobile edge networks.  
}
\label{Section3_Scenario}
\end{figure*}

\subsection{ MoE in Wireless Network scenario}
As illustrated in Figure \ref{Section3_Scenario}, MoE has been integrated into various wireless communication scenarios, including vehicular networks, UAV networks, satellite networks, heterogeneous networks (HetNets), integrated sensing and communications (ISAC), and mobile edge networks. 
The following section explores the application of MoE in these wireless scenarios, with a summary of the integration provided in Table~\ref{table:summary_scenario}.


\subsubsection{Vehicular Networks}
Vehicular networks have become a crucial component of modern transportation systems, enabling real-time data exchange between vehicles and roadside units (RSUs) to enhance traffic efficiency and safety \cite{yuan2023temporal, wu2022deep}. However, the highly dynamic nature of vehicular environments, characterized by rapid topology changes, heterogeneous driver behaviors, and strict latency requirements, poses significant challenges for intelligent decision-making systems \cite{xu2021leveraging, zhao2023autonomous}. 
MoE has emerged as an effective approach to address these challenges by dynamically selecting specialized expert networks tailored to different driving scenarios.
In the urban road scenario, the authors in \cite{vyas2023federated} leverage a multi-gate MoE (MMoE) to process vehicle telematics data with RSU-vehicle collaboration, allowing for more accurate driver behavior prediction in resource-constrained vehicles. The proposed model achieves an accuracy of 98\%, which outperforms traditional centralized models without MoE by 4\%.
In addition, an MoE-enhanced SAC framework incorporates heuristic experts and learning-based experts \cite{yao2025mixture}. Through dynamically activating these two models, MoE can effectively mitigate collision risks and ensure smooth lane transitions, leading to a 13.75\% improvement in average driving speed compared to a conventional DRL method with continuous action spaces.

\subsubsection{UAV Networks}

UAVs are increasingly employed across a wide range of applications, including aerial surveillance, disaster response, and communication relay \cite{stocker2017review}. Nonetheless, UAV networks may face significant challenges, such as real-time state estimation, limited computational resources, and adaptive trajectory optimization in uncertain environments \cite{gupta2015survey, zheng2021uav}. MoE enables UAV networks to improve real-time control, trajectory planning, and task execution by employing expert models adapted to specific flight conditions \cite{zhao2025generative, 10371877}. For example,
To improve UAV position estimations using a single Kalman filter (KF), the authors in \cite{kawamura2023hierarchical} leverage multiple KF experts based on real-time flight conditions, ensuring resilience against sensor noise and enabling UAVs to achieve a 20\% reduction in localization errors. 
Additionally, given the high computational cost of large-scale AI models, MoE-based model partitioning has been proposed to allow UAVs to load only the most relevant experts dynamically, reducing GPU memory usage by 86.4\% while maintaining 98.2\% UAV-onboard inference accuracy \cite{chen2024giant}. 
MoE can be further employed to effectively adapt to complex aerial environments such as line-of-sight (LoS) and non-line-of-sight (NLoS) aerial-ground radio propagation. For instance, a reconfigurable intelligent surface (RIS)-aided UAV network is introduced in \cite{wong2024addressing}, where MoE’s expert selection mechanism enables fast adaptation to diverse communication conditions, leading to a 17\% improvement in UAV trajectory utility evaluated by the cumulative quality of wireless links along the UAV path.

\begin{table*}
\centering
\caption{Summary of Wireless Network Scenarios with Mixture of Experts Integration. \\ \textbf{Note:} Light blue circles indicate the MoE method, green checkmarks and red crosses represent the advantages and challenges of MoE }
\label{table:summary_scenario}
\renewcommand{\arraystretch}{1.19}
{\fontsize{8.5pt}{10pt} \selectfont
\begin{tabular}{|m{0.08\textwidth}<{\centering}|m{0.045\textwidth}<{\centering}|m{0.165\textwidth}<{\centering}|m{0.17\textwidth}|m{0.42\textwidth}<{\centering}|}
\hline  
\textbf{Scenario} & \textbf{Ref.} & \textbf{Network Component} & $~$\textbf{Optimization Variable} & \textbf{MoE Advantages \& Challenges} \\ \hline
\multirow{3}{3cm}{Vehicular \\ Networks} 
& \cite{vyas2023federated} & 
\begin{itemize}[leftmargin=*]
\item[\textcolor{blue}{\resizebox{0.5em}{!}{\ding{108}}}] Connected vehicles
\item[\textcolor{blue}{\resizebox{0.5em}{!}{\ding{108}}}]  Roadside units
\item[\textcolor{blue}{\resizebox{0.5em}{!}{\ding{108}}}] Cloud server
\end{itemize}
& Vehicle speed, location, acceleration, and transmission overhead & 
\multirow{3}{0.42\textwidth}{ 
\vspace{-0.30in}
\begin{itemize}[leftmargin=*] 
\item [\textcolor{blue!40}{ \resizebox{0.6em}{!}{\ding{108}}}] Employ MMoE to learn distinct vehicle behaviors
\item [\textcolor{green}{\ding{51}}] Achieves 98\% RSU-aided behavior prediction (+4pp)
\item [\textcolor{red}{\ding{55}}] Limited computation and storage resource in vehicle
\end{itemize}
}

\\ \cline{2-5}

& \cite{yao2025mixture}  & 
\begin{itemize}[leftmargin=*]
\item[\textcolor{blue}{\resizebox{0.5em}{!}{\ding{108}}}] Connected vehicles
\item[\textcolor{blue}{\resizebox{0.5em}{!}{\ding{108}}}]  Vehicle controller
\item[\textcolor{blue}{\resizebox{0.5em}{!}{\ding{108}}}] V2I communication
\end{itemize}
& Steering angle, acceleration, and lane changing decision  &  
\multirow{3}{0.42\textwidth}{ 
\vspace{-0.3in}
\begin{itemize}[leftmargin=*] 
\item [\textcolor{blue!40}{ \resizebox{0.6em}{!}{\ding{108}}}] Different experts for speed control and lane selection
\item [\textcolor{green}{\ding{51}}] Improves average speed by 13.75\% with zero collision
\item [\textcolor{red}{\ding{55}}] MoE switching requires reliable real-time vehicle data
\end{itemize}
}

\\ \hline

\multirow{7}{3cm}{UAV\\Networks} 
&\cite{kawamura2023hierarchical} & 
\begin{itemize}[leftmargin=*]
\item[\textcolor{blue}{\resizebox{0.5em}{!}{\ding{108}}}] UAVs
\item[\textcolor{blue}{\resizebox{0.5em}{!}{\ding{108}}}] UAV onboard sensor
\item[\textcolor{blue}{\resizebox{0.5em}{!}{\ding{108}}}] Kalman filters
\end{itemize}
& UAV position, angular velocity, flying velocity, and attitude & 
\multirow{3}{0.42\textwidth}{ 
\vspace{-0.3in}
\begin{itemize}[leftmargin=*] 
\item [\textcolor{blue!40}{ \resizebox{0.6em}{!}{\ding{108}}}] Hierarchical MoE for multi UAV states modeling
\item [\textcolor{green}{\ding{51}}] Achieve a 20\% reduction in trajectory localization errors 
\item [\textcolor{red}{\ding{55}}] \spaceskip=0.3em Complex model switching under dynamic UAV conditions
\end{itemize}
}

\\ \cline{2-5}

& \cite{chen2024giant}  & 
\begin{itemize}[leftmargin=*]
\item[\textcolor{blue}{\resizebox{0.5em}{!}{\ding{108}}}] UAVs
\item[\textcolor{blue}{\resizebox{0.5em}{!}{\ding{108}}}] Edge servers 
\item[\textcolor{blue}{\resizebox{0.5em}{!}{\ding{108}}}] Ground base station
\end{itemize}  & UAV-Edge association, expert model selection, UAV memory capacity, and download delay  &  
\multirow{4}{0.42\textwidth}{ 
\vspace{-0.47in}
\begin{itemize}[leftmargin=*] 
\item [\textcolor{blue!40}{ \resizebox{0.6em}{!}{\ding{108}}}] GNN-based MoE to learn UAV-server selection strategy
\item [\textcolor{green}{\ding{51}}] \spaceskip=0.15em  Reduce 86.4\% UAV-onboard GPU memory usage and achieve 98.2\% UAV-onboard inference accuracy 
\item [\textcolor{red}{\ding{55}}] \spaceskip=0.3em Communication latency and storage limitation in UAV 
\end{itemize}
}

\\ \cline{2-5}

& \cite{wong2024addressing} & 
\begin{itemize}[leftmargin=*]
\item[\textcolor{blue}{\resizebox{0.5em}{!}{\ding{108}}}] UAVs
\item[\textcolor{blue}{\resizebox{0.5em}{!}{\ding{108}}}] RIS
\item[\textcolor{blue}{\resizebox{0.5em}{!}{\ding{108}}}] Ground UEs
\item[\textcolor{blue}{\resizebox{0.5em}{!}{\ding{108}}}] Eavesdropper
\end{itemize}
&  UAV and UE trajectories, UAV active beamforming, RIS passive $\quad$ beamforming & 
\multirow{4}{0.42\textwidth}{ 
\vspace{-0.47in}
\begin{itemize}[leftmargin=*] 
\item [\textcolor{blue!40}{ \resizebox{0.6em}{!}{\ding{108}}}] Employ MoE for multi UAV-RIS tasks
\item [\textcolor{green}{\ding{51}}] Imrpove 17\% UAV trajectory utility evaluated by the cumulative quality of wireless links along the UAV path
\item [\textcolor{red}{\ding{55}}] Uncertainty of UAV, UE, and eavesdropper trajectories
\end{itemize}
}

\\ \hline

\multirow{1}{3cm}{Satellite \\ Networks}  &\cite{zhang2024generative} & 
\begin{itemize}[leftmargin=*]
\item[\textcolor{blue}{\resizebox{0.5em}{!}{\ding{108}}}] LEO satellite
\item[\textcolor{blue}{\resizebox{0.5em}{!}{\ding{108}}}] GEO satellite
\item[\textcolor{blue}{\resizebox{0.5em}{!}{\ding{108}}}] Ground UEs
\item[\textcolor{blue}{\resizebox{0.5em}{!}{\ding{108}}}] Ground station
\end{itemize}
& \spaceskip=0.2em Spectrum and channel resource, private and common transit beamforming, energy efficiency, message rate & 
\multirow{5}{0.42\textwidth}{ 
\vspace{-0.62in}
\begin{itemize}[leftmargin=*] 
\item [\textcolor{blue!40}{ \resizebox{0.6em}{!}{\ding{108}}}] MoE-PPO for spectrum, channel, beamforming tasks
\item [\textcolor{green}{\ding{51}}] Optimize satellite communication throughput by  23.4\% and reduce transmission delay by 17.8\% 
\item [\textcolor{red}{\ding{55}}] \spaceskip=0.3em Continuously evolving and highly dynamic satellite network properties
\end{itemize}
}

\\ \hline

\multirow{1}{3cm}{HetNets} 
&\cite{liu2024meta} & 
\begin{itemize}[leftmargin=*]
\item[\textcolor{blue}{\resizebox{0.5em}{!}{\ding{108}}}] Access points
\item[\textcolor{blue}{\resizebox{0.5em}{!}{\ding{108}}}] WiFi systems
\item[\textcolor{blue}{\resizebox{0.5em}{!}{\ding{108}}}] Mobile devices
\end{itemize}
& Multi-access, channel $\quad$ collision probability, $\quad$ network throughput, $\quad$ and fairness  & 
\multirow{3}{0.42\textwidth}{ 
\vspace{-0.32in}
\begin{itemize}[leftmargin=*] 
\item [\textcolor{blue!40}{ \resizebox{0.6em}{!}{\ding{108}}}] Meta-RL based MoE for multi-access control modeling
\item [\textcolor{green}{\ding{51}}] Improve network spectrum efficiency by 21.5\% 
\item [\textcolor{red}{\ding{55}}] \spaceskip=0.3em Real-time adjustment in accessing heterogeneous 5G, WiFi, and IoT networks
\end{itemize}
}

\\ \hline

\multirow{1}{3cm}{ISAC} 
&\cite{wang2024optimizing} & 
\begin{itemize}[leftmargin=*]
\item[\textcolor{blue}{\resizebox{0.5em}{!}{\ding{108}}}] IoT Devices
\item[\textcolor{blue}{\resizebox{0.5em}{!}{\ding{108}}}] Sensing modules
\item[\textcolor{blue}{\resizebox{0.5em}{!}{\ding{108}}}] Datasets
\end{itemize}
& CSI amplitude, phase, Doppler, and target detection accuracy & 
\multirow{3}{0.425\textwidth}{ 
\vspace{-0.3in}
\begin{itemize}[leftmargin=*] 
\item [\textcolor{blue!40}{ \resizebox{0.6em}{!}{\ding{108}}}] MoE experts to characterize different CSI feature
\item [\textcolor{green}{\ding{51}}] Achieve 18\% improvement in multi-target detection accuracy 
\item [\textcolor{red}{\ding{55}}] \spaceskip=0.2em Simultaneously perform sensing and communication tasks
\end{itemize}
}

\\ \hline

\multirow{1}{3cm}{Mobile Edge \\ Networks} 
&\cite{li2024theory} & 
\begin{itemize}[leftmargin=*]
\item[\textcolor{blue}{\resizebox{0.5em}{!}{\ding{108}}}] Edge servers
\item[\textcolor{blue}{\resizebox{0.5em}{!}{\ding{108}}}] Base stations
\item[\textcolor{blue}{\resizebox{0.5em}{!}{\ding{108}}}] Mobile devices
\end{itemize} & Task offloading strategy, data transfer and
computation time & 
\multirow{3}{0.425\textwidth}{ 
\vspace{-0.3in}
\begin{itemize}[leftmargin=*] 
\item [\textcolor{blue!40}{ \resizebox{0.6em}{!}{\ding{108}}}] Employ experts to match task type and server capability
\item [\textcolor{green}{\ding{51}}] \spaceskip=0.22em Reduce test loss in generalization errors by 15.4\%
\item [\textcolor{red}{\ding{55}}] \spaceskip=0.25em Varied MEC features of server availability and capability
\end{itemize}
}

\\ \hline

\end{tabular}
}
\end{table*}

\subsubsection{Satellite Networks}
Satellite networks serve as a fundamental component of global communication systems, particularly for providing connectivity in remote and under-served areas \cite{qin2023service}. However, managing satellite communication resources presents significant challenges due to the heterogeneity of satellite constellations, the dynamic nature of orbital movement, and the need for adaptive power allocation and beamforming \cite{zhang2022cybertwin, ma2022satellite}. To address these challenges, the authors in \cite{zhang2024generative} enhance transmission efficiency by selecting specialized MoE experts for multi-task processing in Low Earth Orbit (LEO) and Geostationary Earth Orbit (GEO) satellite networks, including power control, beamforming, and interference mitigation. Through integrating MoE-based PPO frameworks, satellite networks achieve a 23.4\% increase in throughput while reducing transmission interference by 17.8\% compared to traditional RL-based approaches. Besides, MoE has proven highly effective in onboard satellite data processing. Traditional processing methods require extensive off-line computational resources, whereas the authors in \cite{loyola2002combining} leverage MoE-optimized neural networks to enable near-real-time satellite data processing. The integration of MoE with significantly reduce latency while maintaining high accuracy and minimal memory usage, offering a promising way for the deployment of next-generation spaceborne AI applications.

\subsubsection{Heterogeneous Networks}
HetNets are characterized by the coexistence of multiple communication technologies, such as 5G, WiFi, and IoT networks \cite{agarwal2022comprehensive, zhangcell1, liu2024ga}. Efficient allocation of radio resources and dynamic network access control are critical in these networks due to interference management challenges and network congestion \cite{hui2022digital, jiao2023performance}. MoE has been successfully applied to multiple access control (MAC) protocols to optimize spectrum access decisions dynamically across diverse environments. 
For example, the authors in \cite{liu2024meta} leverage Meta-RL to adjust MAC policies based on real-time network conditions. An MoE-enhanced encoder architecture is established to allow fine-grained access task representation, leading to faster convergence and improved generalization across varying spectrum scenarios. Simulation results demonstrate that MoE-based MAC improves spectrum efficiency by up to 21.5\% compared to DRL approaches without MoE. MoE-driven HetNets achieve robust spectrum sharing, and higher network throughput, providing a promising solution for next-generation heterogeneous wireless access technology.

\subsubsection{Integrated Sensing and Communications}
ISAC has emerged as one of the fundamental technologies in 6G wireless networks, where sensing and communication functionalities are integrated to optimize network intelligence and environmental awareness \cite{zhu2024enabling}. MoE enhances ISAC performance by activating distributed ISAC experts for parallel execution of multiple sensing and communication tasks. For instance, the authors in \cite{wang2024optimizing} leverage different MoE experts for beamforming, spectrum allocation, and objective tracking tasks,  
achieving 18\% improvement in multi-target detection accuracy while enhancing the efficiency of spectrum allocation and signal processing.
In addition, by leveraging MoE-based multi-modal learning, such as characterizing channels with Doppler effect, amplitude, and phase features, ISAC systems can efficiently fuse diverse data sources and ensure real-time adaptability to changing network conditions. This capability is especially advantageous in environments with rapidly fluctuating network conditions, such as smart cities and industrial automation.

\subsubsection{Mobile Edge Networks}
Mobile edge computing (MEC) plays a critical role in low-latency and high-efficiency wireless communications, where computational resources are deployed closer to end-users to reduce the reliance on centralized cloud computing \cite{mehrabi2019device, lin2021model}. MoE improves MEC efficiency by dynamically offloading computational workloads via base stations (BSs), optimizing both inference and training tasks among mobile devices and edge servers \cite{yi2023edgemoe, xu2021cybertwin }. 
Through an adaptive gating network, MoE can select edge experts for each incoming task, ensuring efficient task routing and load balancing \cite{li2024theory}. The employment of MoE in mobile edge networks mitigates the limitations of traditional task offloading strategies, which often lead to high generalization errors and inefficient resource utilization, reducing test loss by up to 15.4\% compared to conventional models.
Furthermore, MoE improves real-time decision-making by enabling edge devices to dynamically adjust model complexity based on energy constraints and communication bandwidth, striking an optimal balance between computational efficiency and task performance
\cite{sarkar2023edge}.


\subsection{Deployment Challenges and Limitations in Wireless Communications}

\subsubsection{Resource and Deployment Constraints}
While MoE architectures offer promising scalability and efficiency across diverse wireless scenarios, their practical deployment in wireless communication systems remains challenging. Real-world wireless environments, such as vehicular networks, UAV-assisted aerial communication, and LEO satellite systems, impose stringent constraints on computational and wireless resources. These constraints limit the applicability of monolithic deep models and pose unique challenges to deploying MoE in a distributed wireless setting.

\begin{itemize}
    \item \textbf{Computational Limitation and Heterogeneity:}
In wireless edge scenarios, devices such as RSUs, UAVs, and LEO satellites are typically equipped with limited onboard processing capabilities, memory storage, and energy consumption. These limitations fundamentally constrain the feasibility of deploying large-scale MoE models in real-world wireless environments. 
Unlike cloud servers with sufficient computational resources to support intensive parallel processing, edge devices often struggle to support even moderately sized inference workloads. Moreover, the computational heterogeneity across different edge nodes, ranging from low-power microcontrollers to advanced AI accelerators, further complicates consistent MoE deployment and performance optimization in wireless systems \cite{wang2025empowering}.


    \item \textbf{Communication Resource Limitation:} In decentralized wireless networks, experts may be distributed across multiple edge nodes with heterogeneous capabilities and variable connectivity. Coordinated MoE execution under such conditions requires efficient model partitioning, inter-node communication, and synchronization. However, limited spectrum, backhaul capacity, and transmission power can severely degrade collaboration performance among MoE experts. These constraints introduce new challenges in expert placement, load balancing, and communication overhead. For example, dynamic wireless links may cause fluctuating delays and partial expert failures, thereby complicating expert routing and increasing inference uncertainty \cite{xue2024wdmoe}. 

    \item \textbf{Trade-offs Between Model Performance and Resource Efficiency: } The expert activation mechanism of MoE provides flexibility to control the number of active experts during inference. However, practical deployment requires a delicate balance between model accuracy and system resource usage. Over-activating experts generally leads to higher prediction performance but incurs increased latency, memory access, and energy consumption, potentially offsetting the overall gains in model efficiency. Conversely, aggressive sparsity may compromise task performance in wireless systems  due to insufficient expert capacity to adapt to rapid variations in dynamic wireless environments \cite{wang2025unified}. Therefore, it remains challenging for adaptive expert scheduling strategies to meet application-specific QoS requirements, such as task priority, channel conditions, or residual battery levels, while also satisfying practical system resource constraints.
\end{itemize}

\subsubsection{Privacy and Security Challenges in Multi-Agent MoE Systems}
The deployment of MoE in distributed wireless environments, such as vehicular networks, multi-UAV coordination, and distributed ISAC, inherently aligns with the architectural paradigm of multi-agent AI systems, where autonomous agents exchange intermediate representations and engage in collaborative inference. However, the openness and decentralization of the MoE framework increase the system's exposure to security vulnerabilities, undermine trustworthy collaboration, and pose significant challenges to the resilience of distributed decision-making under dynamic network conditions.

\begin{itemize}

    \item \textbf{Privacy Leakage through Expert Sharing:} In collaborative multi-agent MoE setups, expert modules may be offloaded, shared, or replicated across heterogeneous nodes (e.g., vehicle-to-vehicle or UAV-to-UAV), raising privacy concerns in systems composed of multiple expert agents \cite{zhang2024survey}. Even without direct data exchange, privacy leakage can occur through latent features and activation patterns that implicitly encode sensitive user behaviors or location traces. Recent studies have demonstrated that collaborative inference and activation inversion attacks can reconstruct user attributes or usage histories, while gradient-based leakage further exposes training-time identities \cite{su2025pwc}. These risks are aggravated in multi-agent MoE due to frequent expert migration and routing synchronization across agents. To mitigate such threats, privacy-preserving communication and federated MoE frameworks can be leveraged, integrating differential privacy, encrypted aggregation, and split learning to protect shared representations while maintaining model accuracy \cite{hou2025split}. Incorporating these mechanisms is essential for constructing trustworthy and privacy-aware multi-agent MoE systems tailored to wireless edge networks.

    \item \textbf{Trust Mechanisms for Multi-Agent MoE:} In decentralized multi-agent MoE deployments, reliable inter-agent trust serves as a critical foundation for secure and coordinated expert collaboration. Traditional approaches such as identity verification are often insufficient in dynamic, decentralized settings where agent behaviors are heterogeneous and unpredictable \cite{ma2023trusted}. To establish operational trust between agents, mechanisms such as reputation-based expert scoring \cite{al2023reputation} to evaluate historical performance, policy-consistent access control \cite{liu2024enablingefficient} to prevent unauthorized expert invocation, and tamper-evident audit trails \cite{wang2024blockchain} to record expert selection and routing can serve as effective means for secure MOE expert collaboration. Consistent with trust requirements in edge intelligence-based MoE, blockchain-assisted ledgers can record model versions and inter-agent transactions, while dynamic trust scores guide expert selection and task delegation under intermittent connectivity \cite{zhu2025moe}. Moreover, multi-agent trust frameworks, such as trust inference and attack detection, offer generalizable patterns that can be seamlessly integrated into MoE expert selection and coordination across heterogeneous agents \cite{wang2024survey}. 
    
    \item \textbf{Assurance and Resilience in Trustworthy Multi-Agent MoE:} Ensuring assurance and resilience in multi-agent MoE systems remains a fundamental and challenging task, requiring that expert routing and inter-agent collaboration be verifiable, explainable, and robust against both system-level failures and adversarial threats. Within the trustworthy assurance layer, providing interpretable explanations and calibrated uncertainty estimates for MoE gating and routing strategies is essential to enable traceable decision-making and support risk management in large-scale, decentralized environments \cite{jiang2023mixture}. Recent research suggests that combining interpretability with uncertainty quantification can enhance trust calibration, particularly when multiple agents jointly select experts under dynamic communication and resource constraints \cite{he2024toward}. Additionally, extending assurance to system-level operation introduces additional complexity,  particularly in wireless edge deployments where achieving runtime integrity becomes a critical challenge. Ensuring trust and resilience across edge agents requires not only remote attestation to verify device integrity but also redundancy-aware task allocation and dynamic trust reconfiguration to sustain reliable cooperation \cite{aljabri2025permutation}. Ultimately, trustworthy and resilient operations form the foundation for multi-agent MoE systems, enabling stable performance under dynamic and uncertain environments.

\end{itemize}


\textbf{Lessons Learned:} 
The integration of MoE into wireless communication systems has demonstrated promising potential in improving adaptability, computational efficiency, and intelligent decision-making across diverse wireless scenarios. Through leveraging sparse expert activation and dynamic selection mechanisms, MoE enables efficient task execution under wireless edge environments, such as vehicular \cite{obando2024mixtures,tao2022double}, UAV \cite{kawamura2023hierarchical, chen2024giant, wong2024addressing}, and satellite networks \cite{zhang2024generative, loyola2002combining}. Furthermore, MoE supports multi-modal data fusion and adaptive resource allocation, which are essential for emerging paradigms such as multi-access control \cite{liu2024meta}, ISAC \cite{wang2024optimizing}, and mobile edge computing \cite{liu2024multimodal, li2024theory}. However, the practical deployment of MoE in wireless networks remains constrained by several challenges, such as increased computational overhead and imbalanced expert utilization under dynamic and uncertain environments. These challenges become especially critical in large-scale, resource-constrained wireless edges, where bandwidth, energy, and storage resources are severely restricted. Future research requires a focus on developing lightweight and generalizable MoE architectures, improving balanced expert activation, and designing resource-aware gating mechanisms that jointly optimize quality, latency, and energy efficiency for next-generation wireless systems.

\section{Applications of Mixtures of Experts in Wireless Networking}

In this section, we explore key applications of MoE in physical layer communication and resource management, including channel prediction and estimation, physical layer signal processing, radio resource allocation, as well as higher-level network optimization, including traffic management, distributed computing, and network security.  A detailed summary of these applications is presented in Tables \ref{SectionIv:summary_fisrt} and \ref{SectionIv:second}.

\subsection{Physical Layer Communication and Resource Management}

\subsubsection{Channel Prediction and Estimation }

\begin{figure*}
\centering
\includegraphics [width=\textwidth] 
{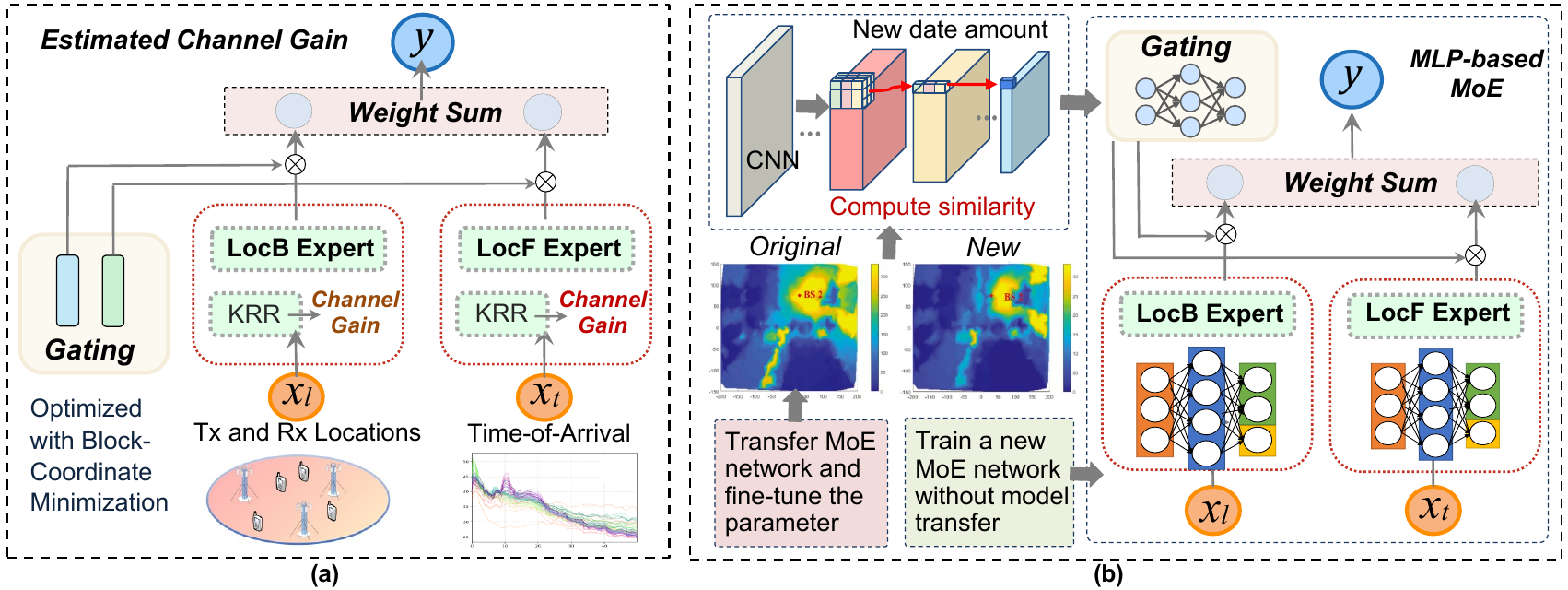} 
\captionsetup{justification=justified,format=plain}
\caption{ Illustration of channel estimation using MoE. (a) The MoE framework in \cite{lopez2020channel} combines a LocB expert that uses the transmitter's and receiver's locations and a LocF expert based on time-of-arrival features, weighted by a gating network optimized through block-coordinate minimization. (b) Exploit transfer learning to fine-tune a pre-trained MoE model or directly train a new MoE network without model transfer \cite{jaiswal2023leveraging}. For the transfer learning scheme, a CNN module is employed to compute the similarity between the source and new environments, determining the amount of data required for fine-tuning the pre-trained MoE.}
\label{LocB_and_LocF}
\end{figure*}

In wireless communication, channel prediction and estimation are critical tasks for ensuring optimal system performance, especially in complex environments involving multiple users and channel conditions \cite{fesl2023learning}. Traditional channel prediction methods typically rely on prior channel models and limited historical data \cite{liu2014channel}. However, these methods may struggle in dynamic environments with rapidly changing channel conditions \cite{senevirathna2005self}. The key advantage of the MoE model lies in its ability to employ multiple experts specialized in different aspects of the channel, which can significantly enhance the accuracy of estimation and prediction under complex channel environments \cite{haihan2016novel}. 


For example, the authors in \cite{senevirathna2006channel} address the critical challenge of adapting to fast time-varying CSI in multi-path fading environments by introducing a self-organizing map (SOM)-based MoE framework. 
Each MoE expert module is implemented using a functional link neural network (FLNN), which offers a computationally efficient alternative to conventional DNN models while maintaining robust approximation capabilities \cite{dehuri2010comprehensive}. The FLNN-based experts serve as nonlinear channel predictors, effectively capturing the complex temporal variations of Rayleigh fading channels. Additionally, a radial basis function (RBF) network is utilized as the gating mechanism, dynamically assigning weights to different experts based on the extracted local channel characteristics. Through extensive simulations in an OFDMA system with a 5-tap Rayleigh fading channel and a maximum Doppler spread of 40 Hz, the proposed predictor demonstrates its capability to forecast CSI five steps ahead (5 ms prediction horizon). This predictive capability significantly enhances adaptive modulation schemes by mitigating the impact of outdated CSI. In addition, simulation results indicate that the MoE-based predictor achieves a 19.2\% improvement in channel prediction accuracy, while simultaneously maintaining a low bit error rate (BER) under real-time prediction constraints.

Building on the strengths of hybrid approaches, the authors in \cite{lopez2020channel} introduce an MoE-based framework for channel gain (CG) estimation, integrating both location-based (LocB) and location-free (LocF) predictive models. As illustrated in Figure \ref{LocB_and_LocF}(a), the LocB expert utilizes estimated locations of transmitter and receiver as input, employing the kernel ridge regression (KRR) method \cite{vovk2013kernel} with a distance-based loss to infer CG variations. In contrast, the LocF expert extracts pilot signal features, such as time-of-arrival (ToA) from the channel impulse response. The KRR method is subsequently employed for CG gain estimation in environments where accurate positioning is unreliable due to multipath propagation and NLoS effects. The key advantage of the MoE-based approach lies in its adaptive gating mechanism, which dynamically assigns weights to the LocB and LocF experts based on localization uncertainty. 
The gating mechanism is optimized using block-coordinate minimization (BCM) \cite{xu2017globally}, which iteratively updates the gating function weights and expert parameters to improve estimation accuracy while mitigating overfitting risks.  Experimental results demonstrate that this hybrid MoE framework achieves a 35\% reduction in location-based estimation errors compared to conventional LocB-only approaches, while significantly improving CG mapping accuracy in NLoS conditions.

Unlike the study in \cite{senevirathna2006channel}, which employs a simple FLNN structure that may struggle to capture complex channel characteristics, and the study in \cite{lopez2020channel}, which relies on BCM for gating optimization but risks suboptimal expert selection due to local minima, the authors in \cite{jaiswal2023leveraging} enhance expert selection and channel estimation by leveraging DNN-based LocB and Loc experts, along with a DNN-based gating network.
In addition, to further extend its capabilities to the new wireless environment, the authors employ a transfer learning approach for wireless environment adaptation. As depicted in Figure \ref{LocB_and_LocF}(b), when it comes to a new wireless environment, the authors either fine-tune a pre-trained MoE model via transfer learning or directly train a new network without model transfer. A CNN module is employed to extract features and compute the similarity between the source and new environments, determining the amount of data required for fine-tuning the pre-trained MoE network. Through the transfer learning scheme, the proposed approach requires only 20\%-40\% of additional data to retrain the prediction model for target wireless environments. Besides, the MoE-based approach outperforms individual LocB and Loc models in mean squared error (MSE) by 26\%-47\%, demonstrating significant effectiveness in improving radio map estimation accuracy under dynamically changing wireless environments.

\begin{figure*}
\centering
\includegraphics [width=\textwidth] 
{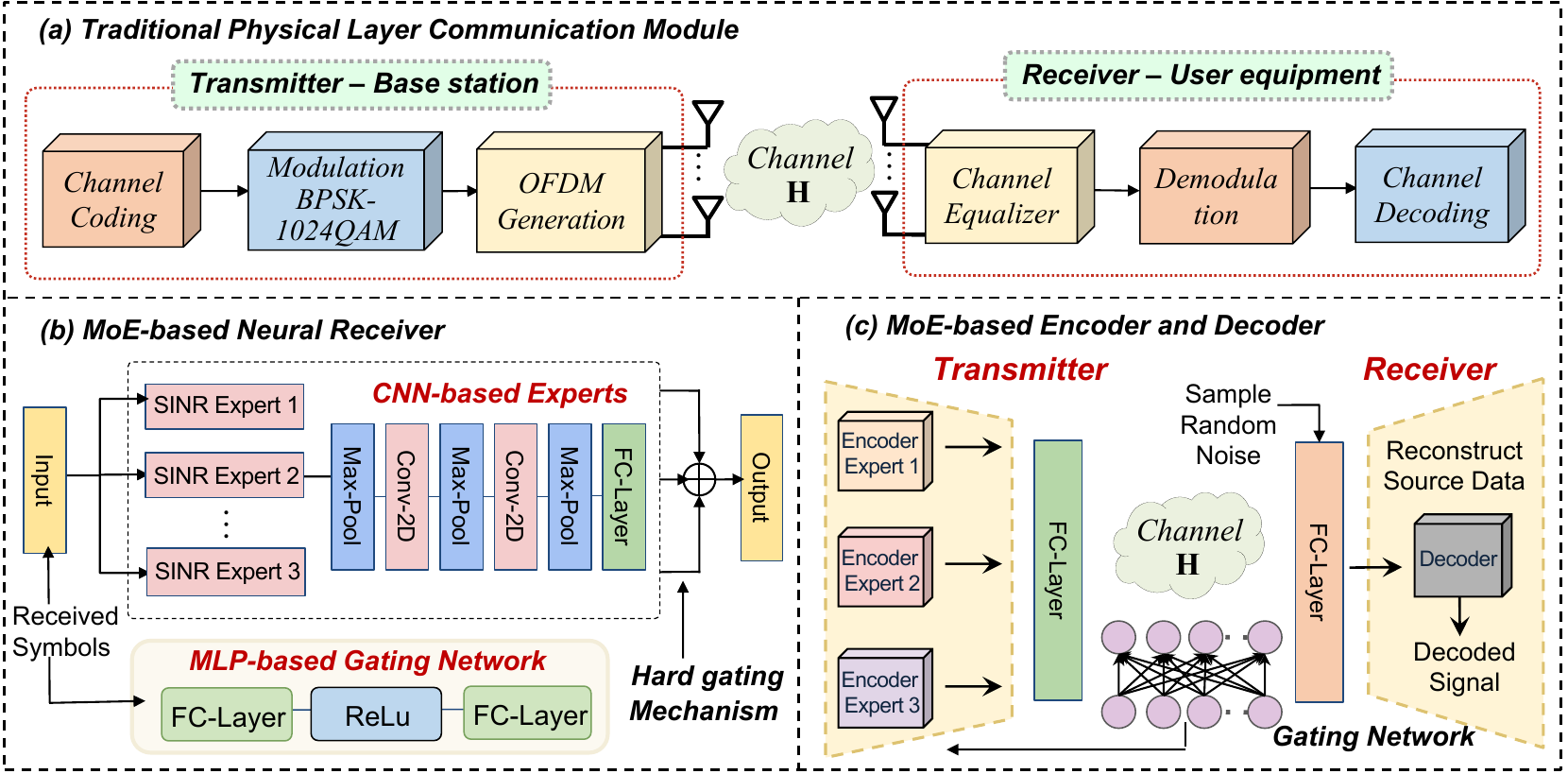} 
\captionsetup{justification=justified,format=plain}
\caption{ Illustration of physical layer transmission for wireless communication. (a) The traditional physical layer modules, consisting of channel coding, modulation, OFDM generation, equalization, demodulation, and decoding. (b) The MoE-based neural receiver employs a hard gating network to select CNN-based experts, each specializing in different SINR conditions \cite{van2024mean}. (c) The MoE-based joint source-channel coding framework employs multiple encoder experts and an MLP gating network to adapt encoding strategies to varying source characteristics \cite{saidutta2021joint}.
}
\label{Section4_physical_layer}
\end{figure*}

\subsubsection{Physical Layer Signal Processing and Communication}

Traditional wireless transceivers employ predetermined signal processing modules encompassing channel coding, modulation, equalization, and demodulation, as depicted in Figure \ref{Section4_physical_layer}(a). While these techniques are foundational to modern communication systems, their rigid structure limit adaptability to dynamic wireless environments and fluctuating channel conditions \cite{fischer2022mixture}. Neural network-based transceivers have emerged as a transformative alternative, offering enhanced signal processing, channel modeling, and error correction capabilities \cite{aoudia2021end}. However, the increasing model complexity and computational demands of neural transceivers pose substantial deployment challenges \cite{hoydis2022sionna}. MoE addresses these challenges by employing multiple experts and dynamically activating specialized experts, significantly reducing parameter overhead while improving adaptability to diverse communication conditions.

In the realm of neural transmitter,
MoE has demonstrated significant efficacy in radio frequency (RF) power amplifier (PA) linearization by effectively mitigating nonlinear distortion effects. Specifically, the authors in \cite{fischer2023sparsely} propose a sparsely gated MoE neural network (MENN) framework, which integrates a real-valued time-delay neural network (RVTDNN) as expert models, coupled with a top-K gating mechanism to optimize computational efficiency. The experts of the proposed MENN are trained to capture distinct nonlinear and memory effects of PAs. 
The gating network is designed as a fully connected neural network, taking the instantaneous signal envelope as input and determining the optimal expert combination via a softmax-based probabilistic weighting function. When applied to 100 MHz and 120 MHz 5G NR OFDM signals, the proposed approach achieves significant reductions in modeling error and adjacent channel leakage ratio compared to conventional NN-based methods. In addition, the MENN reduces half run-time complexity while maintaining superior linearization performance, demonstrating its practical viability for high-throughput real-time implementations.

Different from the study in \cite{fischer2023sparsely} that focuses on enhancing the transmitter-side PA linearization, the authors in \cite{van2024mean} address the receiver-side processing in neural transceivers. As illustrated in Figure \ref{Section4_physical_layer}(b), an MoE-based adaptive neural (MEAN) receiver is introduced for received signal decoding in single-input multiple-output (SIMO) systems. Unlike conventional static neural receivers, which deploy a neural network to handle all channel conditions, the MEAN architecture adopts a dynamic neural network paradigm, leveraging a hard-gated MoE framework to optimize both computational efficiency and adaptability. The hard gating mechanism dynamically activates a CNN-based expert, which is specialized for distinct SINR conditions. Instead of using dense gating mechanisms, the hard gating expert selection ensures that only a single expert is active at any given time. This design significantly reduces runtime computational complexity by up to 50\% compared to static neural receivers, while maintaining competitive demodulation accuracy.  Additionally, MEAN surpasses conventional least-squares (LS)-based estimation techniques, especially in moderate-to-high SINR regimes, underscoring the efficacy of MoE-driven methods in next-generation wireless receivers. 

Extending beyond separate optimizations of the transmitter or receiver, the study in \cite{saidutta2021joint} integrates MoE into a unified end-to-end transceiver framework, addressing both encoding and decoding through joint source-channel coding (JSCC).  
As depicted in Figure \ref{Section4_physical_layer}(c), the authors in \cite{saidutta2021joint} introduce an MoE-based JSCC framework that leverages a mixture of variational autoencoders (MoVAE) to enhance adaptive signal encoding and reconstruction in additive white Gaussian noise (AWGN) channels. Unlike traditional parametric encoding approaches, the proposed system employs multiple specialized VAE-based encoders to handle distinct regions of the source space, along with a universal decoder that ensures robust signal reconstruction. To dynamically assign encoders based on input characteristics, the framework uses an MLP-based gating network with an exponential loss function for encouraging encoder specialization while preventing mode collapse. Experimental results show that MoVAE achieves state-of-the-art rate-distortion performance, surpassing power-constrained vector quantization schemes, with a performance gap of less than 1dB from the Shannon limit in various bandwidth usage scenarios. The incorporation of an MLP-based gating network allows adaptive encoder selection based on real-time channel conditions, leading to a 20\% improvement in peak signal-to-noise ratio (PSNR) compared to single-VAE methods.

\begin{figure*}
\centering
\includegraphics [width=\textwidth] 
{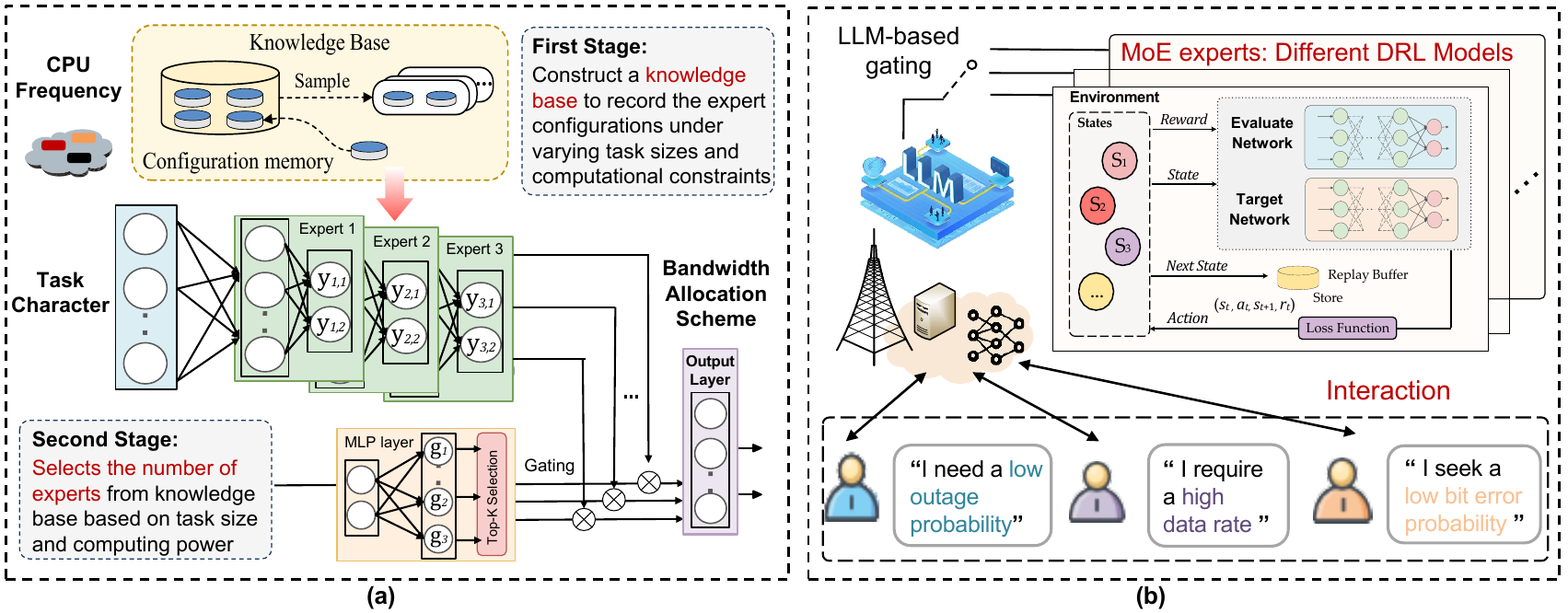} 
\captionsetup{justification=justified,format=plain}
\caption{ Illustration of radio resource allocation and management using MoE. (a) The knowledge-assisted two-stage framework for bandwidth allocation \cite{ma2022demand}. In the first stage, a knowledge base is constructed by training models with different numbers of MoE experts. In the second stage, the system dynamically selects the number of experts based on task size and available computing power, ensuring efficient bandwidth allocation.
(b) The LLM-based MoE framework in \cite{du2024mixture} dynamically assigns DRL experts for network task optimization based on the quality of service (QoS) requirements, such as minimizing outage probability, maximizing data rate, or reducing bit error probability. 
}
\label{Section4_resource_allocation}
\end{figure*}

\subsubsection{Radio Resource Allocation}

Efficient radio resource allocation is fundamental to optimizing wireless network performance, encompassing key tasks such as power, spectrum, and bandwidth allocation \cite{qian2023enabling, liu2024enabling, zhangGAI2}. In next-generation wireless networks, achieving optimal resource distribution is challenging due to dynamic channel conditions, diverse user requirements, and computational constraints \cite{lai2024resource, he2021learning, xue2024cooperative}. Traditional methods, such as iterative convex optimization or static machine learning approaches, often suffer from rigid structure, limited adaptability, and high computational overhead, rendering them unsuitable for rapidly changing environments \cite{du2024age}. 

Recent advancements in resource management have utilized MoE to tackle the challenges of transmit power management.
In \cite{zecchin2020team}, the authors propose a deep MoE (DMoE) framework for decentralized power control, addressing the challenge of adapting to time-varying CSI uncertainty without requiring frequent model retraining. Unlike conventional data-driven power control schemes, which assume a fixed noise level of CSI feedback and require costly retraining when noise statistics change, the proposed DMoE enables each expert to specialize in a different feedback noise regime. A gating network, implemented as a lightweight DNN, dynamically selects the most suitable expert based on the estimated CSI uncertainty level, effectively adapting to varying noise statistics. This selective expert learning strategy ensures robust power allocation policies while significantly reducing retraining overhead. 
Experimental results over a Rayleigh fading channel with multi-user interference show that DMoE achieves a 12\% higher sum-rate than conventional DNNs, particularly under dynamically varying feedback noise conditions.
Compared with weighted minimum mean square error (WMMSE)-based approaches, DMoE improves sum-rate by approximately 20\% while significantly reducing computational complexity, which shows the promising potential for real-time distributed resource allocation in next-generation wireless networks.

\begin{table*}
\centering
\caption{Summary of MoE Applications in Physical Layer Communication and Resource Management. \\  \textbf{Note:} Light blue circles indicate the MoE method, green checkmarks and red crosses represent the advantages and challenges of MoE }
\label{SectionIv:summary_fisrt}
\renewcommand{\arraystretch}{1.2}
{\fontsize{8.5pt}{10pt} \selectfont
\begin{tabular}{|m{0.155\textwidth}<{\centering}|m{0.045\textwidth}<{\centering}|m{0.11\textwidth}<{\centering}|m{2.7cm}|m{0.42\textwidth}<{\centering}|}
\hline
\textbf{Wireless Technology} & \textbf{Ref.} & \textbf{Wireless Task} & \textbf{MoE Framework} & \textbf{MoE Advanvatages \& Challenges} \\ \hline

\multirow{5}{3cm}{Channel Prediction and Estimation } 
&  \cite{senevirathna2006channel} & Channel prediction & 
Functional link neural network (FLNN)-based MoE experts & 
\multirow{1}{0.43\textwidth}{ 
\vspace{-0.27in}
\begin{itemize}[leftmargin=*] 
\item [\textcolor{blue!40}{ \resizebox{0.6em}{!}{\ding{108}}}] Employ FLNN experts to capture channel dynamics
\item [\textcolor{green}{\ding{51}}]  Maintain low BER under real-time prediction constraints
\item [\textcolor{red}{\ding{55}}] Suboptimal expert allocation due to SOM limitations
\end{itemize}
}
\\
\cline{2-5}

& \cite{lopez2020channel} & Channel gain estimation & Location-based (LocB) and location-free (LocF) experts  & 
\rule{0pt}{37pt}\multirow{3}{0.43\textwidth}{ 
\vspace{-0.7in}
\begin{itemize}[leftmargin=*] 
\item [\textcolor{blue!40}{ \resizebox{0.6em}{!}{\ding{108}}}]  Utilize MoE to combine LocB and LocF estimations
\item [\textcolor{green}{\ding{51}}] \spaceskip=0.2em Improve channel gain mapping accuracy under uncertainty
\item [\textcolor{red}{\ding{55}}] Gating function optimization may suffer from local minima
\item [\textcolor{red}{\ding{55}}] Generalization to new environments remains uncertain
\end{itemize}
}

\\ 
\cline{2-5}

&\cite{jaiswal2023leveraging} & Radio map estimation & Transfer learning $~~~$ based MoE & 
\rule{0pt}{37pt}\multirow{3}{0.43\textwidth}{ 
\vspace{-0.70in}
\begin{itemize}[leftmargin=*] 
\item [\textcolor{blue!40}{ \resizebox{0.6em}{!}{\ding{108}}}] Use transfer learning to fine-tune 
LocB and LoC experts
\item [\textcolor{green}{\ding{51}}] Reduces data demand via knowledge transfer
\item [\textcolor{red}{\ding{55}}] Effectiveness of MoE adaption depends on similarity between source and new environments
\end{itemize}
}

\\
\hline

\multirow{6}{3cm}{Signal Processing and Communication} 
& \cite{van2024mean} & Neural receiver $~~~$ design & CNN-based experts $~$ with hard gating $~~$ mechanism & 
\multirow{1}{0.43\textwidth}{ 
\vspace{-0.27in}
\begin{itemize}[leftmargin=*] 
\item [\textcolor{blue!40}{ \resizebox{0.6em}{!}{\ding{108}}}] Select specialized SNR experts for received signal
\item [\textcolor{green}{\ding{51}}]  Improve block error rate in SIMO systems
\item [\textcolor{red}{\ding{55}}] Hard gating may cause expert selection discontinuity
\end{itemize}
}
\\
 \cline{2-5}

& \cite{fischer2022mixture, fischer2023sparsely} & RF power amplifier linearization & MLP-based experts $~$ with sparse gating $~~$ mechanism & 
\multirow{3}{0.43\textwidth}{ 
\vspace{-0.3in}
\begin{itemize}[leftmargin=*] 
\item [\textcolor{blue!40}{ \resizebox{0.6em}{!}{\ding{108}}}] Use Top-K to select relevant experts for signal modeling
\item [\textcolor{green}{\ding{51}}]  Improve power efficiency and maintain signal quality 
\item [\textcolor{red}{\ding{55}}] Unbalanced expert training due to sparse gating
\end{itemize}
}

\\
\cline{2-5}

& \cite{saidutta2021joint} & Joint source-channel coding & VAE encoder-based experts with dense $~$ gating mechanism & 
\multirow{3}{0.43\textwidth}{ 
\vspace{-0.3in}
\begin{itemize}[leftmargin=*] 
\item [\textcolor{blue!40}{ \resizebox{0.6em}{!}{\ding{108}}}] Employ different encoder experts and reconstruct signal
\item [\textcolor{green}{\ding{51}}] Improve adaptation for varying channel conditions
\item [\textcolor{red}{\ding{55}}] \spaceskip=0.2em High training complexity due to dense gating
\end{itemize} 
}

\\
 \hline

 \multirow{6}{3cm}{Radio Resource Allocation} 
& \cite{zecchin2020team}  & Power control & MLP-based MoE experts and gating network & 
\rule{0pt}{20pt}\multirow{3}{0.43\textwidth}{ 
\vspace{-0.45in}
\begin{itemize}[leftmargin=*] 
\item [\textcolor{blue!40}{ \resizebox{0.6em}{!}{\ding{108}}}] Select experts for different noise levels
\item [\textcolor{green}{\ding{51}}] Enable decentralized power control without retraining
\item [\textcolor{red}{\ding{55}}]  Decentralized architecture may impact global efficiency
\end{itemize}
}

\\ \cline{2-5}

& \cite{ma2022demand} & Bandwidth allocation & Knowledge-based Dynamic neural $~~$ network (DyNN)  & 
\rule{0pt}{29pt}\multirow{3}{0.43\textwidth}{ 
\vspace{-0.623in}
\begin{itemize}[leftmargin=*] 
\item [\textcolor{blue!40}{ \resizebox{0.6em}{!}{\ding{108}}}] Adjust model capacity based on resource constraints
\item [\textcolor{green}{\ding{51}}] Balance computation and transmission delay
\item [\textcolor{red}{\ding{55}}] Offline knowledge base may have limited generalization
\end{itemize}
}
\\ \cline{2-5}

& \cite{du2024mixture} & Network task  optimization & LLM-based gating network & 
\rule{0pt}{29pt}\multirow{3}{0.43\textwidth}{ 
\vspace{-0.62in}
\begin{itemize}[leftmargin=*] 
\item [\textcolor{blue!40}{ \resizebox{0.6em}{!}{\ding{108}}}] LLM-based MoE for different optimization tasks
\item [\textcolor{green}{\ding{51}}] Reduce overhead while improving decision efficiency
\item [\textcolor{red}{\ding{55}}] High complexity and inference latency of LLM
\end{itemize}
}
\\ \hline

\end{tabular}
}
\end{table*}

While the study in \cite{zecchin2020team} leverages MoE to enhance power control adaptability, it employs a fixed network architecture that does not dynamically adjust to varying computational constraints and resource availability. To address this limitation, the study in \cite{ma2022demand} extends MoE-based decision-making by incorporating a knowledge-assisted dynamic neural network (DyNN) for bandwidth resource management.
As depicted in Figure~\ref{Section4_resource_allocation}(a), the maximum number of MoE experts is investigated in relation to performance, revealing that the optimal width of DyNN varies with both task size and CPU frequency. 
Instead of using static networks with fixed maximum experts, DyNN dynamically adjusts its network width through a two-stage optimization.
In the first stage, a knowledge base is constructed by training models with different maximum MoE expert numbers. The system analyzes their performance under varying tasks and computational constraints (e.g., CPU frequency) and records the optimal expert configurations.
In the second stage, the maximum number of experts is dynamically selected based on the knowledge base, considering the current task and available computing power. 
Simulation results demonstrate that increasing the maximum number of MoE experts improves bandwidth allocation performance, but at the cost of higher computational complexity. For instance, using only 1 expert achieves 93.6\% of the full 30-expert model performance, with less than 4\% of the total operation costs, while using 5 experts achieves over 98\% performance. 
In addition, by leveraging prior knowledge for adapting network width and computational resources, DyNN realizes a 47.1\% reduction in service delay for small tasks and 12.8\%–20.6\% reduction for larger tasks compared to static neural networks.

However, knowledge-based expert selection may be inherently limited in scenarios with unpredictable network situations, where offline knowledge may not generalize to unseen tasks. To further enhance adaptability, the authors in \cite{du2024mixture} propose an LLM-enabled MoE framework that eliminates the need for predefined expert mappings.
Through leveraging the contextual reasoning capabilities of LLMs to infer user requirements, LLMs themselves can be employed as intelligent gating mechanisms to select experts and assign expert weights accordingly.
As illustrated in Figure \ref{Section4_resource_allocation}(b),
a key application is the utility maximization of a network service provider (NSP), where users exhibit varying quality of service (QoS) demands, such as low outage probability (OP) for uninterrupted voice calls or high data rates for streaming applications. The LLM-based MoE first interprets textual user requests, translating them into formal optimization objectives. It then selects the most relevant DRL experts, such as OP minimization experts or data rate maximization experts, and synthesizes their outputs to determine an optimal transmit power allocation strategy. Unlike traditional gating networks, LLMs can obtain human intent from natural language, enabling better alignment between user requests and optimization goals in diverse user-centric tasks.
Compared to single DRL-based power control, the LLM-based MoE approach achieves over 40\% reduction in computational cost, resulting in a 15\% revenue gain for the NSP under varying network loads and user demands.

\subsection{Network Optimization and Security}

\subsubsection{Traffic Management}

Traffic management involves intricate spatial-temporal dependencies, rapidly evolving vehicular environments, and stringent requirements for real-time decision-making \cite{ke2024interpretable,sun2024generalizing, xue2024sparse}.
MoE frameworks can effectively utilize specialized expert models for distinct tasks such as traffic flow prediction, time series analysis, and driving motion estimation, thereby enhancing performance and adaptability in complex traffic optimization scenarios \cite{john2018estimation}.
To capture the complex spatial structures of road networks and the temporal dependencies in traffic dynamics, the authors in \cite{chattopadhyay2022mixture} propose an MoE-based spatial-temporal graph convolutional network (STGCN) for traffic state prediction. The proposed approach integrates multiple graph neural network (GNN)-based experts, with a neural gating network to dynamically assign expert weights based on real-time traffic inputs. Unlike traditional deep learning models that rely on a single traffic predictor, STGCN effectively learns region-specific traffic patterns through different experts. To prevent expert overfitting and encourage specialization, the authors introduce a novel entropy-based loss function, which ensures that different experts focus on distinct regions of the input space rather than all experts contributing equally to every prediction.
Experiments conducted on 228 road segments with traffic flow data demonstrate that, compared to single GNN model, STGCN more effectively captures diverse spatial-temporal traffic patterns, resulting in a 12.3\% reduction in mean absolute error (MAE) for 15-minute forecasts, 10.6\% for 30-minute forecasts, and 8.8\% for 45-minute forecasts.

\begin{figure*}
\centering
\includegraphics [width=\textwidth] 
{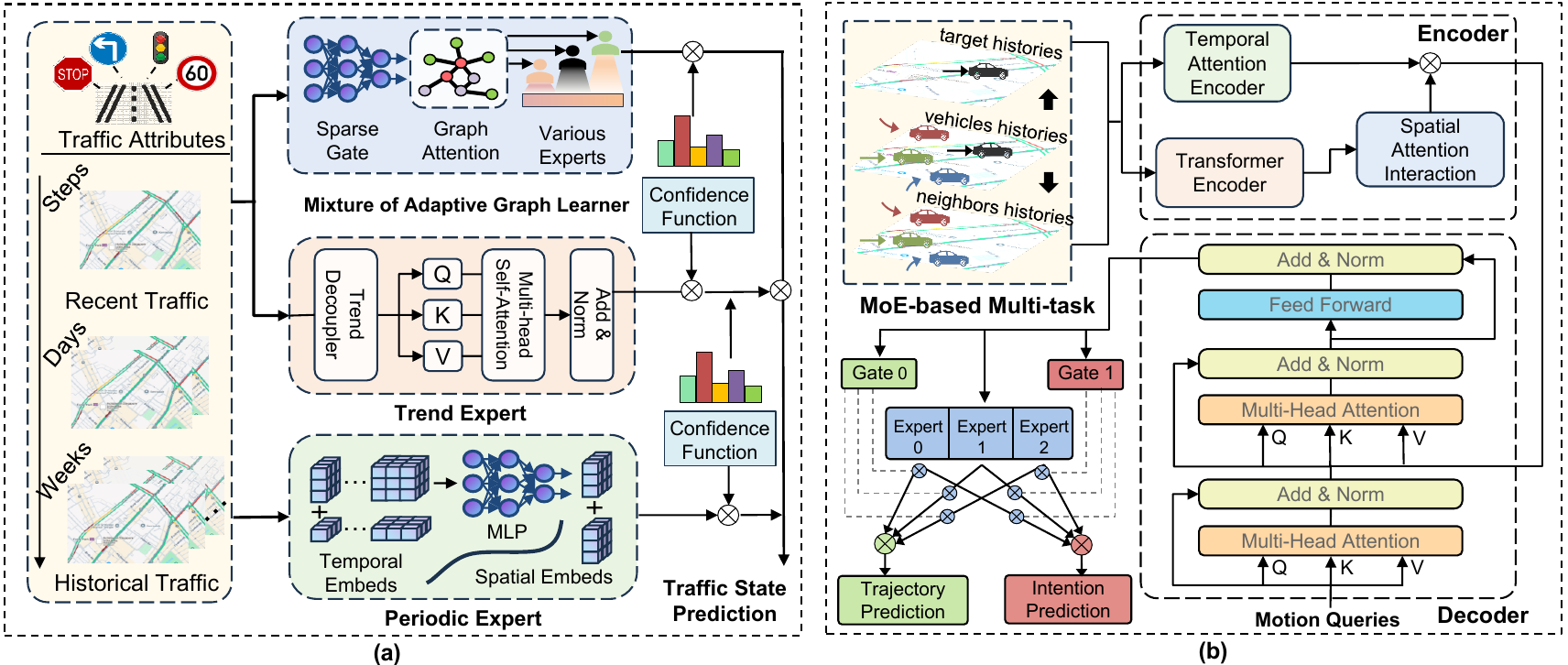} 
\captionsetup{justification=justified,format=plain}
\caption{ (a) The CP-MoE framework in \cite{jiang2024interpretable} introduces a hierarchical gating structure to model distinct temporal patterns, where experts are designated for stable traffic trends and periodic variations. 
(b) The TAME framework in \cite{jiang2024hybrid} addresses simultaneous trajectory forecasting and driver intention inference. A non-autoregressive Transformer with temporal and spatial attention encoders is utilized to capture intra-vehicle and inter-vehicle dependencies, along with an MoE-based decoder for multi-modal trajectory generation and driver behavior prediction.}

\label{Section4_traffic management}
\end{figure*}

However, while the study in \cite{chattopadhyay2022mixture} utilizes the MoE framework to enhance traffic state prediction, it does not explicitly differentiate between long-term stable patterns and short-term non-recurring fluctuations in urban traffic, limiting its adaptability to varying congestion scenarios. To address this challenge, the authors in \cite{jiang2024interpretable} propose a congestion
prediction MoE (CP-MoE) framework, which extends conventional MoE by introducing a hierarchical gating structure that explicitly models different congestion patterns. As depicted in Figure \ref{Section4_traffic management}(a), distinct experts of the CP-MoE framework specialize in learning stable traffic trends (e.g., recurring peak-hour congestion) and periodic variations (e.g., congestion caused by specific events), allowing for more refined temporal modeling and improving adaptability. A hierarchical integration with confidence functions dynamically selects and fuses experts based on traffic stability, enhancing the model’s interpretability through different expert contributions.
Extensive experiments on real-world traffic datasets validate the effectiveness of CP-MoE, achieving a 3.9\% improvement in accuracy and a 5.7\% improvement in congestion forecast tasks.
These results shed light on the applicability of MoE in spatial-temporal pattern modeling for wireless-enabled intelligent transportation systems and broader vehicular network scenarios.

Unlike the previous studies in \cite{chattopadhyay2022mixture} and \cite{jiang2024interpretable}, which focus on modeling temporal patterns and solely predicting traffic flow dynamics, the authors in \cite{jiang2024hybrid} simultaneously address trajectory forecasting and driver intention inference for multi-task processing in autonomous driving. To handle diverse driving scenarios and uncertainty in multi-modal trajectory and intention prediction, this study proposes a non-autoregressive Transformer with attention-based MoE (TAME) framework. As illustrated in Figure \ref{Section4_traffic management}(b), this framework consists of a temporal attention encoder that captures intra-vehicle and inter-vehicle dependencies and a spatial-attention interaction encoder that refines vehicle interaction representations. Besides, a multi-task prediction decoder is employed based on MoE structure to generate multi-modal trajectory predictions while simultaneously inferring driver intentions, which strengthens the model’s representational capacity across diverse motion patterns and predictive tasks. Experimental evaluations demonstrate that the proposed approach outperforms state-of-the-art motion prediction models, achieving a 28.6\% reduction in average displacement error. Additionally, the approach exhibits superior long-term prediction accuracy, with a 68\% improvement in root mean square error (RMSE) for 4-5 second predictions, showcasing its capability in handling complex, long-horizon motion forecasting tasks.

\subsubsection{Edge Computing and Distributed Systems}

\begin{figure*}
\centering
\includegraphics [width=\textwidth] 
{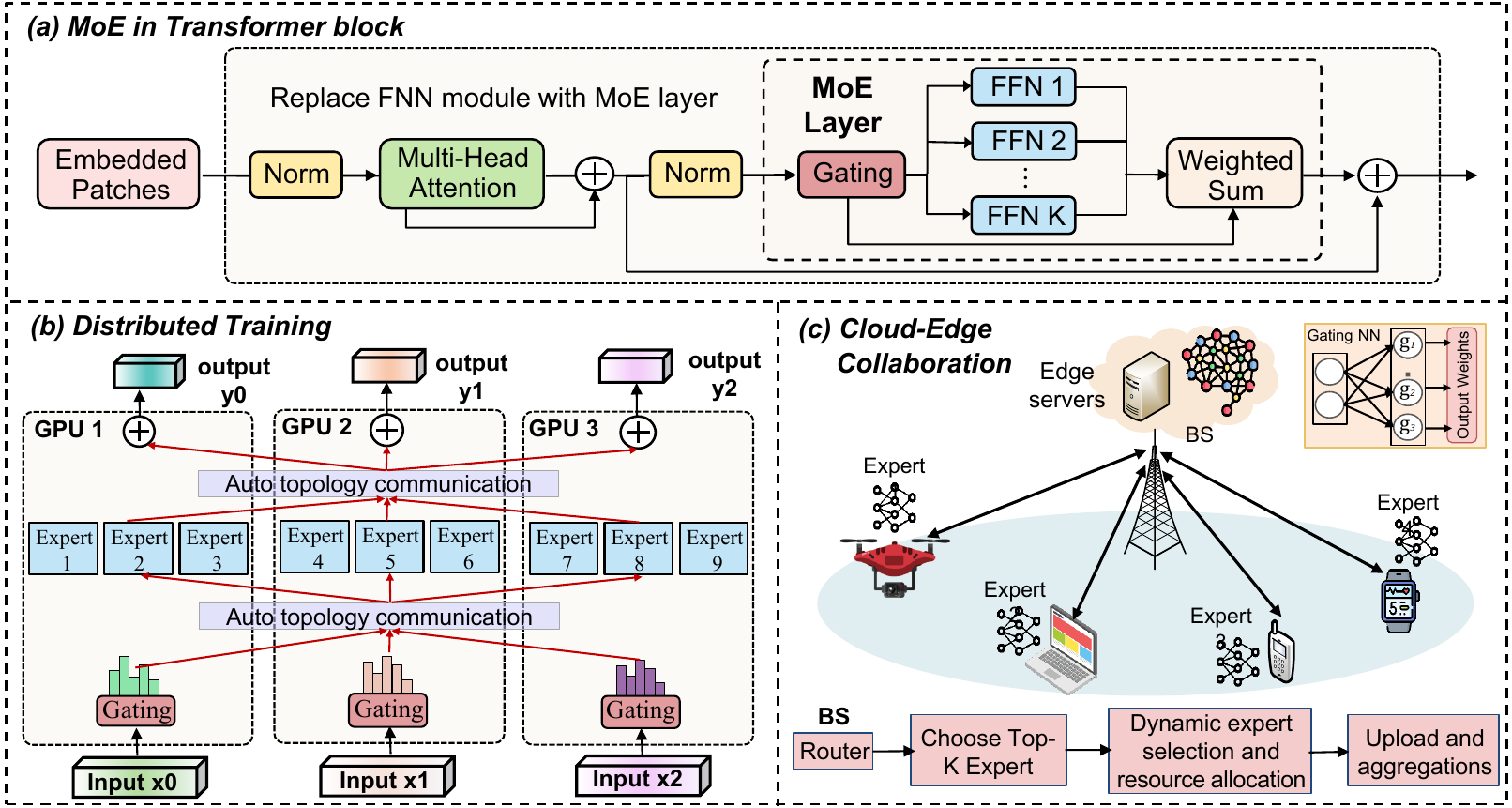} 
\captionsetup{justification=justified,format=plain}
\caption{ Illustration of MoE-based frameworks for the distributed and collaborative computing paradigm.
(a) MoE integration in Transformer blocks replaces traditional FFN with multiple smaller expert sub-networks, coordinated by a coordinated by a gating network to activate a sparse subset of experts \cite{lepikhin2020gshard}. 
(b) FlexMoE \cite{nie2023flexmoe} enables distributed training across multiple GPUs by dynamically assigning expert sub-networks to heterogeneous devices. 
(c) WDMoE architecture \cite{xue2024wdmoe} supports collaborative inference of LLMs across cloud-edge systems. Attention modules and a gating network are executed at the BS, while lightweight expert networks are distributed among edge devices.
}
\label{Section4_distributed_computing}
\end{figure*}

MoE frameworks provide an effective paradigm for parallel and collaborative processing, thereby facilitating their application in edge computing scenarios that are constrained by limited resources and specialized computational demands \cite{chen2024enabling, zhangGAI1, ma2023dynamic}. As illustrated in Figure~\ref{Section4_distributed_computing}(a), MoE was initially applied to large-scale Transformer models by replacing the original FFN layers with multiple smaller expert sub-networks \cite{lepikhin2020gshard}.
With their inherent modularity and parallelism, the frameworks of MoE have enabled large-scale training and inference across multiple devices \cite{liu2024fusion}. For instance, FlexMoE \cite{nie2023flexmoe}, a distributed MoE training framework illustrated in Figure~\ref{Section4_distributed_computing}(b), supports expert parallelism across multiple GPUs of heterogeneous devices based on workload distribution, demonstrating significant scalability of MoE frameworks for distributed model training.

Building upon the development in distributed training, MoE has been explored for inference optimization on edge devices with limited computational resources.
To enable efficient multi-task visual inference on resource-constrained edge platforms, the authors in \cite{sarkar2023edge} propose Edge-MoE, a Vision Transformer (ViT)-based architecture that leverages an MoE mechanism with task-specific experts for edge deployment. Edge-MoE integrates multiple MLP-based expert networks, each specialized for different vision tasks such as depth estimation and semantic segmentation. A compact MLP-based gating network is employed to assign expert weights and selectively activate a subset of experts based on each task and input token, significantly reducing memory usage and computational overhead. In addition,
to support real-time inference on low-memory edge inference, the authors incorporate techniques including attention reordering \cite{yang2022dtqatten}, expert-level pipelining \cite{qian2024eps}, and approximate activation functions \cite{li2017neural}, which collectively reduce memory overhead and computational latency.  
Experiments on autonomous driving benchmarks demonstrate that Edge-MoE achieves up to 18.8× reduction in inference latency and over 4× improvement in energy efficiency compared to traditional ViT baselines. 
These enhancements highlight the effectiveness of Edge-MoE in supporting real-time, resource-efficient inference in wireless edge scenarios.

However, Edge-MoE primarily focuses on computational optimization at individual edge devices, without addressing the broader challenges of distributed coordination and communication constraints in wireless environments.
To enable scalable and low-latency inference of LLMs over wireless edge networks, the authors in \cite{xue2024wdmoe} propose a wireless distributed MoE (WDMoE) architecture that facilitates collaborative execution of MoE-based LLMs across a BS-equipped server and multiple mobile devices. As depicted in the architecture of Figure~\ref{Section4_distributed_computing}(c), the computationally intensive attention modules and the MLP-based gating network for server-device collaboration are deployed at the BS, while the lightweight FNN-based expert sub-networks are distributed among mobile devices. This decomposition leverages the modularity and computational parallelism of MoE frameworks, enabling efficient utilization of distributed and heterogeneous wireless edge resources in terms of both computation and communication. Unlike conventional cloud-based LLM inference, WDMoE leverages heterogeneous wireless link qualities and device capabilities to enable dynamic expert selection and bandwidth-aware collaboration. 
To coordinate this process, an optimization framework is employed to jointly maximize expert utility and minimize overall inference latency under wireless constraints.
Simulations are conducted to compare with the Mixtral deployment model \cite{jiang2024mixtral}, which distributes experts across devices but lacks coordinated expert selection and bandwidth optimization. The results show that WDMoE achieves up to a 45.75\% reduction in latency on the Physical Interaction Question Answering (PIQA) dataset \cite{bisk2020piqa}, while maintaining high accuracy across multiple NLP benchmarks.

\begin{figure*}
\centering
\includegraphics [width=\textwidth] 
{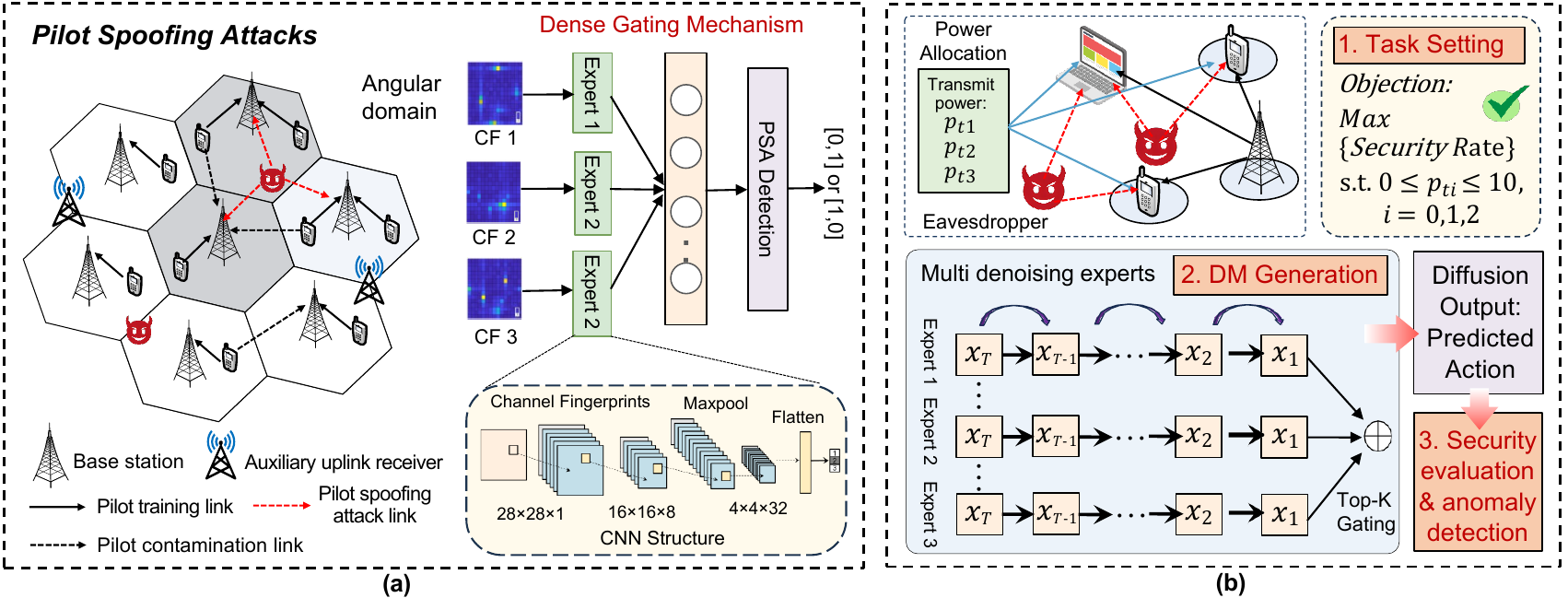} 
\captionsetup{justification=justified,format=plain}
\caption{ Illustration of MoE-based frameworks for wireless security and anomaly detection.
(a) The Distributed MoE-based Neural Network (D-MoENN) \cite{wang2024fighting} addresses pilot spoofing attacks (PSAs) in multi-cell massive MIMO systems by leveraging angular-domain channel fingerprints (CFs) collected from multiple auxiliary uplink receivers. 
(b) A generative MoE-based framework combines modular denoising experts with a diffusion model for signal-level anomaly detection, enabling adaptive denoising process across varying received signals \cite{zhao2024enhancing}. 
} 
\label{Section4_security}
\end{figure*}

While WDMoE effectively enables collaborative inference across wireless edge devices, it does not explicitly incorporate fine-grained storage management or adaptive offloading strategies, both of which are critical for efficient deployment in bandwidth and storage-constrained wireless environments. To address these challenges, the authors in \cite{yuan2024efficient} propose MedMixtral 8x7B, a fine-tuned MoE model designed for resource-efficient collaboration between wireless edge nodes and local servers. MedMixtral introduces a memory-aware inference offloading strategy that leverages heterogeneous storage hierarchies across devices, such as GPU memory, system memory, and local disk storage, to balance memory utilization and reduce inference latency. 
The model adopts a sparse MoE architecture, where 2 out of 8 FNN-based experts are activated for each token and executed on edge devices. This Top-2 gating mechanism, implemented with a single linear projection followed by a softmax function at the BS, computes expert weights based on input representations and system status. 
Experimental evaluations on real-world question and answer (Q\&A) datasets \cite{li2023chatdoctor} show that MedMixtral reduces memory consumption by up to 45.1\% and effectively improves inference latency by 32\% through the proposed offloading strategy, which allows experts to be flexibly distributed across edge devices and BS based on runtime resource availability.


\subsubsection{Security and Anomaly Detection}


Ensuring security and detecting anomalies in communication networks is increasingly challenging due to the dynamic nature of threats and the complexity of modern wireless systems. MoE frameworks offer a powerful approach to enhance security measures such as detection accuracy and secrecy rates by dynamically activating specialized experts to capture heterogeneous wireless characteristics \cite{zhao2024generative}. 
For example, the authors in \cite{wang2024fighting} propose a Distributed MoE-based Neural Network (D-MoENN) framework to address the severe security threats posed by pilot spoofing attacks (PSAs) in multi-cell massive MIMO systems. Traditional detection methods, such as hypothesis testing \cite{wang2019countermeasures} and angular-domain fingerprinting \cite{wang2019pilot}, often suffer from poor performance under low-SNR conditions and lack scalability in distributed settings.
D-MoENN leverages spatial diversity from multiple auxiliary uplink receivers, each extracting angular-domain channel fingerprints (CFs) that capture the angle-of-arrival (AoA) characteristics of both legitimate and spoofed signals. As depicted in Figure~\ref{Section4_security}(a), these CFs are processed by CNN-based local expert models, trained to estimate the number of users sharing a pilot sequence. A hidden gating network at the central BS aggregates the expert outputs via learned soft weights and produce a global prediction of the pilot reuse level used for PSA detection.
Simulation results show that the proposed D-MoENN achieves up to 90\% higher detection accuracy and 10\% improvement under low-SNR conditions (e.g., –10 dB).

While D-MoENN in \cite{wang2024fighting} focuses on wireless anomaly detection and secure transmission, network-level intrusion detection remains a critical component of 5G and beyond wireless infrastructures. 
To handle heterogeneous network traffic and evolving threat models, the authors in \cite{ilias2024convolutional} propose a sparsely-gated MoE-CNN framework for intelligent and adaptive intrusion detection. 
The proposed model first reshapes one-dimensional network flow features into 6×13 matrices, enabling spatial feature extraction through a four-layer CNN module. The resulting representation vector is then passed to an MoE layer composed of 128 fully connected experts, where a gating network dynamically activates the top-32 experts per input. To improve generalization and avoid expert collapse, the framework introduces load balancing regularization for encouraging equitable weight assignment and uniform data allocation among experts.
Evaluated on the real-world 5G-NIDD dataset \cite{samarakoon20225g}, which includes eight types of attacks collected from an operational 5G network, MoE-CNN achieves an overall intrusion detection accuracy of 99.96\%, outperforming strong baselines such as the CNN-LSTM approach. 
These results provide empirical evidence for the MoE framework in addressing diverse network attack types.

\begin{table*}
\centering
\caption{Summary of MoE Applications in Network Optimization and Security  \\ \textbf{Note:} Light blue circles indicate the MoE method, green checkmarks and red crosses represent the advantages and challenges of MoE}
\label{SectionIv:second}
\renewcommand{\arraystretch}{1.2}
{\fontsize{8.5pt}{10pt} \selectfont
\begin{tabular}{|m{0.155\textwidth}<{\centering}|m{0.045\textwidth}<{\centering}|m{0.11\textwidth}<{\centering}|m{2.6cm}|m{0.42\textwidth}<{\centering}|}
\hline
\textbf{Wireless Technology} & \textbf{Ref.} & \textbf{Wireless Task} & \textbf{MoE Framework} & \textbf{MoE Advantages \& Challenges} \\ \hline

\multirow{5}{3cm}{Traffic Management } 
& \cite{chattopadhyay2022mixture} &Traffic state prediction & GNN-based experts with MLP-based gating network & 
\multirow{1}{0.43\textwidth}{ 
\vspace{-0.27in}
\begin{itemize}[leftmargin=*] 
\item [\textcolor{blue!40}{ \resizebox{0.6em}{!}{\ding{108}}}] Use entropy-based loss to improve expert specialization
\item [\textcolor{green}{\ding{51}}] Enhance spatial-temporal relationship modeling 
\item [\textcolor{red}{\ding{55}}] High computational cost due to multiple graph experts
\end{itemize}
}
\\
 \cline{2-5}

& \cite{jiang2024interpretable} & Traffic congestion prediction & Attention experts with hierarchical gating network & 
\multirow{1}{0.43\textwidth}{ 
\vspace{-0.27in}
\begin{itemize}[leftmargin=*] 
\item [\textcolor{blue!40}{ \resizebox{0.6em}{!}{\ding{108}}}] \spaceskip=0.2em Employ multimodal MoE for spatial-temporal dependency
\item [\textcolor{green}{\ding{51}}] Enhance robustness by trend and periodic experts
\item [\textcolor{red}{\ding{55}}] Real-time inference efficiency needs improvement 
\end{itemize}
}
\\
\cline{2-5}

& \cite{jiang2024hybrid} & Vehicle trajectory prediction & MLP-based MoE in Transformer block & 
\multirow{1}{0.43\textwidth}{ 
\vspace{-0.27in}
\begin{itemize}[leftmargin=*] 
\item [\textcolor{blue!40}{ \resizebox{0.6em}{!}{\ding{108}}}] \spaceskip=0.2em Employ MoE within Transformer for trajectory prediction
\item [\textcolor{green}{\ding{51}}] Reduce error accumulation in non-autoregressive decoding
\item [\textcolor{red}{\ding{55}}] High cost due to Transformer and MoE integration
\end{itemize}
}
\\
 \hline

\multirow{6}{3cm}{Edge Computing and Distributed Systems} 
& \cite{sarkar2023edge} & Multi visual task inference & MLP-based experts and MLP-based gating network &   
\multirow{3}{0.43\textwidth}{ 
\vspace{-0.3in}
\begin{itemize}[leftmargin=*] 
\item [\textcolor{blue!40}{ \resizebox{0.6em}{!}{\ding{108}}}] MoE training at individual devices for multi-vision tasks
\item [\textcolor{green}{\ding{51}}] Reduces memory overhead and energy consumption
\item [\textcolor{red}{\ding{55}}] \spaceskip=0.2em Lacks support for distributed coordination among devices
\end{itemize}
}
\\
 \cline{2-5}

& \cite{xue2024wdmoe}  & Wireless distributed LLM inference  & FNN-based experts and MLP-based gating network & 
\multirow{3}{0.43\textwidth}{ 
\vspace{-0.3in}
\begin{itemize}[leftmargin=*] 
\item [\textcolor{blue!40}{ \resizebox{0.6em}{!}{\ding{108}}}] Gating network deployed at BS while experts at devices
\item [\textcolor{green}{\ding{51}}] Achieves collaborative inference on wireless devices
\item [\textcolor{red}{\ding{55}}] Lack storage management and offloading strategy
\end{itemize}
}

\\
 \cline{2-5}

& \cite{yuan2024efficient} &  Memory-aware LLM inference & FNN-based experts with sparse gating mechanism &  
\multirow{1}{0.43\textwidth}{ 
\vspace{-0.27in}
\begin{itemize}[leftmargin=*] 
\item [\textcolor{blue!40}{ \resizebox{0.6em}{!}{\ding{108}}}] \spaceskip=0.2em Top-2 experts based on storage hierarchies across devices
\item [\textcolor{green}{\ding{51}}] Improve memory usage and reduce inference latency
\item [\textcolor{red}{\ding{55}}] LLM inference sensitive to wireless link quality
\end{itemize}
}
\\
 \hline

\multirow{5}{3cm}{Security and Anomaly Detection} 
& \cite{wang2024fighting} & Pilot spoofing attack detection & CNN-based experts with dense gating mechanism & 
\multirow{1}{0.43\textwidth}{ 
\vspace{-0.27in}
\begin{itemize}[leftmargin=*] 
\item [\textcolor{blue!40}{ \resizebox{0.6em}{!}{\ding{108}}}] \spaceskip=0.2em CNN-based experts to process angular channel fingerprint
\item [\textcolor{green}{\ding{51}}] Achieve high detection accuracy in low-SNR regime
\item [\textcolor{red}{\ding{55}}] Performance degrades under severe fingerprint overlap
\end{itemize}
}
\\
\cline{2-5}

& \cite{ilias2024convolutional} & Network Intrusion detection & CNN-based experts with sparse gating mechanism & 
\multirow{1}{0.43\textwidth}{ 
\vspace{-0.27in}
\begin{itemize}[leftmargin=*] 
\item [\textcolor{blue!40}{ \resizebox{0.6em}{!}{\ding{108}}}] Top-32 among 128 experts for handing extracted features
\item [\textcolor{green}{\ding{51}}] Improves detection accuracy across diverse attack types
\item [\textcolor{red}{\ding{55}}] \spaceskip=0.2em Requires extensive labeled data and sensitive to data quality
\end{itemize}
}
\\
\cline{2-5}

& \cite{zhao2024enhancing} & Network Anomaly Detection & MLP-based experts with sparse gating mechanism &
\multirow{1}{0.43\textwidth}{ 
\vspace{-0.27in}
\begin{itemize}[leftmargin=*] 
\item [\textcolor{blue!40}{ \resizebox{0.6em}{!}{\ding{108}}}] \spaceskip=0.2em MoE-based diffusion model for wireless anomaly detection
\item [\textcolor{green}{\ding{51}}] \spaceskip=0.2em 
Improve accuracy through denoising expert specialization
\item [\textcolor{red}{\ding{55}}] Limited evaluation on real-world wireless datasets
\end{itemize}
}
\\
 \hline

\end{tabular}
}
\end{table*}

Complementing the discriminative approaches used in D-MoENN \cite{wang2024fighting} and MoE-CNN \cite{ilias2024convolutional}, generative modeling techniques have been adopted to 
address wireless threats such as pilot spoofing and jamming in a more flexible and data-driven manner.
However, while recent GenAI models have demonstrated potential in reconstructing perturbed signals and identifying anomalies, they often suffer from limited adaptability to heterogeneous attack scenarios and high computational overhead due to monolithic inference structures. To overcome these limitations, the authors in \cite{zhao2024enhancing} propose a modular generative framework that integrates the MoE architecture with a diffusion-based denoising process for signal-level anomaly detection.
In the proposed design illustrated in Figure~\ref{Section4_security}(b), a sparse gating mechanism dynamically activates a subset of specialized denoising expert networks based on the perturbation characteristics of the received signal. 
The activated experts enable the detection of both known and unseen anomalies under diverse SNR and channel conditions, guiding the diffusion model toward accurate signal reconstruction.
Experimental results demonstrate an average accuracy improvement of 12.3\% over the traditional diffusion model baseline, underscoring the effectiveness of combining MoE with generative modeling to achieve adaptive and robust wireless network security.

\textbf{Lessons Learned:} 
The integration of MoE into wireless systems and technologies demonstrates layer-specific strengths and functional alignment across different wireless protocol stacks.
In the foundational physical-layer communication domain, which includes channel prediction \cite{senevirathna2006channel, lopez2020channel, jaiswal2023leveraging}, signal processing \cite{van2024mean, fischer2022mixture, fischer2023sparsely, saidutta2021joint}, and radio resource management \cite{zecchin2020team, ma2022demand, du2024mixture}, MoE supports model specialization under heterogeneous and time-varying conditions, significantly improving both predictive performance and computational efficiency. In the system and network-layer domain, covering traffic management \cite{chattopadhyay2022mixture, jiang2024interpretable, jiang2024hybrid}, distributed computing \cite{sarkar2023edge, xue2024wdmoe, yuan2024efficient}, and network security \cite{wang2024fighting, ilias2024convolutional, zhao2024enhancing}, MoE enables communication-aware expert activation and scalable multi-task inference, which satisfies the growing demand for low-latency and adaptive decision-making in complex wireless environments. These developments underscore MoE’s potential in modularity, adaptability, and resource-aware design.
Nevertheless, ensuring expert diversity, avoiding mode collapse, and developing lightweight, robust gating mechanisms remain key challenges.
A principled exploration of these design aspects is envisioned to be essential for realizing intelligent, flexible, and resilient wireless infrastructures in future 6G and beyond systems.

\section{Case Study and Datasets}

In this section, we first present a case study that integrates MoE into a diffusion model-based DRL framework, demonstrating the performance enhancements of MoE for wireless network optimization. Subsequently, we provide an overview of publicly available datasets that are widely adopted in MoE-based models to support diverse machine learning tasks.

\subsection{Case Study}

In modern wireless networks, multi-BS coordinated reception has emerged as a promising access technology to enhance received signal strength and improve spectral efficiency \cite{xu2025fully}.
However, achieving the potential gains remains challenging due to the substantial computation overhead associated with jointly selecting cooperative BS and precoding schemes for each user equipment (UE) \cite{xu2023federated}. Consequently, we leverage DRL to obtain an optimal cooperative BS set and precoding matrix for multi-BS reception under MIMO transmission.

\subsubsection{Problem Description} 
As depicted in Figure~\ref{Section5_case_study}, we investigate the multi-BS reception scenario, where both the BS and UE are equipped with multiple antennas to enable spatial multiplexing and support simultaneous transmission of multiple data streams. The BS receive antennas are assumed to be a
cross-polarized uniform planar array (UPA), and the UE received antennas are assumed to be a single-polarized uniform linear array (ULA). The wireless signal propagation is characterized by the three-dimensional (3D) channel model that accounts for both horizontal and vertical propagation effects, thereby providing a more accurate representation of real-world wireless environments. 
For multi-BS coordinated reception, each UE must select a cooperative BS set from the available BSs and choose a precoding matrix $W$ from a predefined codebook. This joint selection process is computationally prohibitive, particularly in large-scale networks with dense BS deployments and high-dimensional codebooks.

\subsubsection{Solution Design} 
\begin{figure}[t]
\centering
\includegraphics [width=0.486\textwidth] 
{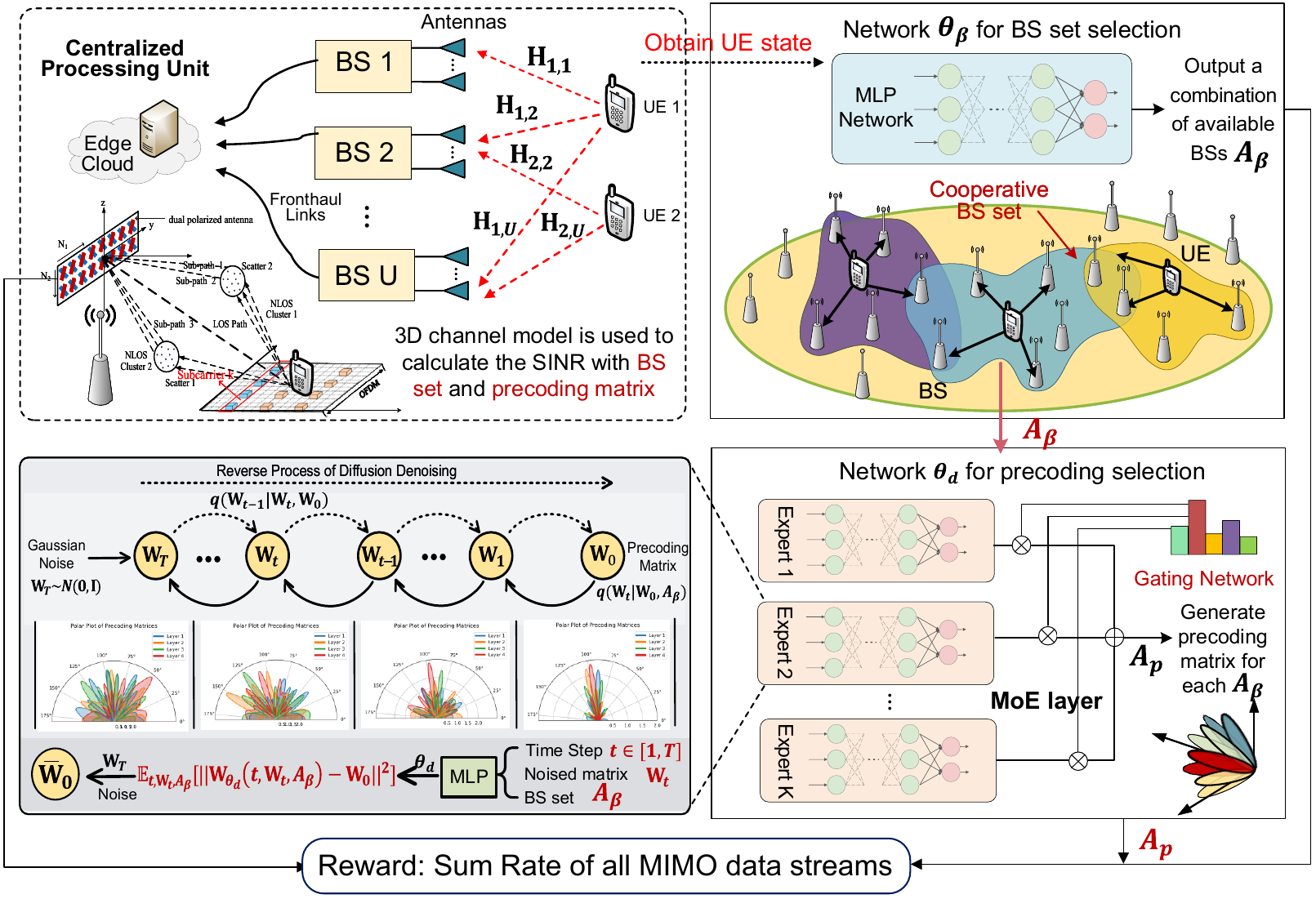} 
\captionsetup{justification=justified,format=plain}
\caption{ Multi-BS reception scenario and the proposed DRL network.
}
\label{Section5_case_study}
\end{figure}

We propose a DRL framework that employs two neural networks to independently generate the cooperative BS set and the precoding scheme as in Figure~\ref{Section5_case_study}, thereby reducing computational complexity and enabling efficient multi-BS coordinated reception. Note that once the cooperative BS set is decided, we can iterate all the satisfied precoding matrices in the codebook to obtain an optimal precoding matrix that maximizes the SINR of all data streams. The optimal precoding matrix is regarded as expert knowledge and approximated using a diffusion model, which allows the DRL framework to generalize under varying BS cooperation schemes \cite{du2024enhancing}. Nonetheless, despite the effectiveness of the diffusion model, BS selection remains a computationally intensive task, as the size of the search space increases combinatorially with the number of candidate coordinated BSs \cite{xu2025fully}. To enhance the model capacity, MoE layers are integrated into the denoising phase of the diffusion process, where a sparse gating mechanism activates appropriate experts to perform adaptive denoising and generate the precoding matrix. Specifically, the state, action, and reward function are defined as follows.

\begin{table}[t]
\small
\centering
\caption{Comparison of Network Architectures}
\begin{tabular}{lcc}
\toprule
\textbf{Method} & \textbf{MoE} & \textbf{Network Architecture}  \\
\midrule
Top-1 Expert  & 4 total experts  & \(20 \rightarrow 26 \rightarrow 32\)  \\
Top-2 Experts & 4 total experts & \(20 \rightarrow 26 \rightarrow 32\)  \\
Baseline  & w/o MoE & \(20 \rightarrow 26 \rightarrow 32\) \\
Deeper Baseline  & w/o MoE & \( 20 {\raisebox{0.19ex}{\small$\rightarrow$}} 23 {\raisebox{0.19ex}{\small$\rightarrow$}} 26 {\raisebox{0.19ex}{\small$\rightarrow$}} 29 {\raisebox{0.19ex}{\small$\rightarrow$}} 32\) \\
\bottomrule
\end{tabular}
\label{tab:network_structures}
\end{table}

\begin{itemize}
    \item State: To achieve multi-BS coordinated reception for each UE, the horizontal and vertical coordinates ${loc}_x$, ${loc}_y$ are used as the state information for the DRL framework. 
    \item Action: The action space of the DRL framework comprises two components for each UE, namely the selection of a cooperative BS set $\mathcal{A}_{\mathcal{B}}$ and the determination of a precoding matrix $\mathcal{A}_{\mathcal{P}}$. For $\mathcal{A}_{\mathcal{B}}$, we use the binomial coefficient $\binom{U}{N_i}$ to indicate the actions of selecting $N_i$ BSs from a total of $U$ available BSs for UE $i$. 
    For $\mathcal{A}_{\mathcal{P}}$, the precoding matrix is generated by the MoE-based diffusion model, which is designed to adapt to different BS selections and varying data stream conditions.
    \item Reward: The reward function is defined as the sum of post-equalized communication rates across all transmit data streams, serving as a key performance metric to evaluate the effectiveness of the set of selected cooperative BSs and the precoding matrix. 
\end{itemize}

\subsubsection{Simulation Setting and Results}

\begin{figure*}[htbp]
    \centering
    \begin{minipage}{0.462\textwidth}
        \centering
        \includegraphics[width=\linewidth]{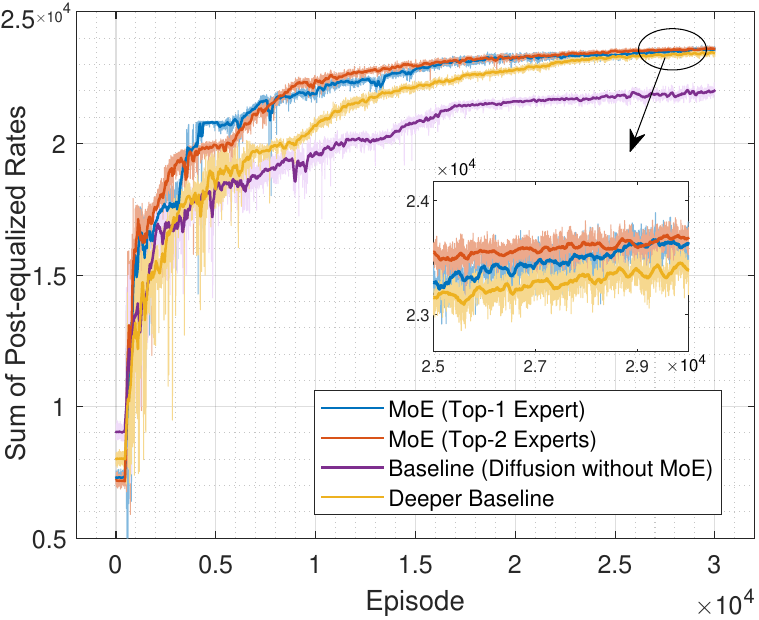}
        \captionof{figure}{Training performance of different network architectures.}
        \label{Section5_fangzhen1}
    \end{minipage}
    \hspace{0.04\textwidth} 
    \begin{minipage}{0.46\textwidth}
        \centering
        \includegraphics[width=\linewidth]{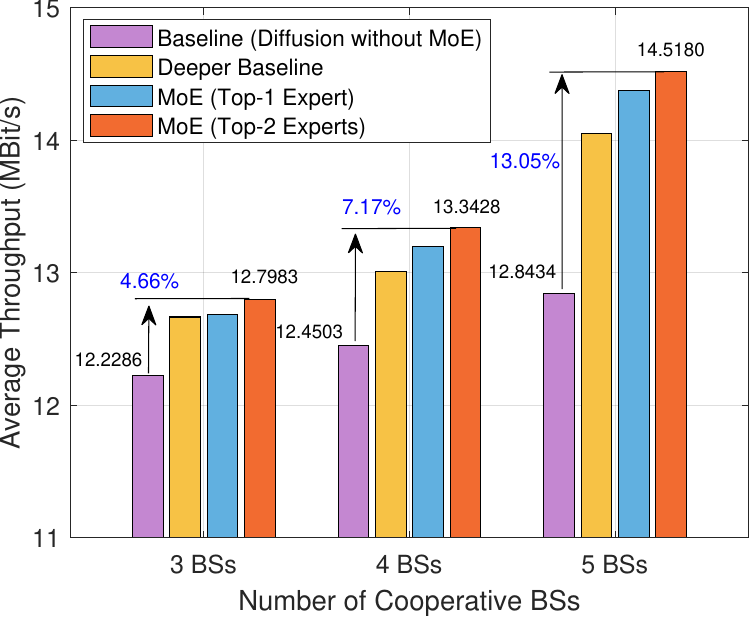}
        \captionof{figure}{Average Throughput of different network architectures.}
        \label{Section5_fangzhen2}
    \end{minipage}
\end{figure*}

To investigate the impact of network architecture on the DRL training performance, we evaluate four models comprising two MoE-based architectures and two baselines without MoE. The architectural specifications are summarized in Table~\ref{tab:network_structures}. Specifically, both the MoE-based Top-1 and Top-2 models incorporate 4 candidate MLP experts to perform the diffusion denoising process. Each expert consists of three hidden layers with dimensions 20, 26, and 32, respectively. A single-layer gating network outputs the selection probabilities for the 4 experts via a softmax function, from which either one (Top-1) or two (Top-2) experts are dynamically selected during each forward pass. In contrast, the other two models do not incorporate expert modules. The standard baseline adopts the same three-layer MLP architecture as the MoE models, whereas the deeper baseline increases the depth to five layers with hidden dimensions of 20, 23, 26, 29, and 32. This deeper configuration is designed to validate whether increased network depth can offset the lack of expert-based specialization.


Figure~\ref{Section5_fangzhen1} presents the training curves of the compared models. The simulation environment is based on the Outdoor $1$ (O$1$) scenario provided by the DeepMIMO dataset \cite{alkhateeb2019deepmimo}, where a total of $18$ BSs are available for coordinated multi-BS reception. 
It is evident from Figure~\ref{Section5_fangzhen1} that both MoE-based architectures achieve faster convergence and higher final reward compared to the other two baselines without MoE modules. The Top-2 Expert model slightly outperforms the Top-1 Expert configuration, demonstrating the benefit of activating more experts in improving learning performance. 
Additionally, for the two baseline models without MoE, the deeper baseline with increased network depth exhibits a noticeable performance improvement compared to the standard baseline. However, it still lags behind both the Top-1 and Top-2 Expert models in terms of overall performance.
Moreover, its training performance exhibits more pronounced fluctuations during the early training stages, suggesting a slower convergence attributed to the larger network scale.
These simulation results highlight the advantage of incorporating MoE to enhance the model representation capacity while reducing the computational overhead associated with DRL-based network optimization.


To further validate the effectiveness of MoE in wireless decision-making, we conducted physical-layer simulations to evaluate the throughput performance (measured in Mbit/s) of the MoE-based DRL framework under multi-BS cooperation scenarios. The simulation pipeline includes channel coding, modulation, channel equalization, demodulation, and decoding, which offers an end-to-end evaluation of real-world transmission performance.

As illustrated in Fig.~\ref{Section5_fangzhen2}, the MoE-based models with Top-1 and Top-2 expert selection consistently outperform both baselines in terms of average throughput across all cooperative BS configurations ranging from 3 to 5 BSs. It is worth noting that the number of possible cooperative combinations increases combinatorially with the number of BSs, with a complexity of approximately $\mathcal{O}(2^n)$ for $n$ cooperating BSs. As the cooperation complexity increases, the performance gains of MoE become more pronounced, rising from 4.66\% with 3 BSs to 13.05\% with 5 BSs, demonstrating MoE’s enhanced effectiveness in more complex network scenarios. These results highlight the capability of MoE to manage high-dimensional decision spaces through scalable and adaptive expert selection, particularly in cooperative transmission tasks.


\subsection{ Open-Source Datasets for MoE }

High-quality data plays a crucial role in training large-scale AI models and evaluating their performance. In this subsection, we provide an overview of the publicly available datasets that are utilized by MoE-based models in the fields of NLP, computer vision, multimodal, and wireless networks. A summary of these datasets is presented in Table~\ref{table dataset:1}.

\subsubsection{Datasets for Natural Language Processing}

In the domain of NLP, datasets such as GLUE \cite{wang2018glue} and C4 \cite{raffel2020exploring} have been widely used within different MoE architectures, e.g., LiteMoE \cite{zhuang2024litemoe} and MoE-based LLM \cite{huang2024mc}, to address diverse language understanding and text generation tasks. GLUE serves as a comprehensive NLP benchmark, supporting sentiment analysis, sentence similarity, and natural language inference, providing a robust ground for various linguistic model generalization and evaluation. Meanwhile, C4, referred to as the Colossal Clean Crawled Corpus, delivers an extensive collection of high-quality text curated from the Common Crawl project \footnote{https://commoncrawl.org}. The hundreds of gigabytes of text from C4 and the diverse language tasks on GLUE allow MoE models to learn specialized expert sub-networks, each proficient in distinct aspects of language representation and usage.
Additionally, domain-specific datasets such as TeleQnA \cite{maatouk2023teleqna}, comprising 10,000 Q\&A instances in telecommunications, 
facilitate the integration of MoE experts to support compositional reasoning and effective domain adaptation \cite{li2024expert}.

\subsubsection{Dataset for Computer Vision}

Computer vision datasets such as ImageNet \cite{deng2009imagenet}, nuScenes \cite{caesar2020nuscenes}, and DOTA \cite{xia2018dota} have been extensively utilized by MoE-based models for large-scale image classification and object detection across diverse scenarios. ImageNet serves as a foundational dataset for generic visual recognition, providing 3.2 million labeled images organized by the WordNet hierarchy \cite{fellbaum2010wordnet} across a broad spectrum of object categories. This large-scale, structured dataset enables MoE-based models to learn diverse hierarchical features in both supervised and transfer learning settings \cite{gross2017hard}. Meanwhile, nuScenes supports 3D detection and tracking using multimodal inputs such as cameras, lidars, and radars. The rich annotations and diverse sensor modalities on nuScenes facilitate the training of specialized experts for distinct perception tasks such as autonomous driving \cite{xia2018dota}. Furthermore, designed for object detection in aerial scenes, the DOTA dataset provides high-resolution aerial imagery with substantial variations in scale, orientation, and scene complexity. Its fine-grained, arbitrarily oriented bounding boxes offer a rigorous benchmark for evaluating advanced MoE-based detection in aerial and remote sensing tasks \cite{li2024sm3det}.

\subsubsection{Dataset for Multimodal}

\begin{table*}
\centering
\caption{Open-Source Dataset Utilized by MoE in NLP, Vision, Multimodal, and Wireless Networks}
\label{table dataset:1}
{\fontsize{8.5pt}{7.5pt} \selectfont
\begin{tblr}{ width = \linewidth,
  colspec = {Q[1.5,c,m] Q[1.5,c,m] Q[5,l,m] Q[3.2,l,m] Q[1.9,c,m]},
  row{1} = {c},
  hlines,
  vline{2-5} = {-}{},
  vline{2} = {2}{-}{},
}
\hline
\textbf{Task Domain} & \textbf{Dataset} & \textbf{Description} & \textbf{Composition} & \textbf{Application} \\
\SetCell[c=5]{c} NLP\\
\SetCell[r=2]{c} Language Processing  &
GLUE \cite{wang2018glue} &
A language understanding benchmark across tasks such as sentiment analysis, paraphrase detection, and natural language inference
&
857,000 samples of sentences or sentence pairs and 9 NLP tasks &
LiteMoE \cite{zhuang2024litemoe} \\


&
C4 \cite{raffel2020exploring}   &
A large-scale corpus derived from web pages and filtered for cleaned  content &
  750GB of clean English text  &
MC-MoE LLM \cite{huang2024mc}  \\

Question and Answer  &
  TeleQnA \cite{maatouk2023teleqna}   &
Knowledge of LLM in lexicon, research overview, publications, standards overview, and specifications &
  10,000 questions distributed across 5 categories  &
ETR-MoE \cite{li2024expert}  \\
\end{tblr}

\begin{tblr}{ width = \linewidth,
  colspec = {Q[1.5,c,m] Q[1.5,c,m] Q[5,l,m] Q[3.2,l,m] Q[1.9,c,m]},
  row{1} = {c},
  hlines,
  vline{2-5} = {-}{},
  vline{2} = {2}{-}{},
}
\SetCell[c=5]{c} Computer Vision\\
Common Image &
%
ImageNet \cite{deng2009imagenet} & 
Large-scale image dataset organized by the WordNet hierarchy for visual recognition tasks &
3.2 million images, 5,247 categories, 500-1,000 images per category
&
  Hard MoE \cite{gross2017hard} \\
Vehicular Image &
nuScenes \cite{caesar2020nuscenes} & 
Autonomous driving dataset including camera, lidar, and radar, annotated for 3D object detection and tracking
&
1.4 million images, 400,000 LiDAR sweeps, 1.3 million radar frames, 23 object classes &
LiMoE \cite{xu2025limoe} \\
%
%
\SetCell[r=1]{c} Aerial Image &
DOTA \cite{xia2018dota} & 
A large-scale aerial image dataset with oriented object annotations for real-world remote sensing applications
 &
2806 images, 188,000 objects, 15 classes  &
 SM3Det \cite{li2024sm3det}   \\
\end{tblr}

\begin{tblr}{ width = \linewidth,
  colspec = {Q[1.5,c,m] Q[1.5,c,m] Q[5,l,m] Q[3.2,l,m] Q[1.9,c,m]},
  row{1} = {c},
  hlines,
  vline{2-5} = {-}{},
  vline{2} = {2}{-}{},
}
\SetCell[c=5]{c} Multimodal\\
\SetCell[r=2]{c}
Image-Text&
%
MS COCO \cite{lin2014microsoft} & 
Natural images with human-written captions and detailed visual annotations including detection boxes, masks, and key points &
330,000 images, 5 captions per image, and annotations for 80 object categories
&
  LIMOE \cite{mustafa2022multimodal} \\
 &
LAION-5B \cite{schuhmann2022laion} & 
A large-scale dataset includes image-text pairs in multiple languages, filtered based on CLIP similarity scores
&
5.85 billion image-text pairs  &
RAPHAEL \cite{xue2023raphael} \\
%
%
\SetCell[r=1]{c} Vedio-Text & 
HowTo100M \cite{miech2019howto100m} &
Video–text data and ASR transcripts collected from narrated YouTube instructional videos
 &
1.22 million videos, 136 million clip-caption pairs, 23,000+ task types
& 
 Automatic speech recognition \cite{wu2024robust}   \\
\end{tblr}

\begin{tblr}{ width = \linewidth,
  colspec = {Q[1.5,c,m] Q[1.5,c,m] Q[5,l,m] Q[3.2,l,m] Q[1.9,c,m]},
  row{1} = {c},
  hlines,
  vline{2-5} = {-}{},
  vline{2} = {2}{-}{},
}
\SetCell[c=5]{c} Wireless Networks\\
  Physical Layer Signal Processing &
%
RadioML 2018.01a \cite{o2018over} & 
Open dataset for automatic modulation classification in wireless signals  &
24 modulation schemes, 26 SNR levels, 1024-sample IQ data per instance
&
 MOE-AMC \cite{gao2023moe} \\
Wireless Channel  &
DeepMIMO \cite{alkhateeb2019deepmimo}  &
A configurable ray-tracing-based dataset for deep learning in mmWave and massive MIMO systems
&
12 base stations, 497,931 users, 1 blockage surface, and 2 reflector surfaces  &
MoE-based mmWave beam selection \cite{isaksson2023mmwave} \\
%
5G Network &
5G NIDD \cite{wang2018glue} &
A labeled intrusion detection dataset collected from a real 5G test network, supporting ML-based security research
&
1.2 million flows, 9 attack types, and 1 benign traffic &
MoE-based intrusion detection \cite{ilias2024convolutional} \\

\SetCell[r=1]{c} Vehicular Network &
CitySim \cite{zhang2023citysim} &
A high-precision vehicle trajectory dataset over diverse road scenarios, designed for intelligent Internet of Vehicle research.
 &
25,000+ vehicle trajectories covering highway, urban, and intersection scenes   &
MoE-based vehicle trajectory prediction \cite{yuan2023temporal}   \\
\hline
\end{tblr}


%

%
}
\end{table*}

In the domain of multimodal research, particularly in image-text and video-text generation tasks, datasets such as MS COCO \cite{lin2014microsoft}, LAION-5B \cite{schuhmann2022laion}, and HowTo100M \cite{miech2019howto100m} have been widely adopted to align visual and textual modalities.
Specifically, MS COCO provides 330,000 images, each annotated with five human-written captions, which is commonly applied for fine-grained analysis of visual scenes in tandem with language annotations \cite{mustafa2022multimodal}. LAION-5B is an open-source large-scale dataset containing 5.85 billion CLIP-filtered image-text pairs \cite{radford2021learning} across multiple languages, offering unprecedented scale and diversity for training and evaluating multimodal models in tasks such as retrieval, captioning, and zero-shot classification \cite{xue2023raphael}. Meanwhile, HowTo100M contains 1.22 million narrated instructional videos with 136 million CLIP-caption pairs and automatic speech
recognition (ASR) transcripts, enabling multimodal learning for video-text alignment across 23,000+ task types. 
Multimodal datasets often present challenges in aligning modalities with different sequence lengths, such as text-image or text-video pairs \cite{yang2020image}.
MoE-based models have demonstrated the ability to perform robust cross-modal matching strategies through specialized experts trained on these multimodal datasets, enabling effective handling of text-image or text-video alignment within a unified framework \cite{wu2024robust}. 

\subsubsection{Dataset for Wireless Networks}

Datasets for wireless networks such as RadioML \cite{o2018over}, DeepMIMO \cite{alkhateeb2019deepmimo}, 5G-NIDD \cite{samarakoon20225g}, and CitySim \cite{zhang2023citysim} serve diverse wireless scenarios ranging from physical-layer signal processing to system-level intrusion detection and vehicular networking. In particular, RadioML offers labeled in-phase/quadrature (IQ) signal recordings, which are utilized by the MoE-based automatic modulation classification model to capture distinctive signal features under varying SINR scenarios \cite{gao2023moe}. 
Meanwhile, DeepMIMO, built on accurate ray-tracing simulations, provides realistic mmWave and massive MIMO channels tailored to beam selection, channel estimation, and large-antenna array optimizations \cite{isaksson2023mmwave}.
For network security, 5G-NIDD presents comprehensive network intrusion detection logs with fully labeled traffic traces derived from operational real-world 5G testbeds. This dataset allows MoE-based intrusion detection systems to differentiate between benign connections, newly emerging threats, and zero-day exploits, reducing false alarms and improving detection agility \cite{ilias2024convolutional}. Additionally, the CitySim dataset focuses on connected and autonomous vehicles, delivering drone-captured vehicular trajectories that include complex mobility patterns, thereby supporting safety-oriented and MoE-based wireless research for connected and intelligent transportation systems \cite{yuan2023temporal}.

Datasets for wireless networks such as RadioML \cite{o2018over}, DeepMIMO \cite{alkhateeb2019deepmimo}, 5G-NIDD \cite{samarakoon20225g}, CitySim \cite{zhang2023citysim},W ViWi-Drone \cite{alrabeiah2020viwi} and DroneRF \cite{allahham2019dronerf} serve diverse wireless scenarios ranging from physical-layer signal processing to system-level intrusion detection, vehicular networking, and UAV-assisted communications. In particular, RadioML offers labeled in-phase/quadrature (IQ) signal recordings, which are utilized by MoE-based automatic modulation classification (AMC) models to capture distinctive signal features under varying SINR scenarios \cite{gao2023moe}. DeepMIMO, built on accurate ray-tracing simulations, provides realistic mmWave and massive MIMO channels tailored to beam selection, channel estimation, and large-antenna array optimizations \cite{isaksson2023mmwave}. For network security, 5G-NIDD presents comprehensive intrusion detection logs with fully labeled traffic traces derived from operational real-world 5G testbeds. This dataset allows MoE-based systems to differentiate between benign connections, newly emerging threats, and zero-day exploits, thus reducing false alarms and improving detection agility \cite{ilias2024convolutional}. CitySim focuses on connected and autonomous vehicles, delivering drone-captured vehicular trajectories with complex mobility patterns to support safety-oriented and MoE-based research for intelligent transportation systems \cite{yuan2023temporal}. Additionally, ViWi-Drone and DroneRF provide realistic air-to-ground and UAV-to-UAV channel measurements, enabling MoE-based models to address dynamic beam prediction, link adaptation, and mobility-aware resource allocation in low-altitude network scenarios \cite{xu2025enhancing}.

\section{Future Research Directions}

MoE has demonstrated promising advances across various wireless scenarios. Nonetheless, several open research directions remain to be explored for fully utilizing its capabilities. 
This section outlines key future directions, categorized into improving existing solutions and exploring new methodologies to extend MoE's capabilities in intelligent wireless systems.
\subsection{Improving Existing Solutions}
\subsubsection{Lightweight and Resource-Efficient MoE Architectures}
While MoE frameworks offer substantial capacity and adaptability, the intensive computation and memory demands remain challenging for edge devices, such as IoT nodes and UAVs that are power or storage constrained. 
To overcome these limitations, parameter-reduction strategies, such as expert pruning \cite{kim2021pqk}, quantization-aware training \cite{huang2023efficient}, and knowledge distillation \cite{kim2019qkd}, present promising solutions to selectively retain critical expert network weights, thereby reducing the computational overhead. In addition, collaborative model execution across edge devices and cloud servers offers a compelling direction, where lightweight expert modules are deployed locally for low-latency processing, and more complex components are offloaded to centralized infrastructure.

\subsubsection{Dynamic Gating Mechanisms}
Dynamic gating represents a promising direction for enhancing the adaptability and efficiency of MoE frameworks in wireless communication systems. Future developments may focus on context-aware gating, where expert selection is conditioned on environmental variables such as user location, mobility patterns, spectrum availability, and signal-to-noise ratio, enabling adaptive prioritization of experts under congestion or energy constraints \cite{song2025mixture}. Reinforcement learning-based gating can further optimize expert activation by dynamically adjusting policies according to real-time feedback from network performance metrics such as throughput, latency, and energy consumption \cite{wu2025mixture}. In addition, multi-task and multi-modal adaptive gating can facilitate shared experts across heterogeneous tasks and modalities while maintaining specialization \cite{guo2024dynamic}. For example, in ISAC scenarios, expert participation can be adapted depending on whether the task involves spectrum allocation, beamforming, or target detection. Finally, meta-gating and other self-evolving architectures, empowered by meta-learning, allow the gating policy itself to adapt efficiently to new operational contexts without retraining from scratch \cite{zhong2022meta}. Collectively, these techniques can yield more intelligent and responsive MoE systems, particularly under the stringent performance and adaptability requirements anticipated in 6G and beyond.

\subsubsection{Communication-Efficient MoE in Distributed Wireless Networks}
Real-time processing is a critical requirement for distributed MoE systems deployed in time-sensitive applications such as vehicular networks and ISAC systems. However, the inherent latency of inter-agent coordination, expert selection, and output aggregation poses significant challenges to meeting stringent timing requirements. Emerging 6G technologies offer promising solutions to mitigate these constraints. Terahertz (THz) communications and extremely-large scale MIMO (XL-MIMO) provide ultra-high data rates and ultra-low transmission delays, enabling near-instantaneous exchange of gating signals and expert outputs. Cell-free massive MIMO eliminates the need for frequent handovers and supports seamless expert migration across distributed nodes, thereby reducing coordination delays in highly dynamic network topologies. Furthermore, advanced ISAC-enhancement techniques, such as joint beamforming for sensing and communication, wideband sensing, and predictive channel estimation, improve real-time responsiveness by delivering rapid environmental awareness and enabling context-driven expert activation. Leveraging these technologies allows distributed MoE frameworks to reduce expert selection latency, accelerate inter-agent communication, and deliver inference results within the strict quality requirements of 6G-enabled mission-critical services.


\subsubsection{Multimodal MoE for Integrated Wireless Networks}
While MoE frameworks have exhibited remarkable performance in multimodal learning tasks involving image-text or video-text data, their application in wireless networks has predominantly remained within unimodal scenarios, typically confined to tasks such as radio frequency signal processing \cite{gao2023moe}. 
By incorporating auxiliary data sources such as environmental imagery, inertial sensor measurements, or high-level semantic user intents, expert networks can be designed to extract complementary features across distinct data patterns. In parallel, adaptive gating mechanisms can enable expert selection based on different modality availability, thereby supporting more accurate channel estimation, adaptive beamforming, and interference suppression. Wireless multimodal synthesis holds significant potential to enhance situational awareness and advance the development of communication networks.

\subsection{Exploring New Methodologies}

\subsubsection{Cross-Layer Coordination and Optimization}
Current utilization of MoE in wireless protocol stacks conventionally focuses on isolated protocol layers, such as physical‑layer communication or MAC‑layer access control. A promising research direction lies in developing cross‑layer MoE architectures that jointly optimize different layer protocol tasks, including link adaptation, congestion control, traffic scheduling, and routing strategy, within a unified decision‑making framework. Such cross‑layer integration facilitates holistic optimization and ensures that decisions made at individual layers align with global system goals, enhancing overall performance such as spectral efficiency, latency reduction, and reliability.

\subsubsection{Hardware Acceleration and Deployment of MoE}
Beyond the domain of algorithmic design, advancing MoE for wireless applications requires addressing challenges related to hardware compatibility and system‑level integration. Exploring co‑optimized hardware architectures, such as AI accelerators with embedded MoE gating logic \cite{sarkar2023edge} or CPU/GPU design pipelines tailored for sparse computation \cite{dai2024deepseekmoe}, provides efficient support for expert selection and conditional MoE execution. Scalable deployment of MoE under such conditions demands intelligent runtime orchestration frameworks capable of adapting to heterogeneous hardware capabilities, dynamic network topologies, and evolving wireless infrastructures.


\subsubsection{Novel Paradigms in IoT, Digital Twins, and Beyond}
As wireless networks evolve toward emerging paradigms such as digital twins, cognitive IoT, and zero‑touch network management \cite{benzaid2020ai}, MoE frameworks are envisioned to serve as a fundamental approach for dynamic adaptability and low‑latency responsiveness. By tailoring gating mechanisms to contextual factors such as user mobility, service‑level agreements, and energy‑efficiency constraints, next‑generation wireless systems can more effectively coordinate heterogeneous operational objectives with minimal resource overhead, thereby advancing toward context‑aware and data‑driven network intelligence.


\section{Conclusion}

This survey has comprehensively reviewed the integration of MoE in wireless networks, demonstrating its substantial advantage in enhancing adaptability and improving efficiency across various wireless scenarios. By systematically exploring fundamental MoE methodologies, including various gating mechanisms and their integration with GenAI and RL, the survey has extensively explored applications across key wireless domains such as vehicular networks, UAVs, satellite networks, HetNets, ISAC, and mobile edge computing. Furthermore, critical wireless network tasks, including physical layer communications, radio resource management, network optimization, and security, have been thoroughly discussed. A case study integrating MoE into a diffusion-based DRL framework has revealed its empirical advantages in wireless optimization tasks. 
Additionally, an overview of open-source datasets has been presented to support ongoing MoE-related experimentation. Future directions should focus on lightweight architectures, balanced expert allocation, and integration with emerging paradigms, facilitating MoE as a pivotal component for next-generation wireless systems.

\bibliographystyle{IEEEtran}
\bibliography{IEEEabrv,Ref}

\end{document}